\documentclass{aa}  
\usepackage{graphicx}
\usepackage{txfonts}
\usepackage{color}
\usepackage{natbib}
\begin{document}
    \title{A derivation of nano-diamond optical constants: \\
    }
    \subtitle{Here be nano-diamonds}

    \author{A.P. Jones,  
           \and
           N. Ysard,   
             }

    \institute{Universit\'e Paris-Saclay, CNRS,  Institut d'Astrophysique Spatiale, 91405, Orsay, France.\\
               \email{anthony.jones@universite-paris-saclay.fr}
              }

    \date{Received ? : accepted  ?}

   \abstract
{Nano-diamonds are an enticing and enigmatic dust component yet their origin is still unclear. They have been unequivocally detected in only a few astronomical objects, yet they are the most abundant of the pre-solar grains, both in terms of mass and number.}
{Our goal is to derive a viable set of nano-diamond optical constants and optical properties to enable their modelling in any type of astrophysical object where, primarily, the local (inter)stellar radiation field is well-determined.} 
{The complex indices of refraction, $m(n,k)$, of nano-diamonds, constrained by available laboratory measurements, were calculated as a function of size, surface hydrogenation, and internal (dis)order, using the THEMIS a-C(:H) methodology optEC$_{\rm (s)}$(a).}
{To demonstrate the utility of the optical properties (the efficiency factors $Q_{\rm ext}$, $Q_{\rm sca}$, and $Q_{\rm abs}$), calculated using the derived $m(n,k)$ data, we show that nano-diamonds could be abundant in the interstellar medium (ISM) and yet remain undetectable there.}
{The derived optical constants provide a means to explore the existence and viability of nano-diamonds in a wide range of astronomical sources. Here we show that up to a few percent of the available carbon budget could be hidden in the form of nano-diamonds in the diffuse ISM, in abundances comparable to the pre-solar nano-diamond abundances in primitive meteorites.}

    \maketitle
%

\section{Introduction}

Pre-solar nano-diamonds with median radii of $1.3-1.5$\,nm \citep{1996GeCoA..60.4853D} have been extracted from primitive meteorites in abundances of up to $\simeq 1400$\,ppm \citep{1995GeCoA..59..115H}. Their isotopically anomalous Xe content (Xe-HL) is considered characteristic of the nucleosynthetic processes in supernovae \citep{Lewis_etal_1987},  and their  $^{15}$N depletion and low C/N ratios are consistent with carbon-rich stellar environments \citep{1997AIPC..402..567A}. The observation of a tertiary carbon CH stretching mode at $3.47\,\mu$m towards dense regions was originally thought to be an indicator of the presence of such nano-diamonds in the interstellar medium (ISM) \citep{1992ApJ...399..134A} but this band most likely has another origin \citep{1996ApJ...459..209B}. However, today we have conclusive observational evidence for the presence of nano-diamonds in proto-planetary discs where they are identified by their characteristic CH stretching modes at $3.43$ and $3.53\,\mu$m 
\cite[{e.g.},][]{1999ApJ...521L.133G,2002A&A...384..568V,2004ApJ...614L.129H}. 

In order to be able to model, analyse, and interpret the spectra of nano-diamonds observed in circumstellar proto-planetary discs we require well-determined wavelength- and size-dependent optical constants: the complex indices of refraction $m_{(n,k)}(\lambda, a,b,c)$ (where $a$, $b$, $c$ are the semi-major axes of an ellipsoidal particle). These can then be used to calculate their optical properties ($Q_i$, $i =$ ext,  sca, abs or pr) and thence their extinction, scattering, absorption, and radiation pressure cross-sections ($\sigma_i = \pi [abc]^{\frac{2}{3}}Q_i$), which in turn can be used to derive their temperatures and the radiation pressure force acting on them in a given stellar radiation field.  Here we concern ourselves with approximately spherical nano-diamond particles ({i.e.}, with $a \simeq b \simeq c$) of radius $a_{\rm nd} = [abc]^{\frac{1}{3}}$.   

It is evident that the particle optical properties are affected by the inherent particle structure ({e.g.}, perfect diamond lattice, surface re-structured or irradiated bulk diamond lattice, nitrogen heteroatom content, specific density, etc.) and so here we consider the most probable nano-diamond shape, the corresponding surface hydrogenation states and the effects of irradiation upon the particle bulk matter. In an accompanying paper the nano-diamond surface CH and CH$_2$ group abundance ratios, [CH]/[CH$_2$], were derived as a function of  particle shape and size \citep{2020_Jones_nd_CHn_ratios}. There it was concluded that, in order to be consistent with observations, the most probable nano-diamond shape must be quasi-spherical.

The products of this nano-diamond modelling are the real and imaginary parts of the complex index of refraction, $n$ and $k$, respectively, as a function of the wavelength, $\lambda$, particle radius, $a_{\rm nd}$, the nature of the surface-passivating CH and CH$_2$ groups, and bulk (dis)order. These are then used to investigate whether nano-diamonds could be present and (un)detectable in the diffuse ISM. In a follow-up paper \cite{2020_Jones_nds_in_discs} we use these same data to derive the likely temperatures, drift velocities, and lifetimes of nano-diamonds in the limited number of astronomical objects where nano-diamonds have been undeniably observed to date. 

This paper is structured as follows: 
Section \ref{sect_properties} describes some key nano-diamond physical properties, 
Section \ref{sect_optEC_model} gives the basis of the model, 
Section \ref{sect_nk} details how the optical constants are derived,
Section \ref{sect_surf_results} gives the results for fully and partially hydrogenated nano-diamonds and also for fully dehydrogenated nano-diamonds, both with and without irradiated cores, 
Section \ref{sect_nd_ISM} considers the possibility of nano-diamonds in the diffuse ISM,
Section \ref{sect_nd_proc} briefly looks at the likely thermally activated processes acting on nano-diamonds, 
Section \ref{sect_results} discusses the results and speculates upon some of their consequences,
Section \ref{sect_C_budget} gives our current best estimates of the partition of carbon across the major interstellar dust components, and
Section \ref{sect_conclusions} presents the conclusions.

\section{Some nano-diamond physical properties}
\label{sect_properties}

The diagnostic potential of interstellar and circumstellar dust species, informed by their laboratory and modelling analogues, is principally driven by their characteristic infrared spectra. In order to fully utilise this potential in the case of aliphatic-rich hydrocarbon materials, {i.e.}, a-C:H nano-particles, and particularly in the case of nano-diamonds, we need to understand how particle size, composition, and morphology determine their spectral characteristics. For nano-diamonds there is a wealth of laboratory data to aid us \citep[{e.g.},][]{1989Natur.339..117L,1994A&A...284..583C,1995MNRAS.277..986K,1995ApJ...454L.157M,Reich:2011gv,1998A&A...330.1080A,2000M&PS...35...75B,1998A&A...336L..41H,2002JChPh.116.1211C,2002ApJ...581L..55S,2004A&A...423..983M,Jones:2004fu,2007ApJ...661..919P,2011ApJ...729...91S,Usoltseva:2018er,1997PhRvB..55.1838Z}. The hope is that we can, at the very least, use these data to enable us to determine nano-diamond sizes in interstellar media through the ratio of the $3.43\,\mu$m and $3.53\,\mu$m band strengths \citep[{e.g.},][]{2002JChPh.116.1211C,2002ApJ...581L..55S,2007ApJ...661..919P,2020_Jones_nd_CHn_ratios} and, as these latter works show, this ratio is indeed size dependent. However, this band ratio is also morphology-dependent \citep[{e.g.},][]{2007ApJ...661..919P,2020_Jones_nd_CHn_ratios} and does depend upon the nature and degree of the surface hydrogen coverage \citep{2020_Jones_nd_CHn_ratios}. Further, and given that nano-diamonds are observed in emission close to hot stars, this ratio will be temperature-dependent because of the underlying thermal emission continuum and  the possibility of differential de-hydrogenation rates from CH and CH$_2$  surface sites. 

In the absence of dehydrogenation effects, (dis)proportionate or otherwise, \cite{2020_Jones_nd_CHn_ratios} studied the structures and  surface abundances of CH and CH$_2$ groups on the fully-hydrogenated surfaces of (semi-)regular euhedral nano-diamond particles\footnote{The studied euhedral particle shapes were: (semi-)regular tetrahedra, truncated tetrahedra, octahedra, truncated octahedra, cuboctahedra, cubes, and truncated cubes.} with well-defined \{111\} and \{100\} crystalline facets. This study also encompassed `spherical' nano-diamonds and derived the surface CH$_n$ abundance ratios ([CH]/[CH$_2$]) for all particle types as a function of size. It was found that euhedral nano-diamonds exhibit a huge dispersion in the [CH]/[CH$_2$] ratio and that this dispersion increases with size. There is also a large, but more limited, size-to-size dispersion in this ratio for small spherical nano-diamonds ($a < 2$\,nm), which appears to converge to a limit that falls well within the observed range for larger sizes ($a > 2$\,nm). It was therefore concluded that spherical particles appear to be the best match to the laboratory and astronomical data and that, further, the best hope of determining nano-diamond sizes from observations is by assuming spherical nano-diamonds and adopting a statistically-averaged, simple analytical expression for the size-dependence of their [CH]/[CH$_2$] ratios, {i.e.}, 
\begin{equation}
\frac{{\rm [CH]}}{{\rm [CH_2]}} = 2.265 \, \left( \frac{ a_{\rm nd}}{ 1\,{\rm nm} }\right)^{0.03} - \frac{ 1 }{ 2.5 } \left( \frac{ a_{\rm nd}}{ 1\,{\rm nm} } \right)^{-1}, 
\label{eq_ratio_fit}
\end{equation}
which is a by-eye fit to the inevitable `ups and downs' of the exact diamond network calculations at small sizes \citep[$a < 5$\,nm, see Figs. 9 and 10 in ][]{2020_Jones_nd_CHn_ratios}. The diamond network calculations are based on the standard C$-$C bond lengths and so are restricted to assuming the $3.52$\,g\,cm$^{-3}$ bulk density for diamond \citep{2020_Jones_nd_CHn_ratios}.


Given that in excited regions the nano-diamond surface hydrogenation may be less than complete, {i.e.},  close to hot stars where the nano-diamonds may undergo extreme heating, a fractional surface H atom coverage factor, $f_{\rm H}$, is introduced, with $0 \leq f_{\rm H} \leq 1$, which is assumed to be the same for both \{100\} and \{111\} facets, {i.e.}, there exist no disproportionate dehydrogenation effects. The calculations presented here consider three (de)hydrogenation cases, $f_{\rm H} = 0, 0.25$, and 1, corresponding to complete dehydrogenation, 75\% dehydrogenation, and fully hydrogenated surfaces, respectively.

\section{Nano-diamond optical properties} 
\label{sect_optEC_model}

The absorbance spectra or mass absorption coefficients of nano-diamonds have been measured in many studies \citep{1998A&A...330.1080A,2000M&PS...35...75B,1994A&A...284..583C,1998A&A...336L..41H,Jones:2004fu,1995MNRAS.277..986K,1995ApJ...454L.157M,Reich:2011gv,2002ApJ...581L..55S,2011ApJ...729...91S,Usoltseva:2018er,1997PhRvB..55.1838Z}. However, the detailed modelling of astrophysical objects requires that the complex indices of refraction must be well-determined over as wide a wavelength range as possible ({i.e.}, ideally from EUV to mm). To this end the optical constants of pre-solar nano-diamonds have been determined in the laboratory over the wavelength ranges $\lambda = 0.1 - 1\,\mu$m by \cite{1989Natur.339..117L} and $\lambda = 0.12 - 100 \,\mu$m by \cite{2004A&A...423..983M}. Nevertheless, these data do not sufficiently cover the wide wavelength range required for interstellar nano-diamond modelling. 

In the following section we describe and develop our approach to the determination of the complex indices of refraction ($n$ and $k$) of hydrocarbonaceous materials that is now applied to nano-diamonds. As per the optEC$_{\rm (s)}$(a) modelling \citep{2012A&A...540A...1J,2012A&A...540A...2J,2012A&A...542A..98J}, this model should be considered as something that can be partially, or even totally, re-constructed as and when more-constraining laboratory and astrophysical observations on (nano-)diamonds become available. 

The range of amorphous carbons encompassed by the optEC$_{\rm (s)}$(a) model includes the sp$^3$-rich, large band gap ($E_{\rm g} \simeq 2.7$\,eV), hydrogenated amorphous carbons (a-C:H), which can be considered as the closest approach to nano-diamonds within that framework. Their optical properties ($n$ and $k$) are predicted over more than seven  orders of magnitude in energy, $2.5 \times 10^{-6}$ -- $56$\,eV (i.e., $0.022\,\mu$m to 50\,cm in wavelength) and we do the same here for nano-diamonds. 

The aim of this modelling is to provide a set of self-consistent data that can be used to test the size- and composition-dependent effects of nano-diamond evolution within the astrophysical context. To this end we provide ASCII data files of $n$ and $k$ for nano-diamond particles for a limited set of particle radii (0.5, 1, 3, 10, 30, and 100\,nm $\equiv 80$, 600, $2 \times 10^4$, $7 \times 10^5$, $2 \times 10^7$, $7 \times 10^8$ C atoms per particle, respectively), which covers particles from the large molecule domain to bulk materials.The six data files are for three different hydrogenation states ($f_{\rm H} = 0, 0.25$, and 1) and, in each case, for non-irradiated and irradiated nano-diamond particle cores.\footnote{The nano-diamond $n$ and $k$ data are available from the following website: https://www.ias.u-psud.fr/themis/}  As per the optEC$_{\rm (s)}$(a) model, these data are founded upon the laboratory-measured properties of (nano-)diamonds available at this time.

\section{Optical constant methodology and determination}
\label{sect_nk}

In order to determine the EUV to mm optical constants for nano-diamonds we use the optEC$_{\rm (s)}$(a) methodology developed by \cite{2012A&A...540A...1J,2012A&A...540A...2J,2012A&A...542A..98J} for hydrogenated amorphous hydrocarbon materials, a-C(:H). However, in this case we can use exact calculations for the particle structures by assuming a `perfect' diamond lattice rather than the statistical extended Random Covalent Network (eRCN) and Defective Graphite (DG) descriptions that were developed for amorphous carbonaceous solids. 

The optical spectra of wide band gap hydrogenated amorphous carbons (a-C:H) show two clear and separated peaks: a $\pi-\pi^\ast$ peak at $\sim 4$\, eV and $\sigma-\sigma^\ast$ at $\sim 13$\, eV \citep[{e.g.},][]{1986AdPhy..35..317R}. An additional peak at $\sim 6.5$\, eV has been attributed to C$_6$, `benzene-like' aromatic clusters in the structure. As per the optEC$_{\rm (s)}$(a) model we adopt these three characteristic band energies ({i.e.}, 4.0\,eV for $\pi-\pi^\ast$, 6.5\,eV for C$_6$ and 13.0\,eV for $\sigma-\sigma^\ast$), and their annealing-dependent behaviour. 

We derive a single-parameter model for the evolution of the imaginary part, $k(E)$, of the complex refractive index, $m(E)=n(E)+ik(E)$, as a function of energy, where the critical characterising parameter is the particle radius. This derivation is based upon the optEC$_{\rm (s)}$(a) model \citep{2012A&A...540A...1J,2012A&A...540A...2J,2012A&A...542A..98J} and therefore requires and input band gap for the material, $E_{\rm g}$. Given that we are here only concerned with wide band gap diamond we fix the equivalent of the optEC$_{\rm (s)}$(a) band gap to its maximum allowed value within optEC$_{\rm (s)}$(a) ($E_{\rm g} = 2.67$\,eV) and adjust the other material parameters to best-fit the laboratory-measured diamond data. We then use the derived $k(E,a_{\rm nd})$ to calculate $n(E,a_{\rm nd})$ using the Kramers-Kronig Fortran toolbox (KKTOOL) provided by V\"olker Ossenkopf \footnote{Available for download from the following website http://hera.ph1.uni-koeln.de/~ossk/Jena/pubcodes.html \label{footnote2}}. The KKTOOL fortran code was slightly updated and modified to allow for more wavelength coverage and more materials but was otherwise used {\it as-is}.

\subsection{Band profiles}

The optEC$_{\rm (s)}$(a) model was built within the limiting framework of diamond and graphite \citep{2012A&A...540A...1J,2012A&A...540A...2J}  and therefore  encompasses wide band gap diamond-like carbonaceous materials within its realm ($E_{\rm g} \simeq 2.7$\,eV), albeit that bulk diamond has a significantly wider band gap ($E_{\rm g} \simeq 5.47$\,eV). 

In order to allow for the likelihood that the properties of diamond at nanometre sizes may exhibit more hydrogenated aliphatic-rich material-like behaviour we also include the $\pi-\pi^\ast$,  C$_6$ and $\sigma-\sigma^\ast$ bands, as per optEC$_{\rm (s)}$(a), but modify their intensities for consistency with the measured diamond properties. As per the optEC$_{\rm (s)}$(a) model, these bands can be empirically well-fit with a log-normal profile in energy, $g_i(E)$, of the form 
\begin{equation}
g_i(E) = {\rm exp} \Bigg\{ - \left[ {\rm ln} \left( \frac{E}{E_{0,i}} \right) \right]^2 \frac{1}{2 \delta_i^2} \Bigg\}, 
\label{eq_log_normal}
\end{equation}
where the assumed band widths, $\delta_i$, and band-centre energies, $E_{0,i}$, are given in Table~\ref{nk_nano_d_params}. The adopted values enable a fit to the pre-solar nano-diamond $k$ data of \cite{1989Natur.339..117L} in the $0.2-2\,\mu$m wavelength range for particles with radii $\ll 100$\,nm (see Fig. \ref{fig_k_nanod}).\footnote{{\it N.B.}, Even though the high-energy values of $\sigma_{i}$ and $S_i(E_{\rm g+})$ for  the $\pi-\pi^\ast$ band are zeroed (Table~\ref{nk_nano_d_params}) this band does nevertheless contribute because of the mixing of the high and low $E_{\rm g}$ cases inherent to the optEC$_{\rm (s)}$(a) methodology.} At high and low energies the adopted log normal band profiles are modified to take account of the appropriate energy-dependencies of the laboratory-measured diamond optical constants \citep{1985HandbookOptConst...665,1989Natur.339..117L}.

\begin{table}
\caption{The optEC$_{\rm (s)}$(a) model-based input parameters.}
\begin{center}
\begin{tabular}{lccc}
                                         &                             &                   &                    \\[-0.35cm]
\hline
\hline
                                         &                             &                     &                  \\[-0.35cm]
  Parameter                      &  $\pi-\pi^\ast$       &  C$_6$         &  $\sigma-\sigma^\ast$      \\[0.05cm]
\hline
                                         &                             &                    &                   \\[-0.35cm]
$i$                                    &             1               &          2          &           3           \\
E$_{0,i}$ [ eV ]                 &             4.0            &        6.5          &        13.0        \\
$\sigma_{i}$ \, [ eV ]        &               0             &        0.4          &        0.3        \\
$S_i(E_{\rm g+})$             &             0              &       0.01         &         1.50        \\
$S_i(E_{\rm g-})/0.47$     &          1.65             &       0.60         &         1.00        \\
\hline
                       &                             &                   &                    \\[-0.25cm]
\end{tabular}
\end{center}
\label{nk_nano_d_params}
\end{table}

\subsubsection{High energy behaviour}

For energies beyond 16\,eV ($E_{\rm EUV}$) we modified and extended the $\sigma-\sigma^\ast$ band using a power law behaviour determined by the expected photo-electron emission cross-section at EUV to x-ray wavelengths,  $k(E,a_{\rm nd}) \propto E^{-2.5}$, {i.e.}, 
\begin{equation}
g_i^\prime(E) = g_i(E) \ \times \Bigg\{ \frac{ E }{ E_{\rm EUV} } \Bigg\}^{-2.5} \ \ \ \ \ \ \ \ \ \ \ \ {\rm for} \ \ E \geq E_{\rm EUV}. 
\label{eq_high_E_fit} 
\end{equation}
This results in asymmetric $\sigma-\sigma^\ast$ band profiles essentially identical in form to the 2-TL (two Tauc-Lorentz oscillators) formalism used by \cite{2007DiamondaRM...16.1813K} in their optical property derivation. 

As the size of a nano-diamond particle decreases the band gap has been observed to increase, {i.e.}, the EUV absorption peak effectively shifts to higher energies. Hence, and as per \cite{1999PhRvL..82.5377C}, and as implemented in the optEC$_{\rm (s)}$(a) model \cite[See Appendix C.3,][]{2013A&A...558A..62J} for particle radii less than 10\,nm, we impose a shift in the $\sigma-\sigma^\ast$ band peak energy position, $E_{0,3}$, given by
\begin{equation}
E_{0,3} =  E_{0,3} \, \times \Bigg\{ \, 1 + 0.06 \, \left( 10 - \frac{a_{\rm nd}}{{\rm [ 1\,nm ] }} \right) \Bigg\} \ \ \ \ {\rm [ eV ] }
\label{eq_SS_Eshift}
\end{equation}
which corresponds to a shift by a factor of $\simeq 1.5$, 1.4, 1.3 for $a_{\rm nd} = 1$, 3 and 5 nm, respectively.

We extend the photo-electron-emission energy-dependence out to energies well beyond 60\,eV in order to allow a proper determination of $k(E,a_{\rm nd})$, using KKTOOL, for $E \leq 55$\,eV. In the tabulations of $n$ and $k$ we therefore only present complex refractive index data for energies $\leq 56$\,eV (wavelengths $\geq 22$\,nm), {i.e.}, from the EUV to the dm domain. 

\subsubsection{Low energy behaviour}

The optEC$_{\rm (s)}$(a) low energy contribution to $k$ is unnecessary in this determination of $k$, as is the conductivity term \cite[see section 4.4,][]{2012A&A...540A...2J}. The long wavelength behaviour of $k$ is dominated by that of bulk diamond\footnote{This study considers both non-irradiated and irradiated N-poor diamond bulk materials, see section \ref{sect_add_bulk_diamond}.} and, for surface-hydrogenated nano-diamonds, by the absorption bands of CH$_n$ and aliphatic CC bonds (see Fig. \ref{fig_k_nanod}).\footnote{For the interested reader the full details of the behaviour of $k$ at low energies and long wavelengths can be found in section 4.1.2 of \cite{2012A&A...540A...2J} and will not be repeated here.} In this approach we must additionally include the long wavelength properties of bulk diamond (see section \ref{sect_add_bulk_diamond}). 

The electrical conductivity of interstellar dust analogue materials is nevertheless an issue for their long-wavelength behaviour. However, within the astrophysical context it is not clear how important this could be for small, isolated, hydrogen-passivated nano-diamond particles. Based upon the experimental evidence and our earlier work \citep{2012A&A...540A...2J}: we now briefly discuss this. 

H-poor, a-C materials exhibit electrical conductivities as a result of their aromatic character, {i.e.}, enhanced but locally delocalised (within the $\pi$ cluster) electrons resulting from the $\pi-\pi^\ast$ band contribution. Given that nano-diamonds normally exhibit no aromatic sp$^2$ carbon domains,\footnote{In the absence of the surface reconstruction of \{111\} diamond facets to aromatic structures \cite[{e.g.},][]{Barnard:2005dt}.} there are no available conduction electrons. The electrical conductivities of nano-diamonds must therefore be extremely small ($\leq 10^{-2}$\,$\Omega^{-1}$\,cm$^{-1}$ at $T \leq 1000$\,K) and, at the temperatures of interest for interstellar and planetary system dust ({\it viz.}, $T \simeq 10-300$\,K) their electrical conductivity must be $\ll  10^{-2}$\,$\Omega^{-1}$\,cm$^{-1}$. For ISM dust studies $T_{\rm dust}$ is generally $\lesssim 50$\,K, which implies electrical conductivities $\ll 10^{-10}$\,$\Omega^{-1}$\,cm$^{-1}$ for a-C:H-like nano-diamonds \citep{2012A&A...540A...2J}. As in this earlier study we therefore conclude that we can ignore any contribution of the electrical conductivity to the optical properties of non-surface-reconstructed nano-diamonds at long wavelengths ($\lambda > 100$\,$\mu$m) and low temperatures ($T < 300$\,K). 

The fact that the electrical conductivity of nano-diamonds is low, and especially so at low temperatures, implies that their thermal conductivities must also be very low, which has interesting consequences for their temperatures, especially for their peak temperatures during stochastic-heating in circumstellar and diffuse interstellar regions.

\subsection{Derivation of the imaginary part of the refractive index}

The optEC$_{\rm (s)}$(a) methodology achieved satisfactory fits to the laboratory-measured optical properties of a-C(:H) materials, at visible to UV wavelengths, for the a-C and a-C:H end-members by scaling the 4, 6.5 and 13\,eV band strengths, for the limiting a-C ($E_{\rm g+}$) and a-C:H ($E_{\rm g-}$) cases, and here we maintain this approach but use different band scaling factors, $S_i$ (see Table \ref{nk_nano_d_params}). 

The imaginary part of the refractive index for surface-hydrogenated nano-diamonds can now be derived as a function of energy, $E$, with the particle size as the single input parameter, and the fixed and `effective' value of $E_{\rm g}$ set to 2.67\,eV, as a linear combination of the end-member compositions, {i.e.}, 
\begin{equation}
k(E, E_{\rm g}) =  \sum_{i=1}^3 \Bigg\{ f(E_{\rm g+}) S_i(E_{\rm g+}) + f(E_{\rm g-}) S_i(E_{\rm g-}) \Bigg\} \ g_i(E)
\label{eq_optEC_1 }
\end{equation}
where $g_i(E)$ is the log normal band profile defined in Eq.~(\ref{eq_log_normal}). The fractions of the high [low] band gap material $f(E_{\rm g+})$ [f($E_{\rm g-}$)] are given by a simple linear interpolation between the limiting values for the band gap, {i.e.}, 
\begin{equation}
f(E_{\rm g+}) = \frac{(E_{\rm g}-E_{\rm g-})}{(E_{\rm g+}-E_{\rm g-})}, \ \ \ \ \ f(E_{\rm g-}) = 1-f(E_{\rm g+}). 
\label{eq_optEC_2 }
\end{equation}

The surface CH and CH$_2$ group abundances, representative of \{111\} and \{100\} facets respectively, are calculated using the diamond network model and assuming spherical diamond particles as per \cite{2020_Jones_nd_CHn_ratios} and the C$-$C modes are assumed to come from only the outer four atomic layers, with the bulk of the particle assumed to follow the bulk diamond properties and optical behaviour (see following section).

\begin{table*}[ht]
\caption{The adopted nano-diamond C-H and C-C band parameters.}
\begin{center} 
\begin{tabular}{ccccll}
                       &              &               &          &     \\[-0.35cm]
\hline
\hline
                       &                       &              &        &        \\[-0.35cm]
                      &  $\nu_0$                            &  $\delta$           &  $\sigma$         &         Band             &           \\
         no.        &  [ cm$^{-1}$ ( $\mu$m ) ]   &  [ cm$^{-1}$ ]   &   [ $\times 10^{-18}$ cm/bond ]   &        assignment  &  Notes \\[0.05cm]

\hline
                       &                      &                       &              &      \\[-0.2cm]
                       
 & \multicolumn{4}{l}{\underline{C-H stretching modes}}                         &         \\[0.2cm]
       0    &    2976  ( 2.976 )      &      ---     &   ---   &    sp$^2$ CH$_2$                     &    \\
       1    &    2953  ( 3.386 )      &      ---     &   ---   &    sp$^3$ CH$_3$                     &   \\
       2    &    2943  ( 3.398 )      &    20.0    &    2.20   &    sp$^3$ CH$_2$  \{100\}    &  J+04   \\ 
       3    &    2931  ( 3.412 )      &    20.0    &    2.60   &    sp$^3$ CH$_2$  \{100\}    &  J+04   \\
       4    &    2925  ( 3.419 )      &    20.0    &    0.50   &    sp$^3$ CH$_2$                 &  J+04  \\
       5    &    2915  ( 3.431 )      &    20.0    &    0.60   &    sp$^3$ CH          \ \ \{111\}  &  J+04    \\
       6    &    2901  ( 3.447 )      &    20.0    &    0.50   &    sp$^3$ CH$_2$                  &  J+13, modified $\delta$ \\
       7    &    2884  ( 3.467 )      &    30.0    &    0.60   &    sp$^3$ CH                          &  J+13, modified $\sigma$ \\
       8    &    2874  ( 3.480 )      &    15.0    &    1.80   &    sp$^3$ CH$_2$                  &   J+13, modified $\delta$ \& $\sigma$ \\
       9    &    2858  ( 3.499 )      &    15.0    &    4.00   &    sp$^3$ CH$_2$                  &   J+04, modified $\sigma$ \\
     10    &    2845  ( 3.515 )      &    10.0    &    2.20   &    sp$^3$ CH$_2$                   &   J+04, modified $\sigma$ \\
     11    &    2833  ( 3.529 )      &    10.0    &    2.20   &    sp$^3$ CH          \ \ \{111\}    &  J+04, modified $\sigma$  \\[0.2cm]
                       
 & \multicolumn{4}{l}{\underline{C-H bending modes}}                                    &   \\[0.2cm]
      12    &    1450  ( 6.897 )    &    3.0    &    1.20     &    sp$^3$ CH$_2$        &  J+13  \\[0.2cm]
                       
 & \multicolumn{4}{l}{\underline{C-C modes}}                                                   &  \\[0.2cm]
      13    &    1328  ( 7.530 )    &    120.0  &    0.10      &    sp$^3$ C$-$C        &  J+13 \\
      14    &    1300  ( 7.692 )    &    120.0  &    0.10      &    sp$^3$ C$-$C        &  J+13 \\
      15    &    1274  ( 7.849 )    &    120.0  &    0.10      &    sp$^3$ C$-$C        &  J+13 \\
      16    &    1163  ( 8.600 )    &      90.0  &    0.10      &    sp$^3$ C$-$C         & J+13 \\[0.2cm]   
           
\hline
                       &                      &                      &       & \\[-0.2cm]

 \multicolumn{5}{l}{\underline{Reconstructed \{111\} and \{100\} diamond facet sp$^2 $ modes}}  & \\[0.2cm]
                             
 & \multicolumn{4}{l}{\underline{C-H stretching modes}}                               &   \\[0.2cm]
      17    &     3078 ( 3.249 )    &   22.5  &    1.4      &   olefinic \ \ CH$_2$ asym.  &  J+13, not used here \\
      18    &     3050 ( 3.279 )    &   53.1  &    1.5      &   aromatic CH                      &   J+13 \\
      19    &     3020 ( 3.311 )    &    50.0  &    0.5      &      olefinic \ \ CH                 &   J+13 \\
      20    &     3010 ( 3.322 )    &   47.1  &    2.50      &      olefinic \ \ CH                &   J+13 \\
      21    &     2985 ( 3.350 )    &   17.7  &    1.15      &      olefinic \ \ CH$_2$ sym. & J+13, not used here \\[0.2cm]
     
 & \multicolumn{4}{l}{\underline{C-H bending modes}}                                     &  \\[0.2cm]
      22    &    1430  ( 6.993 )    &     60.0  &    0.40      &     aromatic CH            &  J+13 \\
      23    &    1410 ( 7.092 )     &      30.0  &    1.00      &     olefinic \ \ CH$_2$   &  J+13, not used here \\
      24    &    890   ( 11.236 )    &      20.0  &    0.20      &     aromatic CH             &  J+13  \\
      25    &    880   ( 11.363 )    &    40.0  &    0.50      &     aromatic CH              &  J+13  \\
      26    &    790   ( 12.658 )    &    50.0  &    0.50      &     aromatic CH              &  J+13  \\[0.2cm]
      
 & \multicolumn{4}{l}{\underline{C-C modes}}                                                &     \\[0.2cm]
      27    &   1640   ( 6.098 )    &    40.0  &    0.10      &     olefinic \ \ CC       &  J+13  \\
      28    &   1600   ( 6.250 )    &    60.0  &    0.76      &     aromatic CC        &   J+13 \\
      29    &   1500   ( 6.667 )    &    40.0  &    0.15      &    aromatic CC         &   J+13 \\[0.2cm]                             
\hline
\hline
                       &                      &                      &    &    \\[-0.15cm]        
\end{tabular}
\tablefoot{The CH and CC band parameters (upper entries) and the assumed sp$^2 $ modes for surface-reconstructed \{111\} and \{100\} diamond facets are: band centre ($\nu_0$), width ($\delta$), and integrated cross-section ($\sigma$). Bands 0 and 1 have been associated with (nano-)diamond surfaces and are included for completeness but are not used in our analysis. The bands indicated: J+04 were estimated using \cite{Jones:2004fu}; J+13 are from \cite{2013A&A...558A..62J}, and references therein \citep[{i.e.},][and E. Dartois private communication]{1986AdPhy..35..317R,2001A&A...372..981V,2004A&A...423..549D,2004A&A...423L..33D,1994A&A...281..923J,1998JAP....84.3836R,1967ApSRv...1...29W,2008ApJ...682L.101M}. sym. and asy. indicate symmetric and asymmetric modes, respectively.}
\end{center}
\label{spectral_bands}
\end{table*}

\subsection{The addition of IR bands into the $k$ determination}
\label{sect_add_IR_bands}

\begin{figure*}
\begin{center} $
\begin{array}{cc}
   \includegraphics[width=9.0cm]{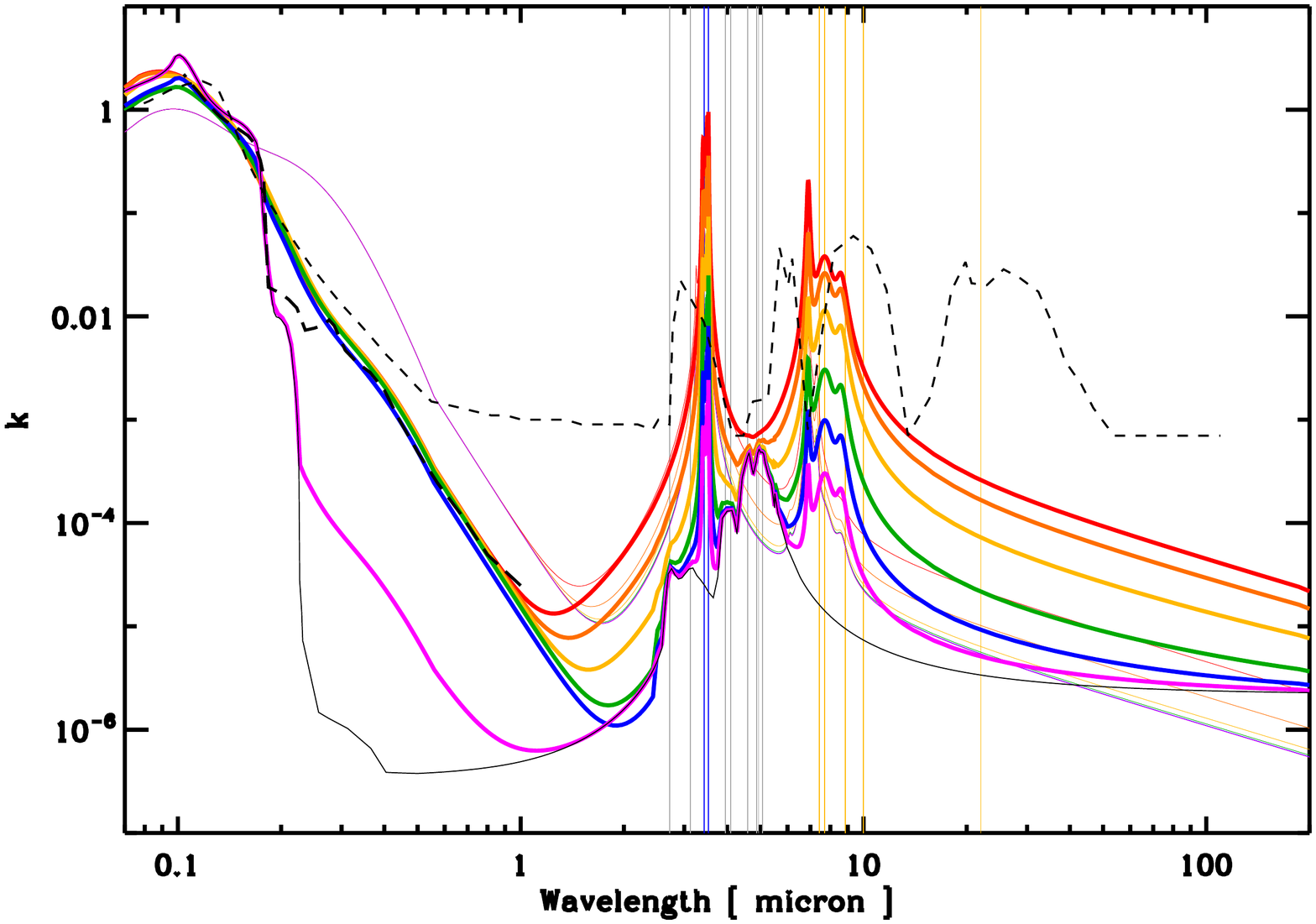}
   \includegraphics[width=9.0cm]{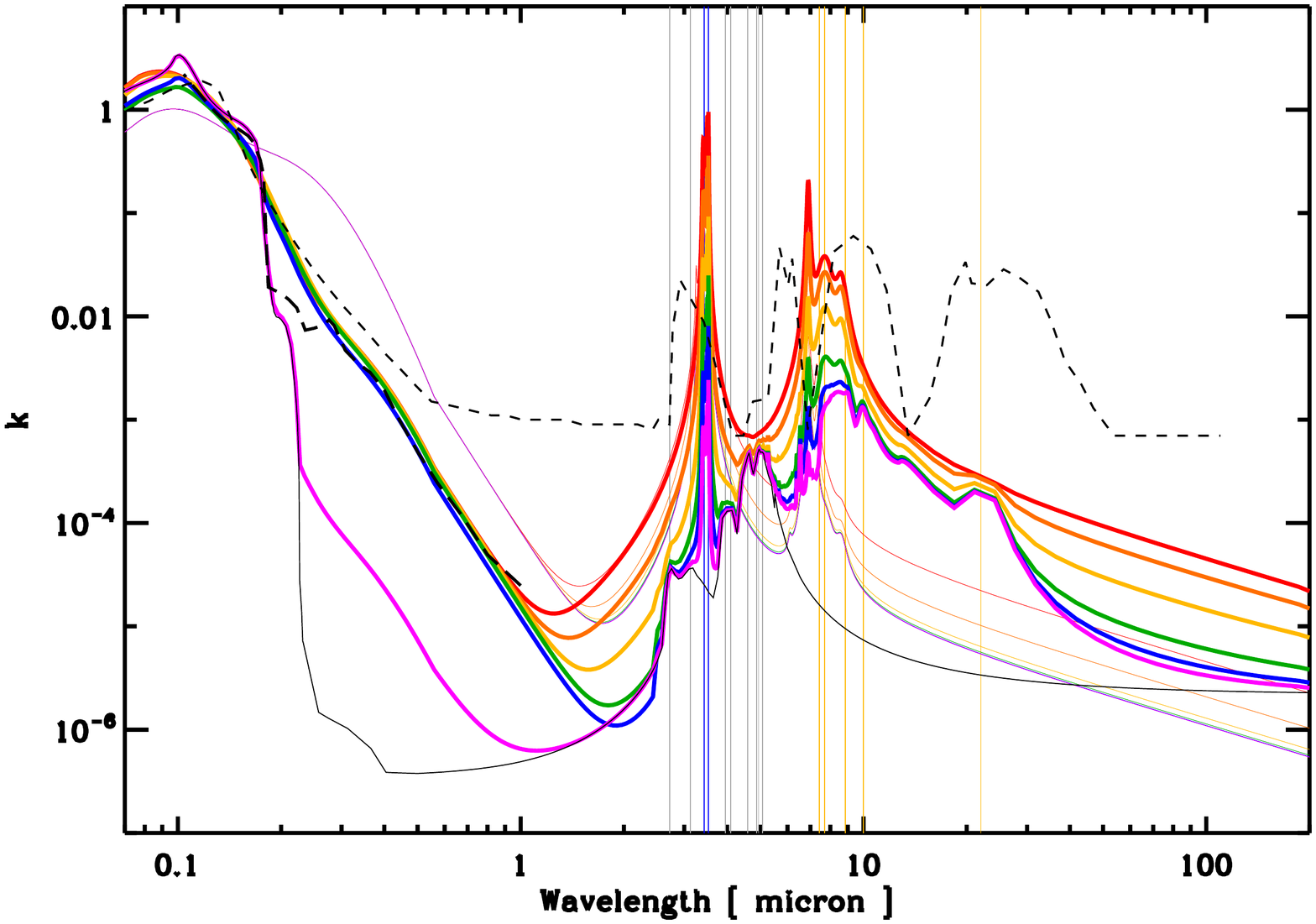}
\end{array} $
\end{center}
\vspace{-1.75cm}
\hspace*{-0.9cm}
\begin{center} $
\begin{array}{cc}
   \includegraphics[width=9.0cm]{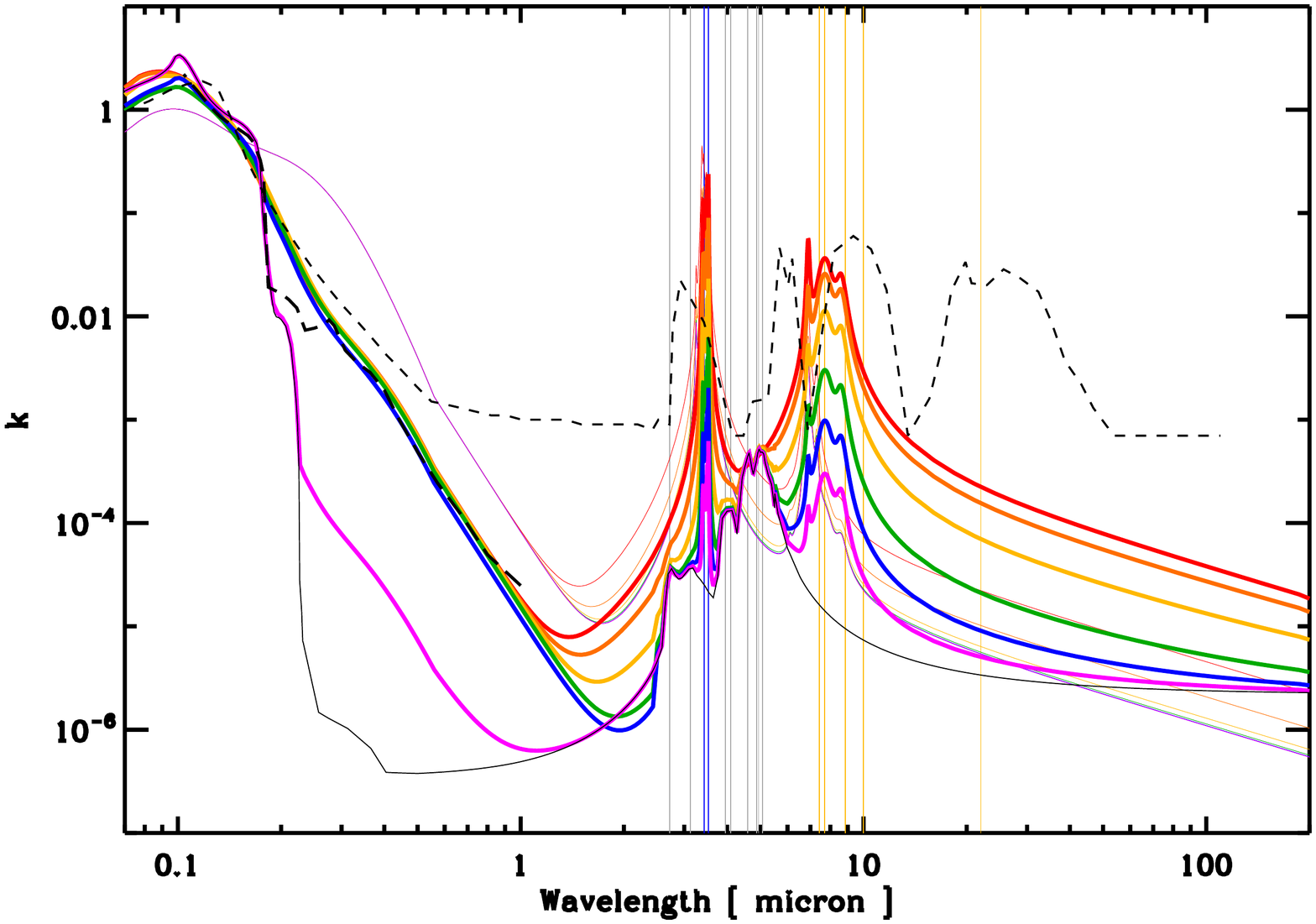}
   \includegraphics[width=9.0cm]{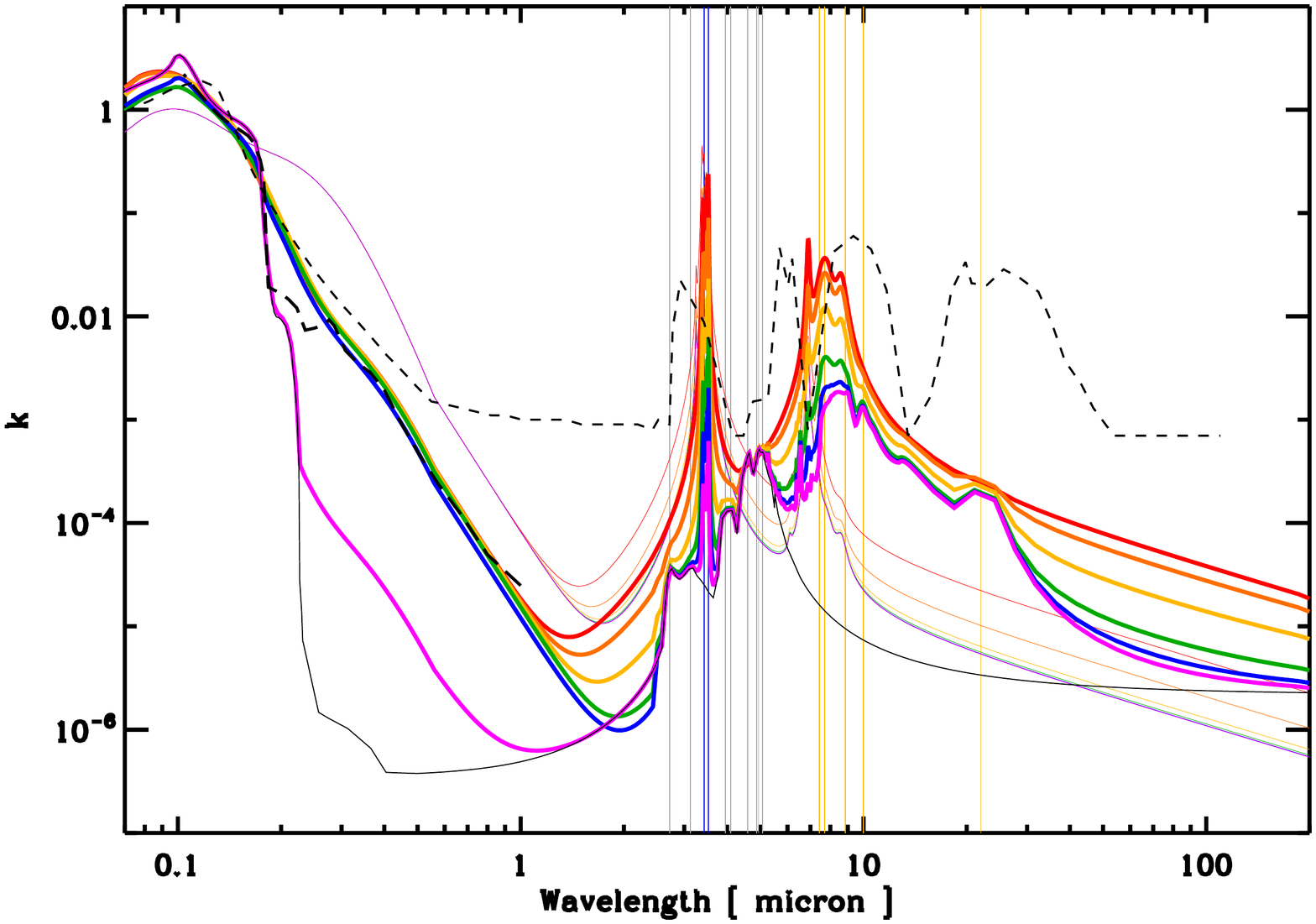}
\end{array} $
\end{center}
\vspace{-1.75cm}
\begin{center} $
\begin{array}{cc}
   \includegraphics[width=9.0cm]{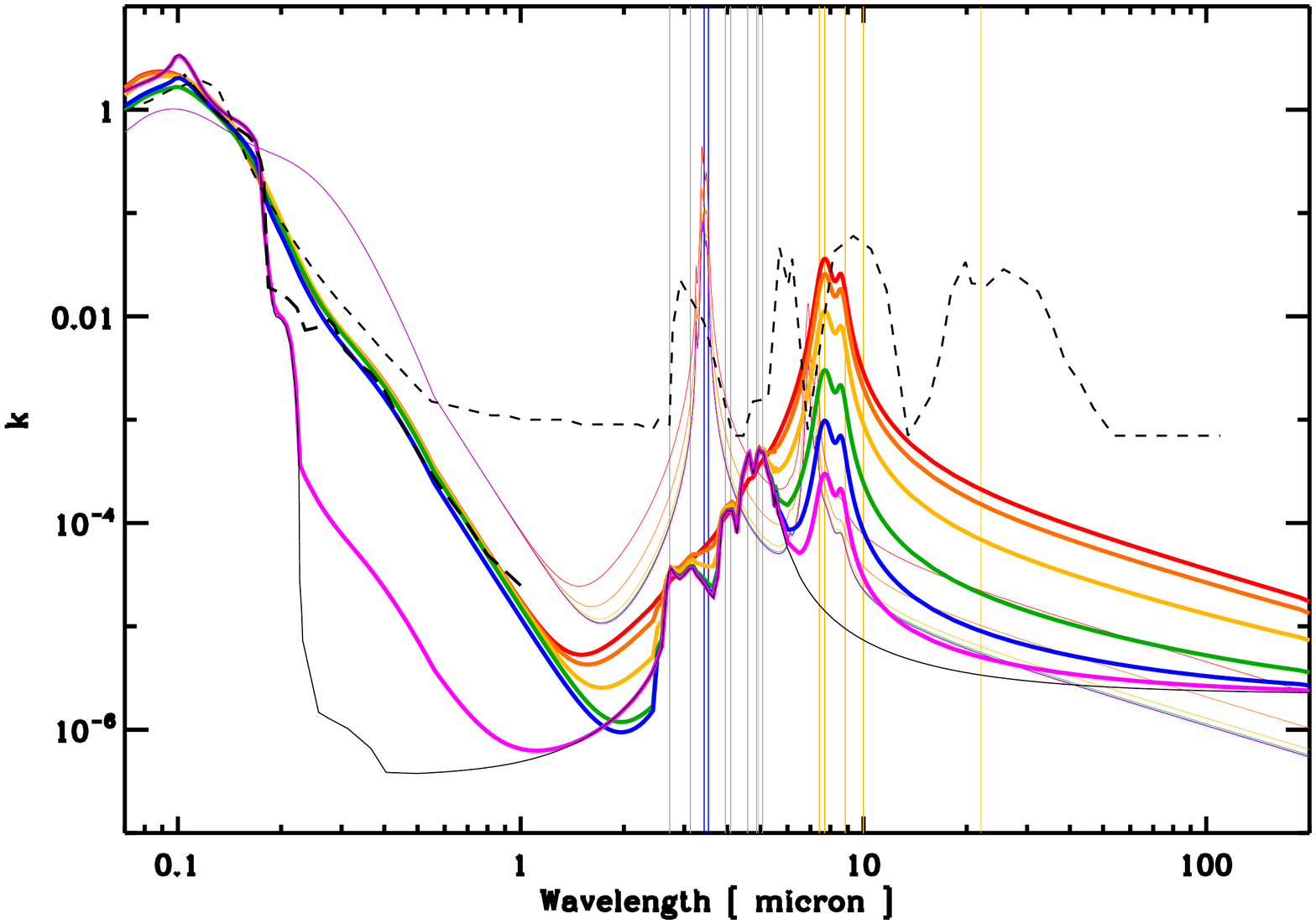} 
   \includegraphics[width=9.0cm]{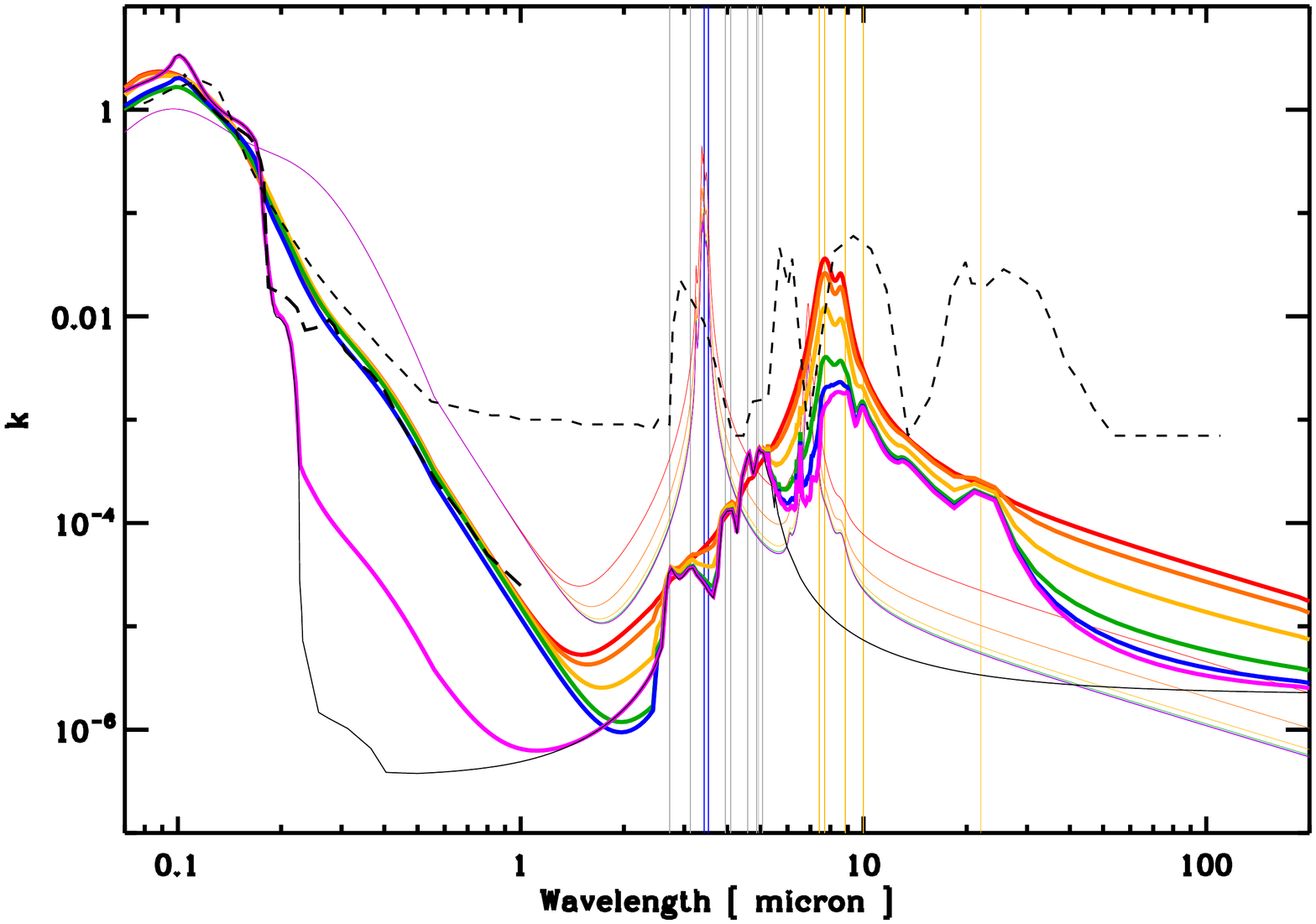}
\end{array} $
\end{center}
\vspace{-1.0cm}
      \caption{The derived imaginary part, $k$, of the nano-diamond complex indices of refraction as a function of radius (0.5, 1, 3, 10, 30, and 100\,nm: thick red, orange, yellow, green, blue, and violet, respectively). The thin lines show the same radii optEC$_{\rm (s)}$(a) data for aliphatic-rich ($E_{\rm g} = 2.67$\,eV) particles (size colour-coding as for nano-diamonds). The thin black line shows the bulk diamond data \citep{1985HandbookOptConst...665} and the short-dashed and long-dashed black line shows the data for pre-solar nano-diamonds of \cite{2004A&A...423..983M} and \cite{1989Natur.339..117L}, respectively. We note the 100\,nm $7-30\,\mu$m data (violet line) delineates the input neutron-irradiated, N-poor bulk diamond absorbance data. The thin grey vertical lines indicate the diamond two and three phonon mode peak positions, the thin blue vertical lines mark the hydrogenated nano-diamond $3.43$ and $3.53\,\mu$m bands and the thin yellow vertical lines approximately indicate the characteristic type Ib diamond IR band positions and also highlight the $\simeq 22\,\mu$m neutron-irradiated N-poor diamond band. The fractional surface hydrogen coverage, $f_{\rm H} = 1, 0.25$ and 0, decreases from the upper to lower panels, respectively. The data in left panels assumes `pristine' bulk diamond and that in the right panels neutron-irradiated bulk diamond \citep{1998A&A...336L..41H}. A zoom into the mid-IR region of these data is shown in Fig.~\ref{fig_k_zoom_nanod}.}
      \label{fig_k_nanod}
\end{figure*}

As for a-C(:H) materials \citep{2012A&A...540A...1J,2012A&A...540A...2J,2012A&A...542A..98J} the IR band profiles, with their characteristic positions, widths, and intensities are added into the determination of $k$, where  $k = \sigma_{\rm C} \, N_{\rm C} \, \lambda / (4 \pi) = h\, c\, \sigma_{\rm C} \, N_{\rm C}  / (4 \pi \, E) $.  This is done by summing over the $n$ contributing bands, {i.e.}, 
\begin{equation}
k_{\rm IR}(E,a_{\rm nd} ) = k(E,a_{\rm nd} ) + N_{\rm C}({\rm V})  \sum_{j=1}^n \Bigg\{ \frac{ h\  c\ \sigma_{{\rm C},j}(E)}{ 4 \pi E } \Bigg\}, 
\label{eq_IR_bands_k}
\end{equation}
where $N_{\rm C}({\rm V})$ is the number of carbon atoms per unit volume in the material under consideration. The adopted band positions, widths, and intensities are shown in Table \ref{spectral_bands}. The contribution of the $j^{\, \rm th}$ band to $k$ is determined from
\begin{equation}
\sigma_{{\rm C},j}(E) = \sigma_{0,j}(E) \ X_j \ g_j(E)
\label{eq_sigma_C}  
\end{equation}
where the cross-section $\sigma_{0,j}$ is derived from the integrated cross-sections in Table \ref{spectral_bands}, $X_j$ is the C atom fractional abundance for the participating C$_p$H$_q$ ($p\geq1$, $q\geq0$) functional group.  We do not include all of the indicated CH$_n$ ($n = 1$-3) bands listed in Table \ref{spectral_bands} because we do not include any methyl (CH$_3$) or CH$_2 $ ethyl (=C$<^{\rm H}_{\rm H}$) groups. Further, we have slightly modified the CH$_n$ ($n = 1$,2) bands numbered $6-11$ in Table \ref{spectral_bands} with respect to the standard THEMIS values \citep{2013A&A...558A..62J,2017A&A...602A..46J}, in order to fine-tune the model to the measured nano-diamond bands \citep[{e.g.},][]{2002JChPh.116.1211C,2002ApJ...581L..55S,Jones:2004fu}. The parameters of all the other bands remain unchanged from the usual THEMIS values. For surface hydrogenated nano-diamond we therefore consider only CH$_2$ and CH bonds on \{100\} and \{111\} facets, respectively. For the band shapes, $g_j(E)$, we assume Drude profiles \cite[as per][]{2012A&A...540A...1J} and the resultant effects of adding the infrared bands to the refractive index data at wavelengths $> 3\,\mu$m are clearly evident in Fig.~\ref{fig_k_nanod}.

\begin{figure*}
\begin{center} $
\begin{array}{cc}
   \includegraphics[width=9.0cm]{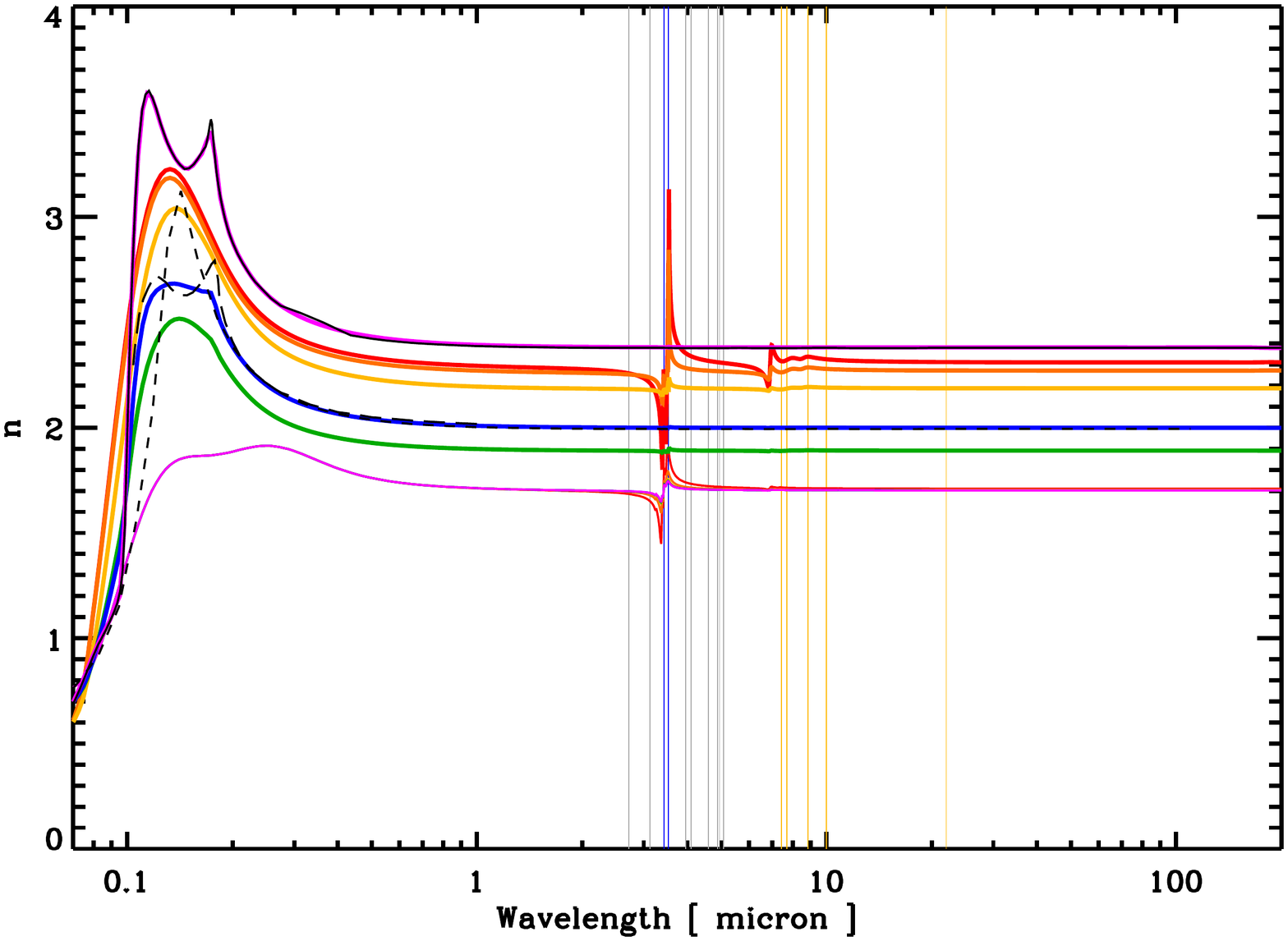}
   \includegraphics[width=9.0cm]{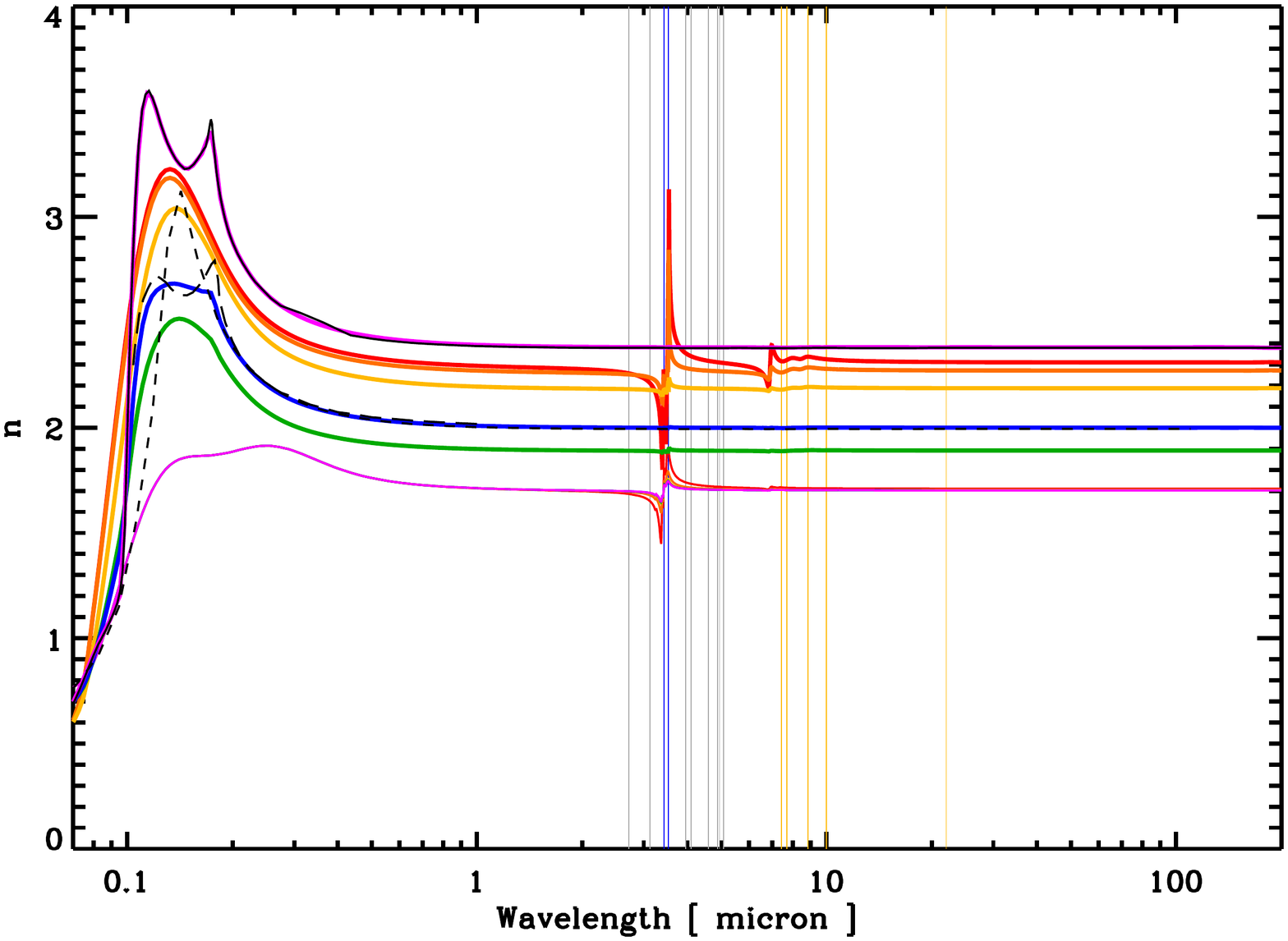}
\end{array} $
\end{center}
\vspace{-1.75cm}
\hspace*{-0.9cm}
\begin{center} $
\begin{array}{cc}
   \includegraphics[width=9.0cm]{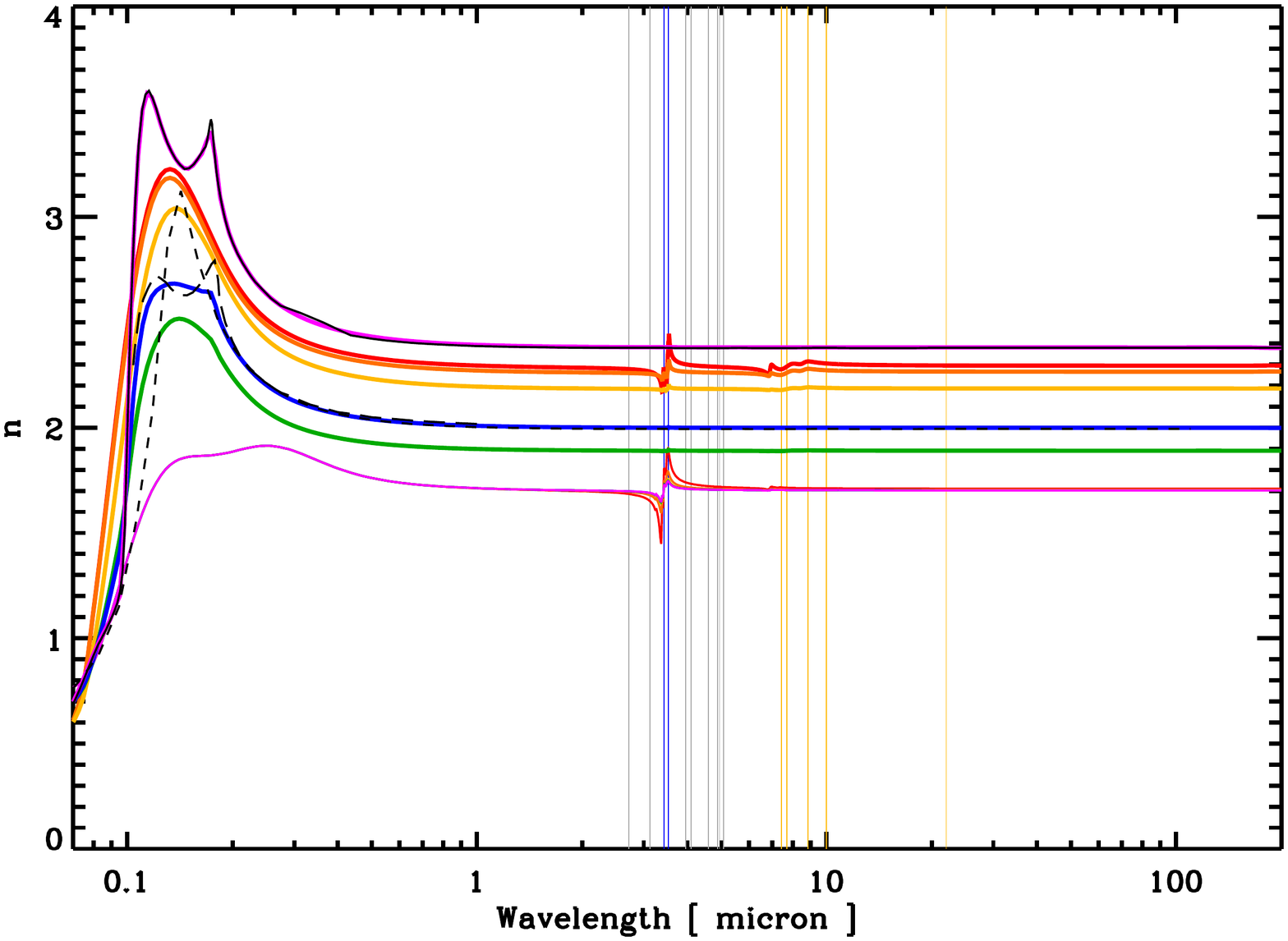}
   \includegraphics[width=9.0cm]{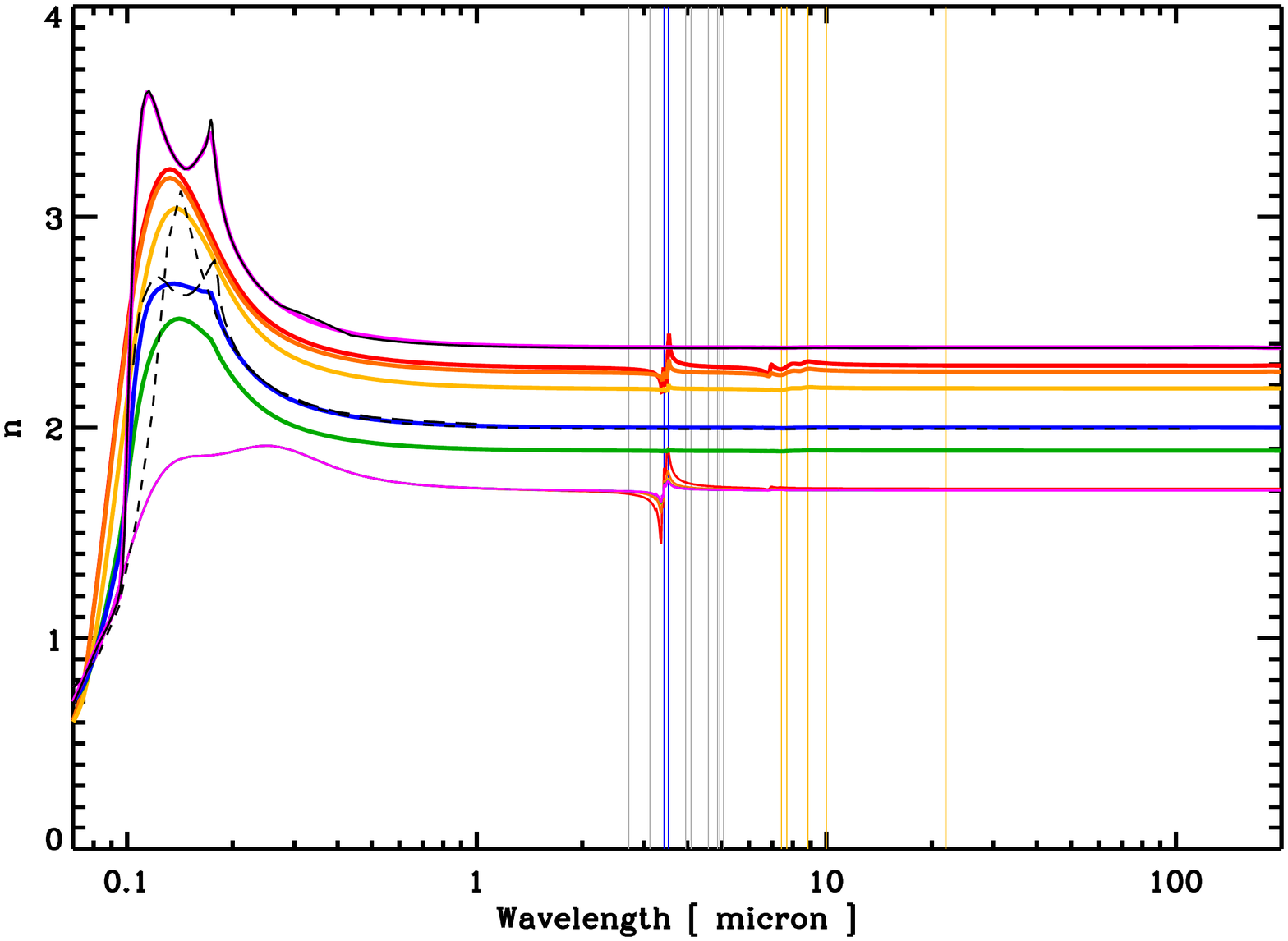}
\end{array} $
\end{center}
\vspace{-1.75cm}
\begin{center} $
\begin{array}{cc}
   \includegraphics[width=9.0cm]{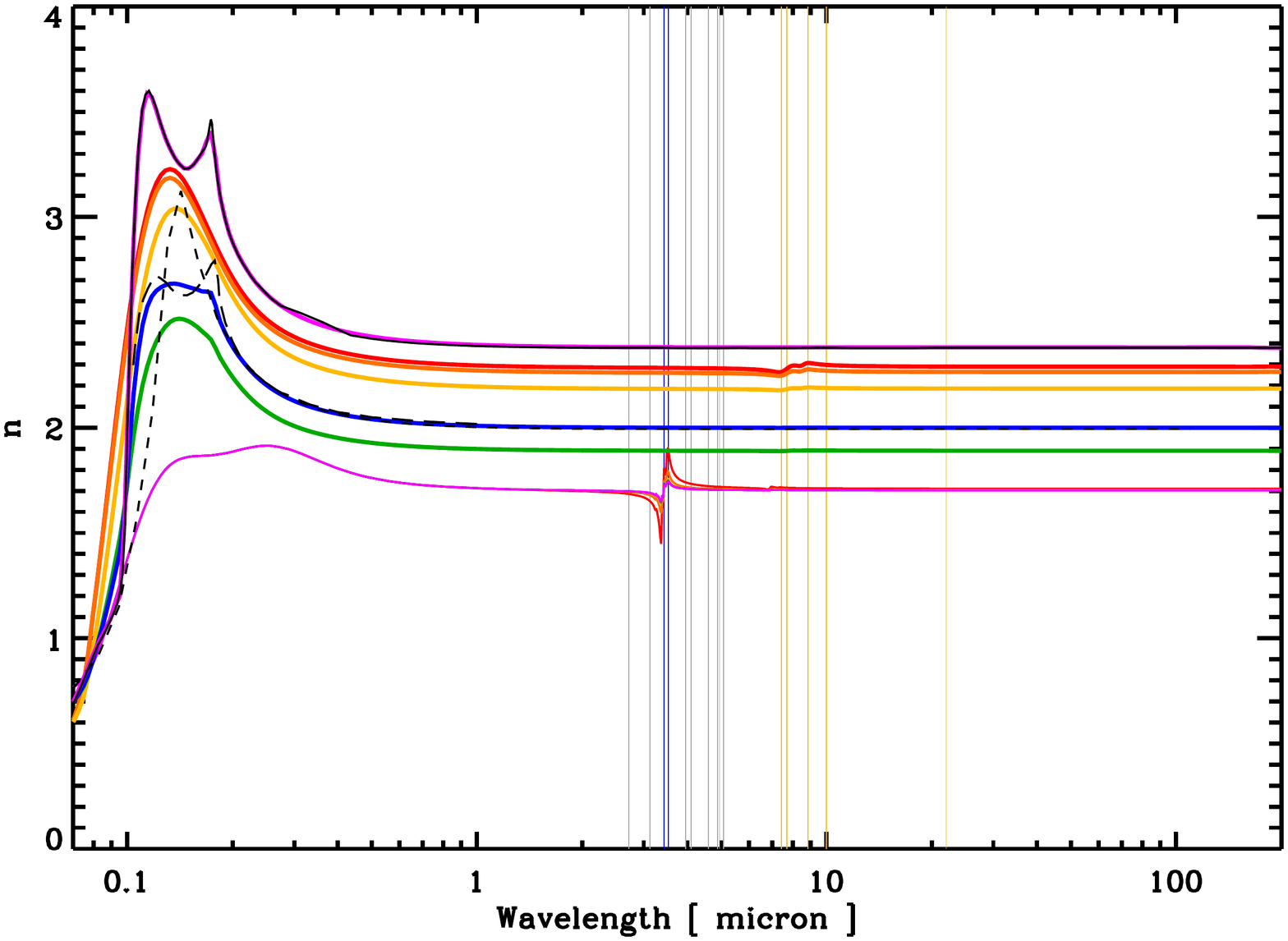} 
   \includegraphics[width=9.0cm]{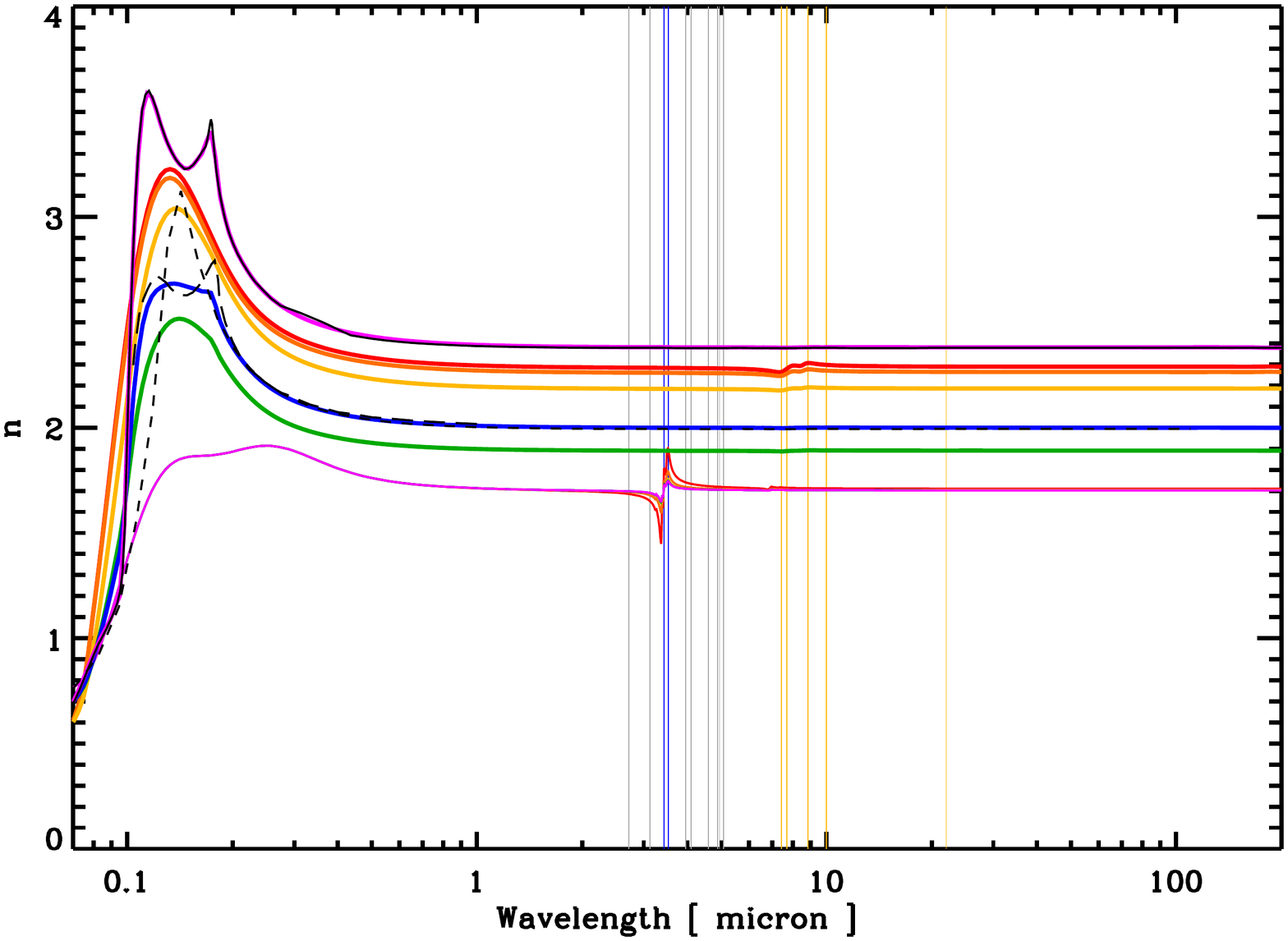}
\end{array} $
\end{center}
\vspace{-1.00cm}
      \caption{The KKTOOL-derived real part, $n$, of the nano-diamond complex indices of refraction (lines and colour coding are the same as in Fig. \ref{fig_k_nanod}). The fractional surface hydrogen coverage $f_{\rm H} = 1, 0.25$ and 0, from top to bottom, respectively. The left panels show the `pristine' bulk diamond data and the right panels that for neutron-irradiated bulk diamond \citep{1998A&A...336L..41H}.}
      \label{fig_n_nanod}
\end{figure*}

\subsection{The addition of the bulk diamond data into the $k$ determination}
\label{sect_add_bulk_diamond}

Given that interstellar nano-diamonds are subject to the harsh interstellar environment where they are exposed to the effects of shock waves and cosmic ray irradiation we choose data appropriate for an irradiated diamond. For this, and as a limiting case, we use the laboratory-measured neutron-irradiated, N-poor bulk diamond material of \cite{1998A&A...336L..41H}  as our reference. These absorbance data, over the wavelength range $5.1-24.2\,\mu$m, are scaled to the $k$ values of \cite{1985HandbookOptConst...665} in the two and three phonon mode region ($\simeq 2.5-6\,\mu$m) with the long wavelength wing of the $\approx 22\,\mu$m band extrapolated (for $\lambda > 24.2\,\mu$m) using a Drude profile with a band centre at 450\,cm$^{-1}$ ($22.2\,\mu$m) and a width parameter of 100 (see Fig. \ref{fig_k_nanod}). In order to bracket our results we also consider the other limiting case of a N-poor un-irradiated diamond with low mid-IR emissivity, which shows no spectral bands in the $7-30\,\mu$m region (see the thin, solid black line in Fig.~\ref{fig_k_nanod}). We do not undertake any blending of the irradiated and un-irradiated cases  but use them as illustrative upper and lower limits on the bulk diamond material properties. 

At visible to EUV wavelengths ($\lambda < 2.5\,\mu$m) the values of $k_{\rm EUV-vis}$ determined with the optEC$_{\rm (s)}$(a) methodology, $k_{\rm nd}$, differ in shape, but not significantly in absolute value, from those for bulk diamond, $k_{\rm bulk}$. We assume particles with radii $> 100$\,nm are completely bulk-like \cite[{e.g.}, as][found for a-C(:H) materials]{2012A&A...542A..98J} and that this behaviour decreases significantly with size. In order to impose a smooth transition we mix the two sets of $k$ values using a bulk diamond weighting fraction, $f_{\rm bulk}$, normalised at $a = 100$\,nm  ($a_{\rm ref}$), and defined by
\begin{equation}
f_{\rm bulk} = {\rm min} [ \ (a_{\rm nd} / a_{\rm ref}) \ , \ 1 \ ],  
\end{equation}
which implies a dependence on the particle volume/surface area ratio, which seems reasonable. The resultant $k$ is then derived with the following simple mixing rule 
\begin{equation}
k_{\rm EUV-vis}  = ( 1 - f_{\rm bulk} ) \ k_{\rm nd}({\rm EUV-vis}) \ + \ f_{\rm bulk} \ k_{\rm bulk}({\rm EUV-vis}), 
\end{equation}
The loss of diamond lattice symmetry with decreasing grain size leads to a weakening, and eventual disappearance, of the bulk diamond two and three phonon modes ($\simeq 2.5-6\,\mu$m), which may be replaced by one phonon modes \cite[{e.g.},][]{1998A&A...336L..41H}.   We here assume that this weakening also affects the neutron irradiation-induced bands at longer wavelengths. 

In keeping with this requirement and at wavelengths $> 2.5\,\mu$m, we empirically and in a physically-reasonable manner, merge the derived nano-diamond $k_{\rm nd}(\lambda > 2.5\,\mu{\rm m})$ with that for bulk (neutron-irradiated, N-poor) diamond  $k_{\rm bulk}(\lambda > 2.5\,\mu{\rm m})$ as a function of radius to derive $k$ at  wavelengths $> 2.5\,\mu$m ($k_{\rm IR-mm}$). We do this by assuming that the surface or non-bulk properties extend four carbon atom layers deep into the particle. This ensures that the distinctive bulk diamond two and three phonon modes, in the $2-6\,\mu$m region, are suppressed in the smaller nano-diamonds considered here ($a \lesssim 3$\,nm) in order to ensure consistency with nano-diamond laboratory data \citep{1998A&A...336L..41H,2004A&A...423..983M}. We then mix the two sets of $k$ values using the following mixing rule
\begin{equation}
k_{\rm IR-mm} = k_{\rm nd}({\rm IR-mm}) \ + \ ( 1 - F_{\rm V,CC} ) \ k_{\rm bulk}({\rm IR-mm}), 
\end{equation}
where $F_{\rm V,CC}$ is the fraction of the volume in the {\it at-surface} four carbon atom deep layer. Note that in this case the $k_{\rm nd}({\rm IR-mm})$ are not weighted because these surface-dependent properties are exactly determined for the defined particle surfaces. In contrast, the bulk-like material contribution does decline as it is progressively replaced by the sp$^3$ C$-$C surface modes with decreasing particle radius. 

At the longest wavelengths ($\lambda > 200\,\mu{\rm m}$), and only for  particles with radii larger than 2\,nm, we impose a power-law extrapolation in order to be consistent with laboratory data \citep{1985HandbookOptConst...665}, {i.e.}, 
\begin{equation}
k_{\rm long} = \left( \frac{ \lambda_{\rm long}{\rm [\mu m]} }{ 200 } \right)^{-\frac{3}{4}} \ k_{\rm IR-mm}(\lambda_{\rm long}). 
\end{equation}

Once the size- and energy-dependent imaginary part of the complex index of refraction, $k(E,a_{\rm nd})$, has been derived, using the above-described methodology, the corresponding real part, $n(E,a_{\rm nd})$, is calculated using KKTOOL (see the beginning of this section).

\section{Results for hydrogenated and dehydrogenated nano-diamonds}
\label{sect_surf_results}

The wavelength dependencies of $k(E,a_{\rm nd})$ for nano-diamonds as a function of size, surface hydrogenation, and irradiation state are plotted in the six panels of Fig.~\ref{fig_k_nanod}. For reference these panels also show the data for  bulk diamond \citep{1985HandbookOptConst...665} and for pre-solar nano-diamonds from \cite{1989Natur.339..117L} and \cite{2004A&A...423..983M}. The equivalent data for the most diamond-like, aliphatic-rich, hydrogen-richest hydrocarbon (a-C:H) particles with $E_{\rm g} = 2.67$\,eV, generated with optEC$_{\rm (s)}$(a), are also plotted. Fig.~\ref{fig_n_nanod} shows the complimentary KKTOOL-derived real part of the complex indices of refraction, $n(E,a_{\rm nd})$.

\begin{figure*}
\begin{center} $
\begin{array}{cc}
   \includegraphics[width=9.0cm]{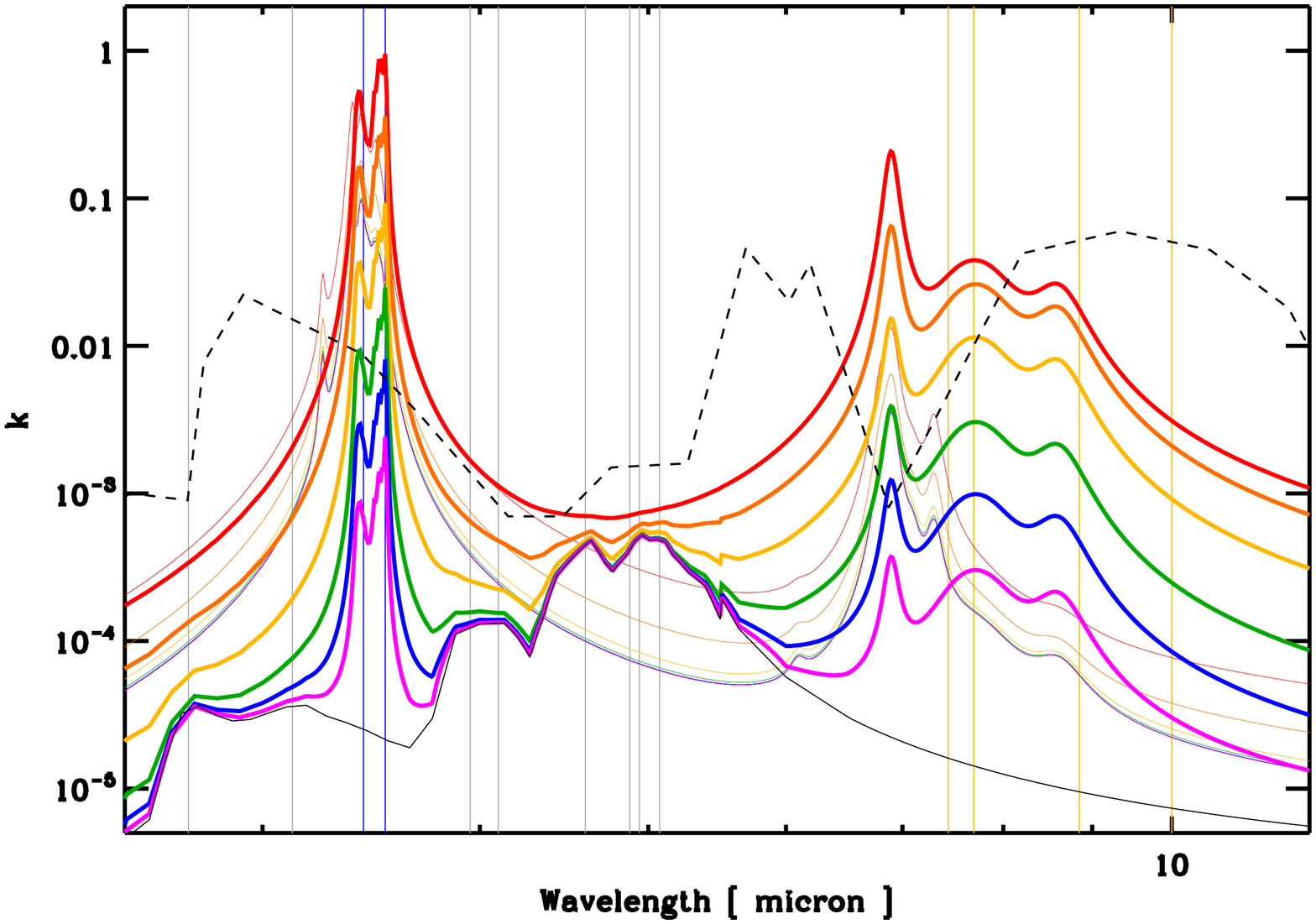}
   \includegraphics[width=9.0cm]{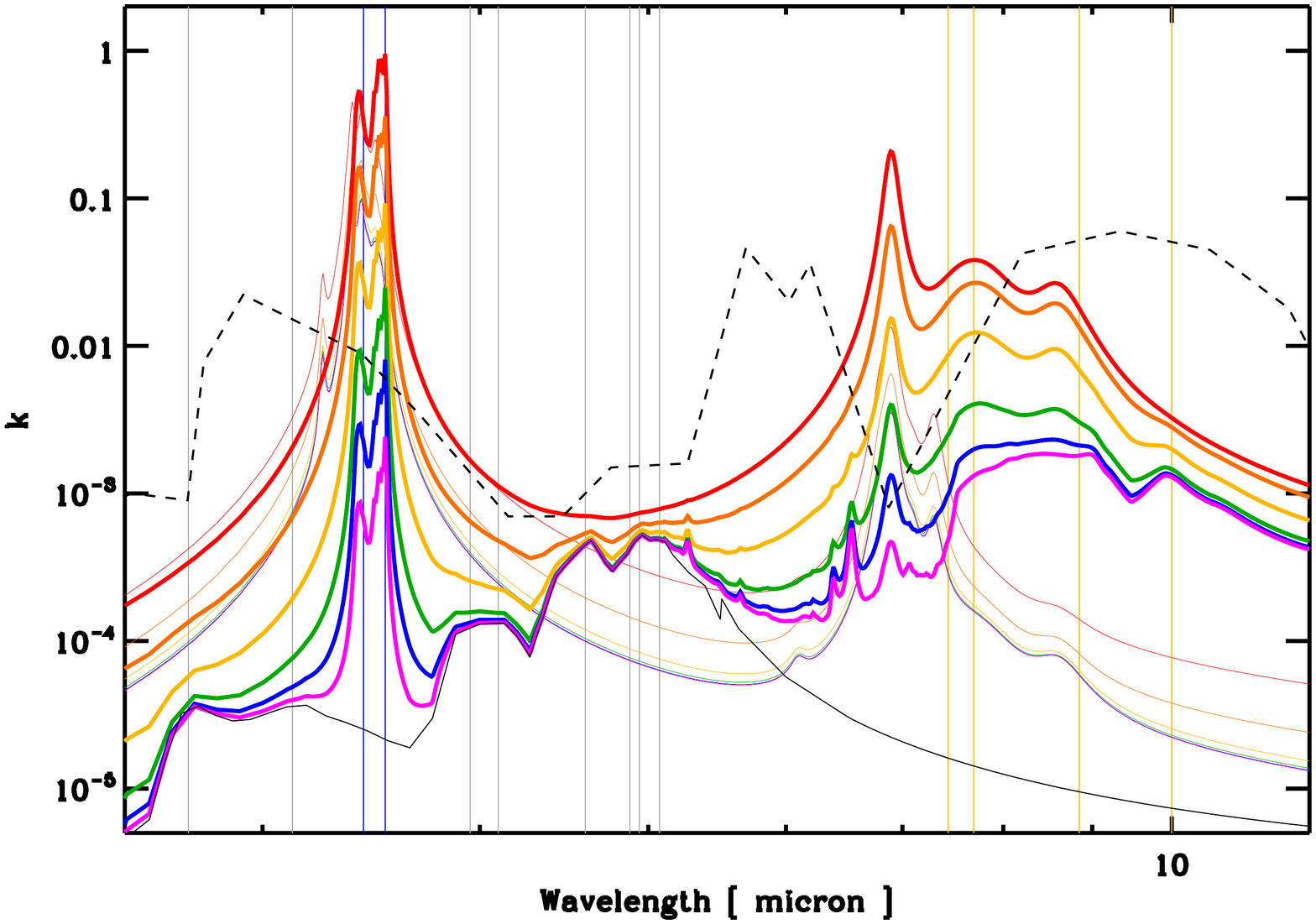}
\end{array} $
\end{center}
\vspace{-1.75cm}
\hspace*{-0.9cm}
\begin{center} $
\begin{array}{cc}
   \includegraphics[width=9.0cm]{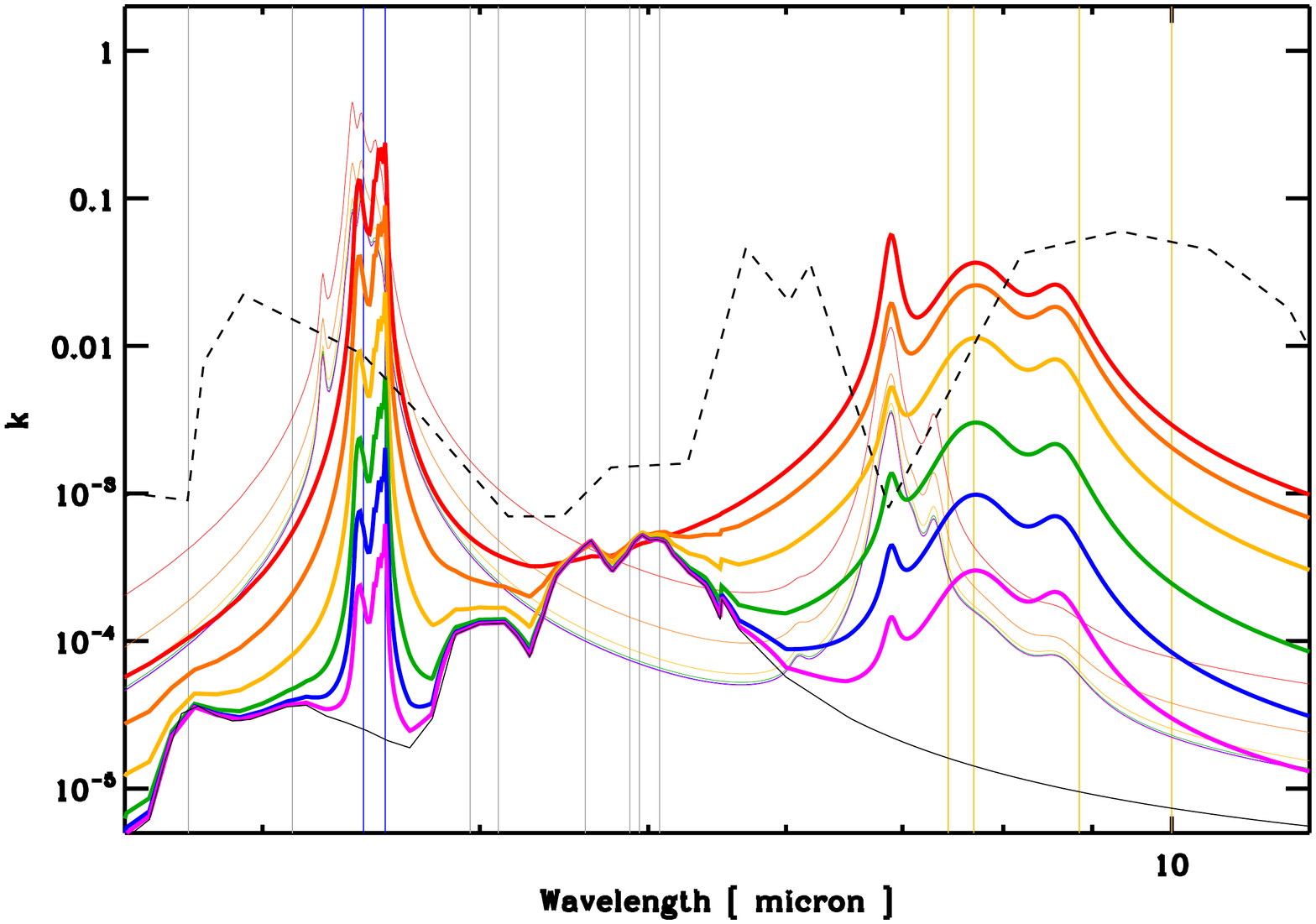}
   \includegraphics[width=9.0cm]{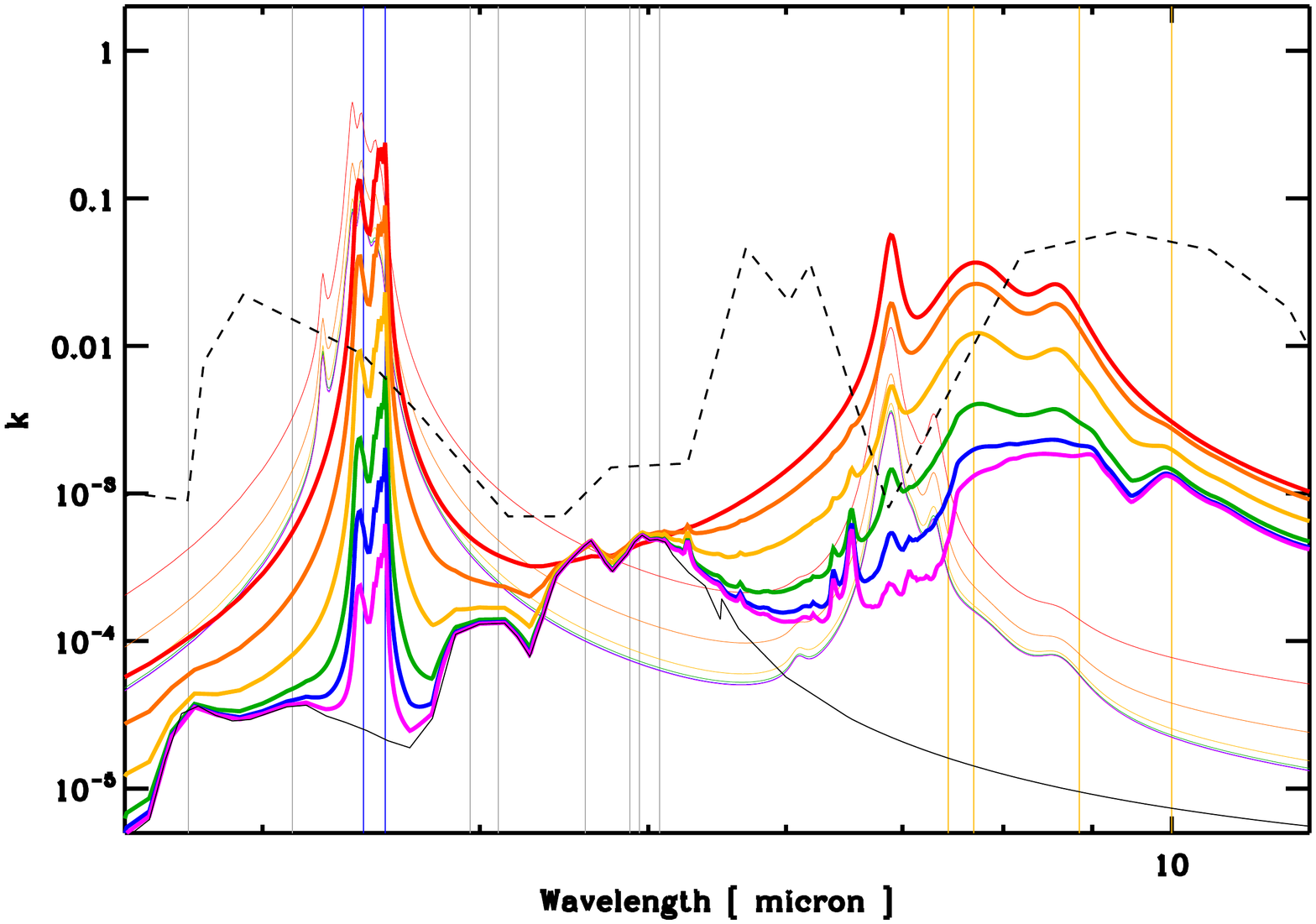}
\end{array} $
\end{center}
\vspace{-1.75cm}
\begin{center} $
\begin{array}{cc}
   \includegraphics[width=9.0cm]{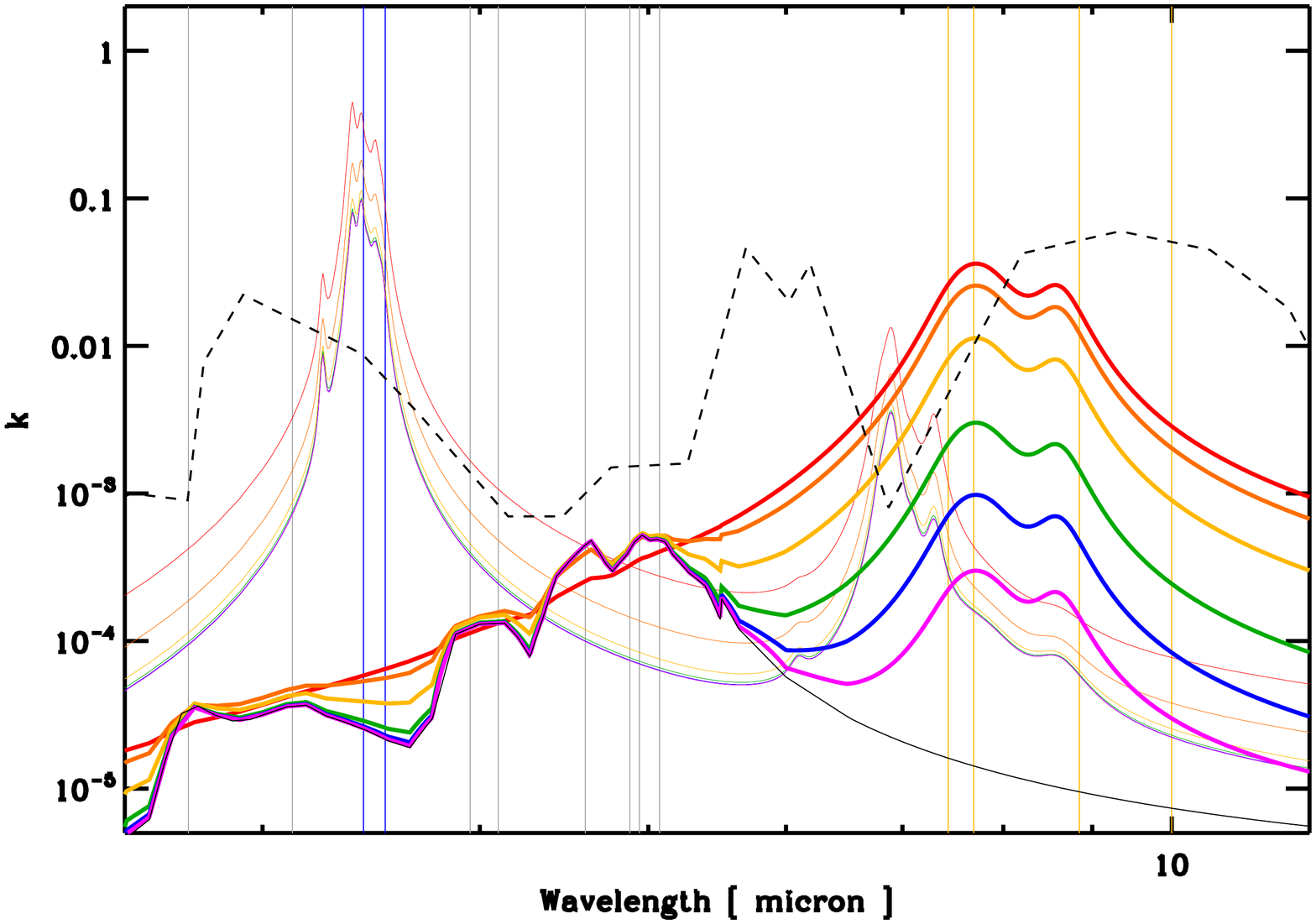} 
   \includegraphics[width=9.0cm]{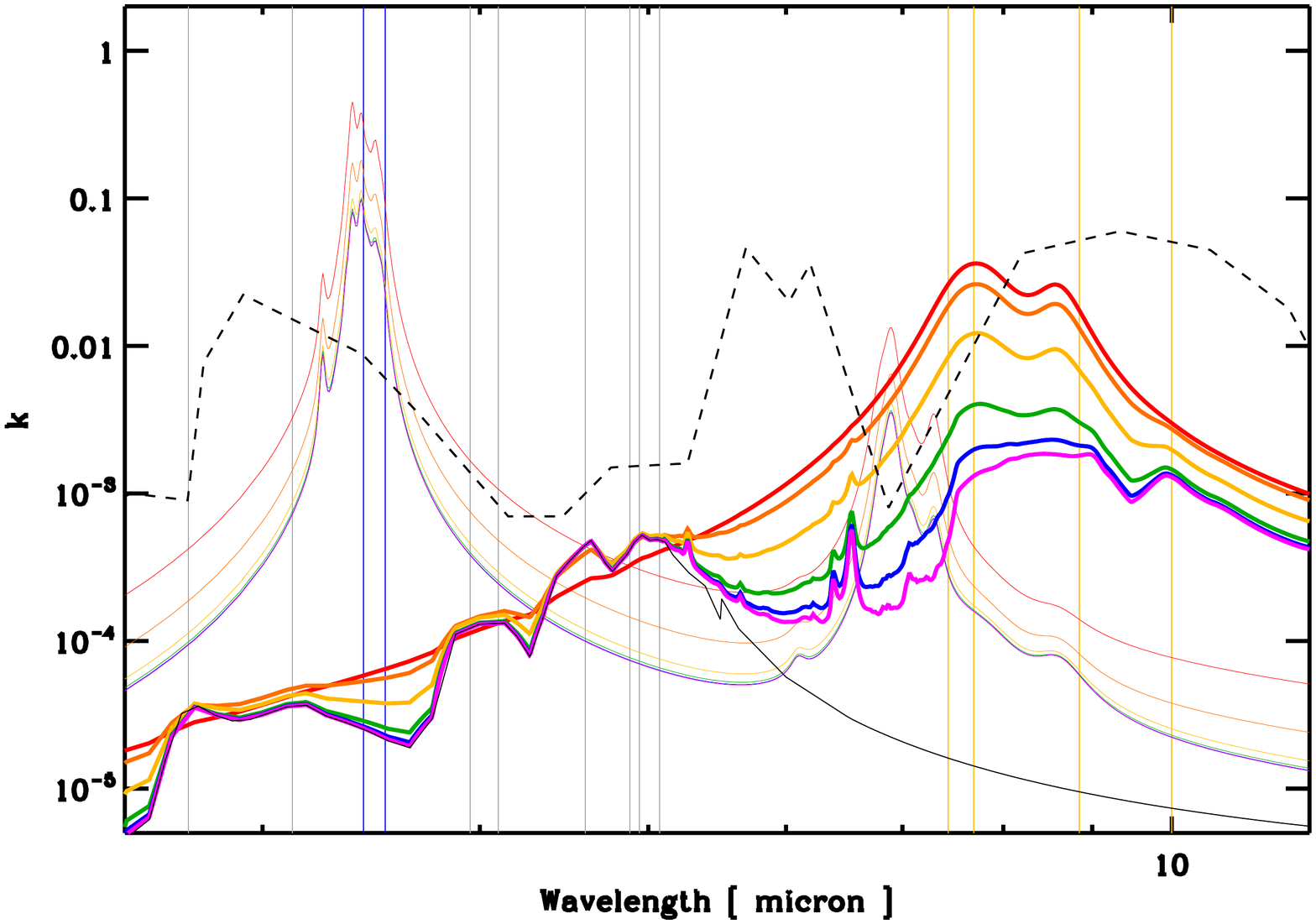}
\end{array} $
\end{center}
\vspace{-1.0cm}
      \caption{A zoom into the IR band region of $k$ for nano-diamonds. The lines and colour coding are the same as in Fig. \ref{fig_k_nanod}. The fractional surface hydrogen coverage $f_{\rm H} = 1, 0.25$ and 0, from top to bottom, respectively. The left panels show the `pristine' bulk diamond data and the right panels that for neutron-irradiated bulk diamond \citep{1998A&A...336L..41H}.}
      \label{fig_k_zoom_nanod}
\end{figure*}

\begin{figure*}
\centering
\begin{center} $
\begin{array}{cc}
   \includegraphics[width=9.0cm]{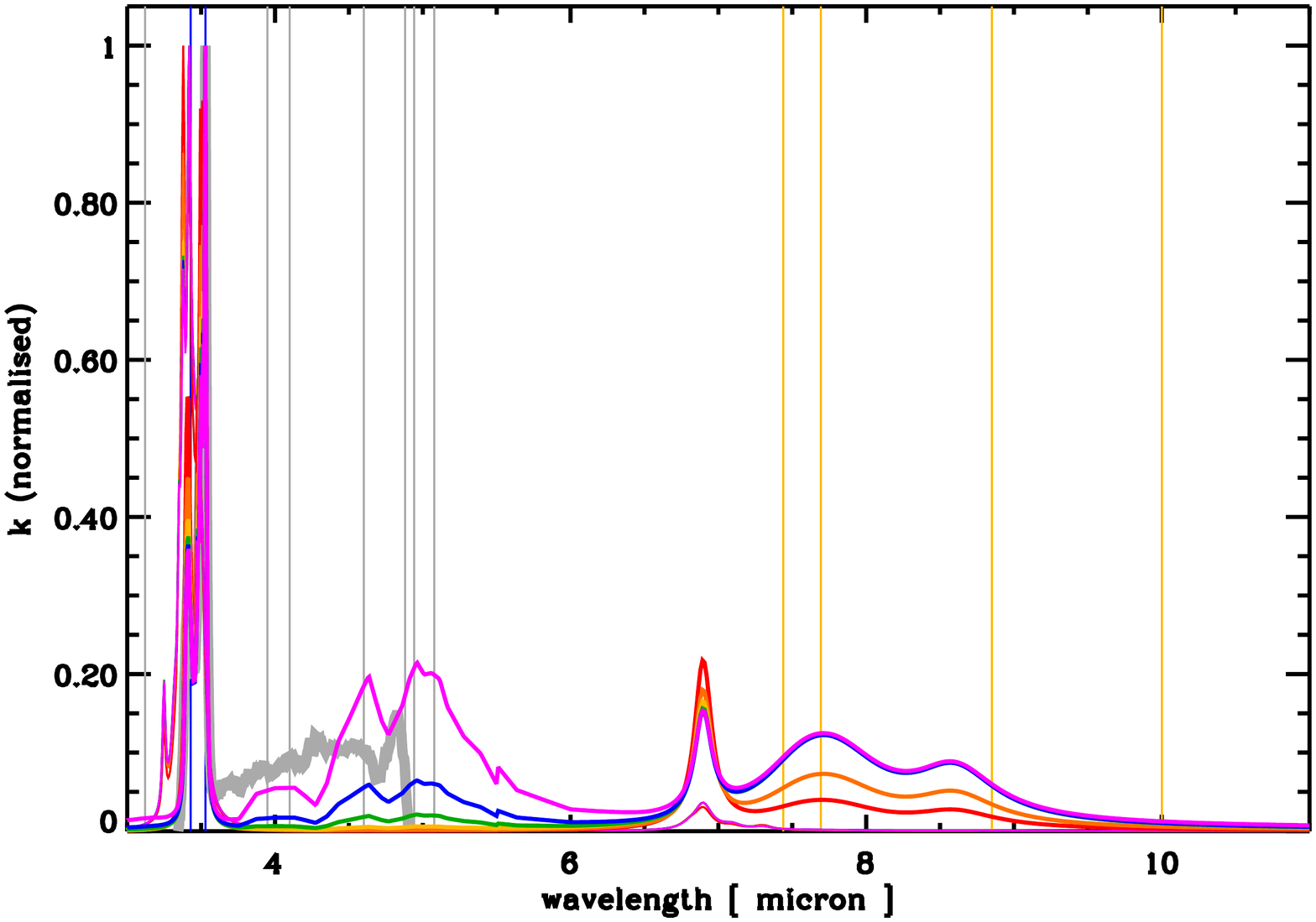}
   \includegraphics[width=9.0cm]{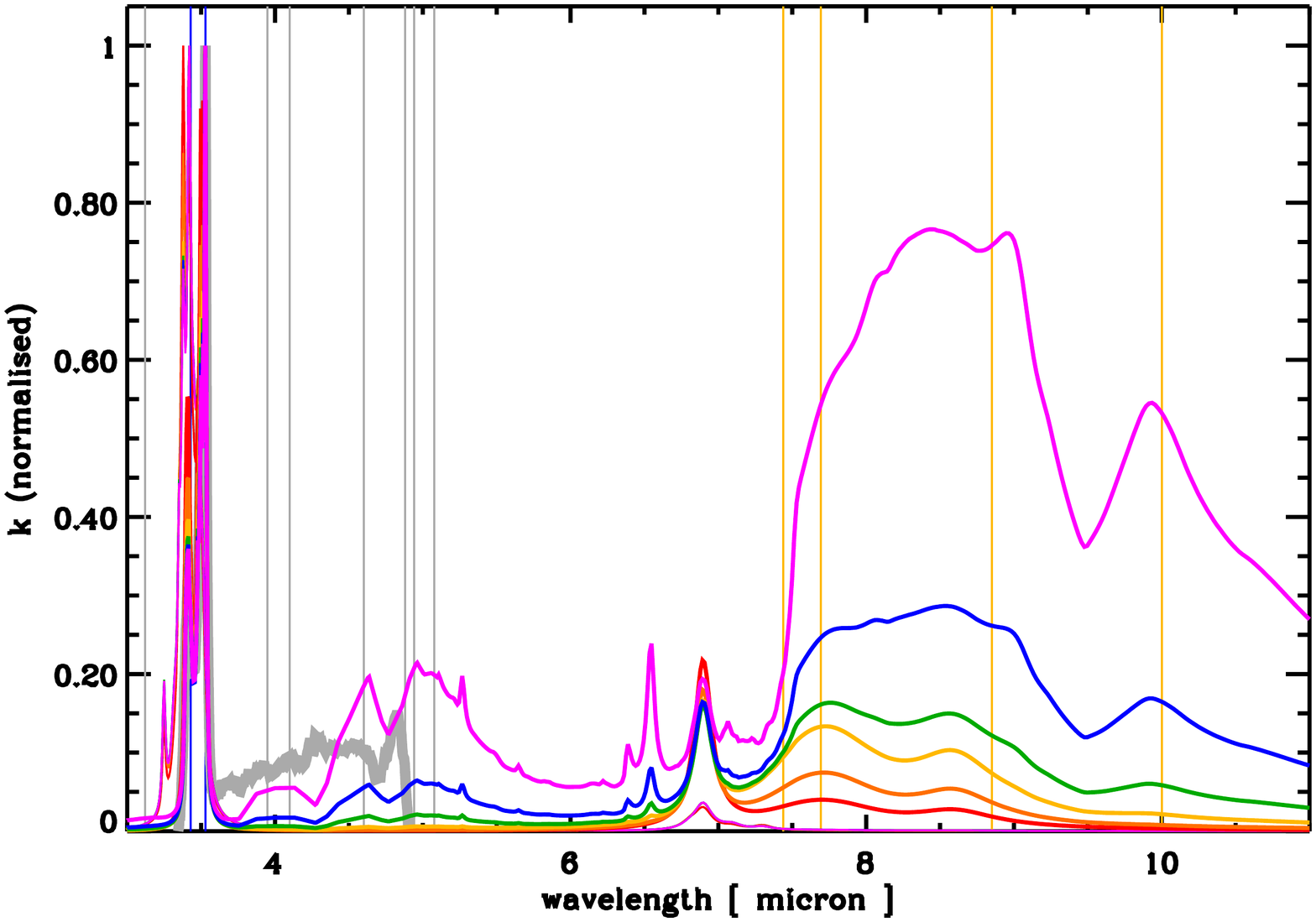}
\end{array} $
\end{center}
\vspace{-1.75cm}
\hspace*{-0.9cm}
\begin{center} $
\begin{array}{cc}
   \includegraphics[width=9.0cm]{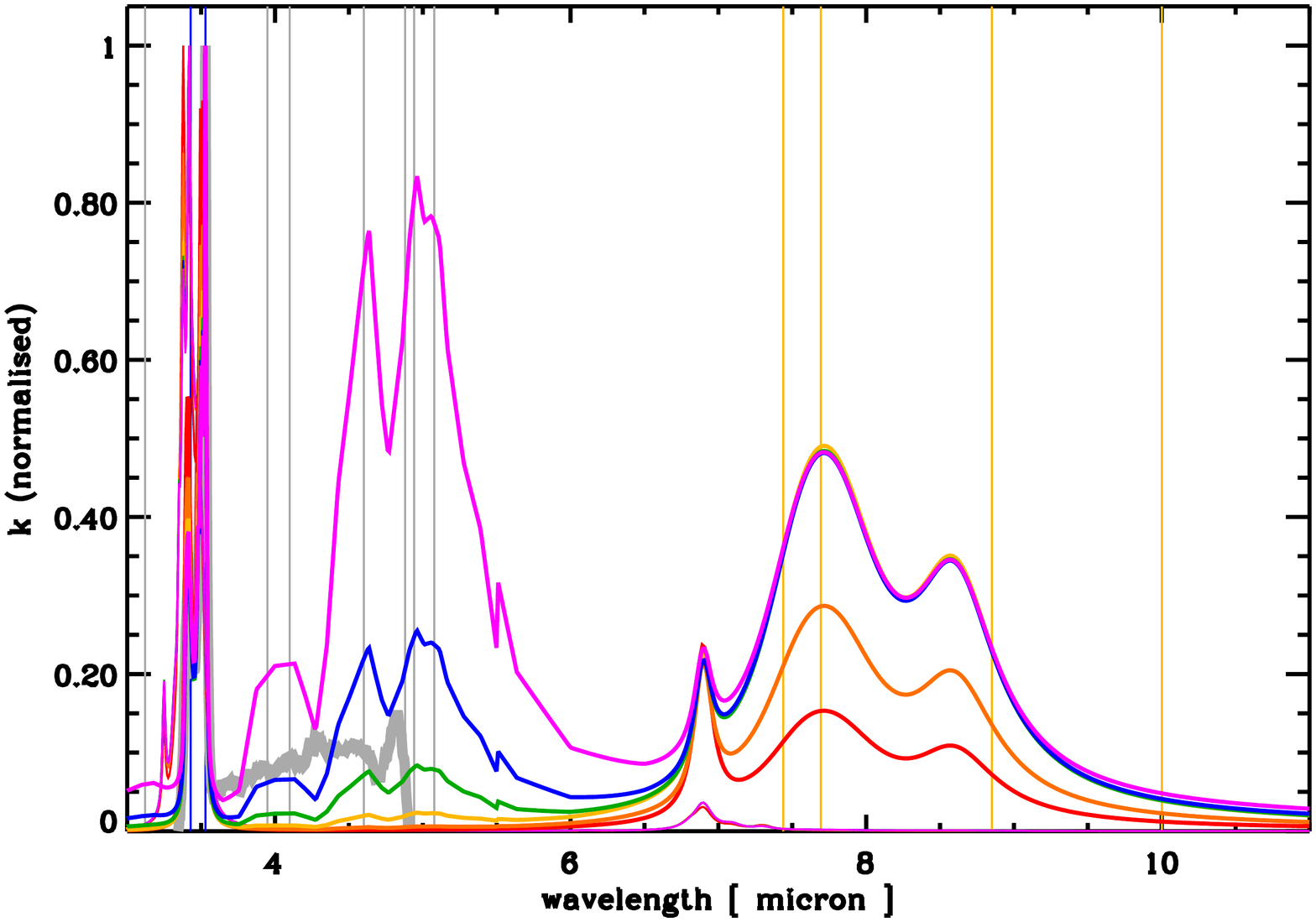}
   \includegraphics[width=9.0cm]{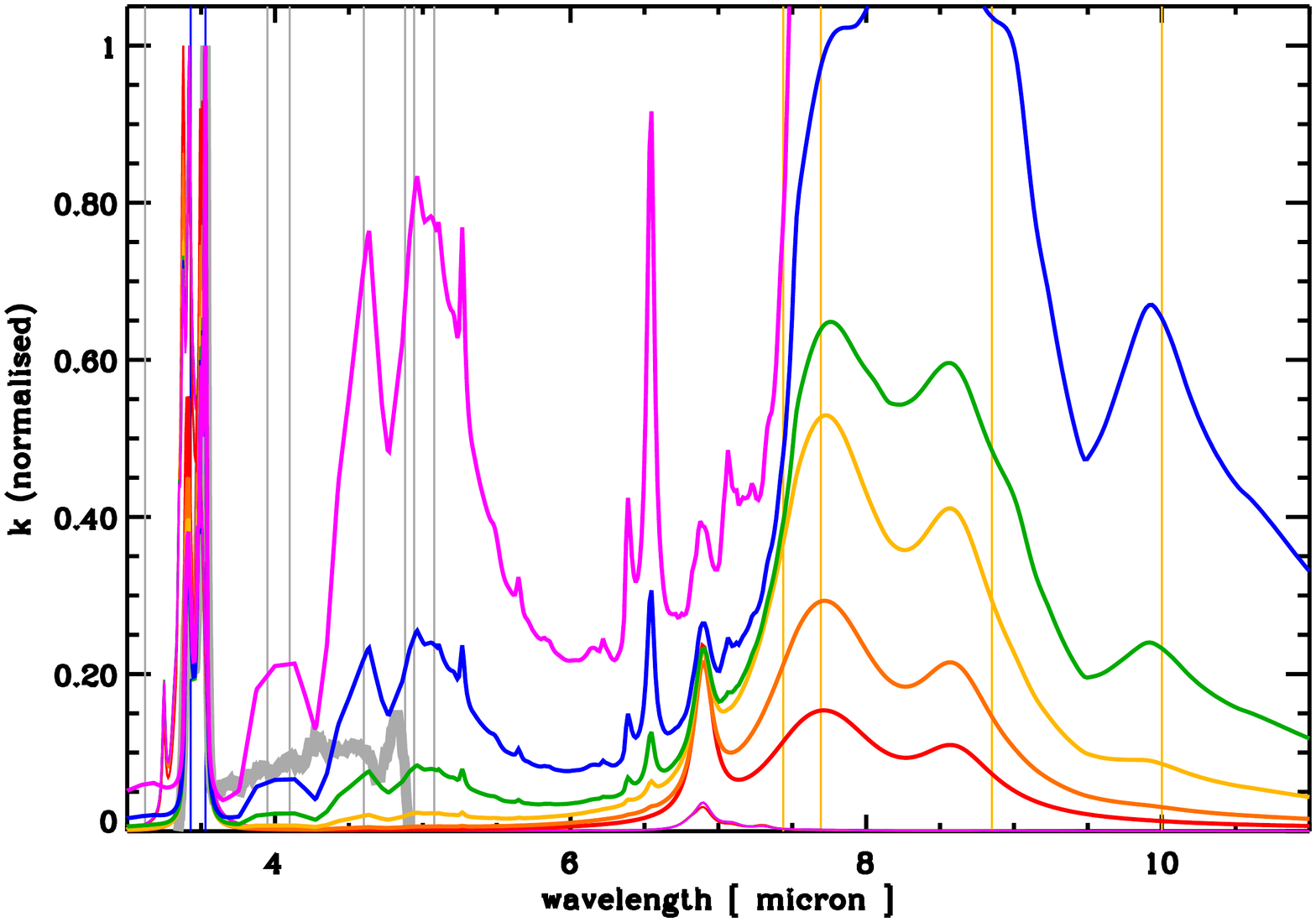}
\end{array} $
\end{center}
\vspace{-1.0cm}
       \caption{The IR band region of $k$ for nano-diamonds in the $3-11\,\mu$m wavelength region, normalised to the peak of the bands in the $3-4\,\mu$m region on a linear-linear scale. The lines and colour coding are the same as in Fig. \ref{fig_k_nanod}. Here the non-hydrogenated nano-diamond data, $f_{\rm H} = 0$, is not shown. $f_{\rm H} = 1$ (top), 0.25 (bottom). The left panels show the `pristine' bulk diamond data and the right panels that for neutron-irradiated bulk diamond \citep{1998A&A...336L..41H}.}
      \label{fig_k_spect_nanod}
\end{figure*}

In order to better show the IR spectral properties derived for nano-diamonds Fig.~\ref{fig_k_zoom_nanod} shows the IR band region for the derived nano-diamond properties. Fig.~\ref{fig_k_spect_nanod} shows these same data normalised to the peak of the CH$_n$ bands in the $3-4\,\mu$m region and on a linear-linear scale in order to better show the transition from bulk-like, non-irradiated, and irradiated, N-poor 100\,nm particles (violet line) to molecule-like species ($a = 0.5$\,nm) dominated by the surface-originated CH$_n$ and CC bonds (red line). 

Note that the form of the spectrum in the $3-4\,\mu$m wavelength region, and in particular, the ratio of the intensities of the $3.43\,\mu$m  and $3.53\,\mu$m bands is dependent upon the particle size because the CH and CH$_2$ surface abundances for spherically-approximated and statistically-averaged nano-diamond particles are size-dependent \citep{2020_Jones_nd_CHn_ratios}. The IR band ratios  will be addressed in more detail in a follow-up paper. 
It should also be noted that these data, for simplicity, assumed that dehydrogenation from CH and CH$_2$ groups was proportionate, {i.e.}, the [CH]/[CH$_2$] ratio is independent of the degree of dehydrogenation. However, as \cite{2014Nanot..25R5702B} point out tertiary C$-$H bonds ($>$\hspace*{-0.25cm}$-$C$-$H) are more easily deprotonated than secondary C$-$H bonds ($>$C$<^{\rm H}_{\rm H}$) on facets, edges, and vertices. Thus, deprotonation, {i.e.}, hydrogen abstraction or dehydrogenation by H$^+$ loss, is not actually proportionate but somewhat more likely to occur from facets.  However, it is not clear that this trend also holds for (quasi-)spherical particles where facets are likely to be small with respect to the particle `radius', whereas for euhedral particles the facet dimensions are necessarily of the same linear extent as the particle's effective radius \citep[{e.g.},][]{2020_Jones_nd_CHn_ratios}.

\subsection{Optical properties $Q_{\rm ext}$ and $Q_{\rm abs}$}
\label{sect_Qs}

\begin{figure*}
\centering
\begin{center} $
\begin{array}{cc}
   \includegraphics[width=9.0cm]{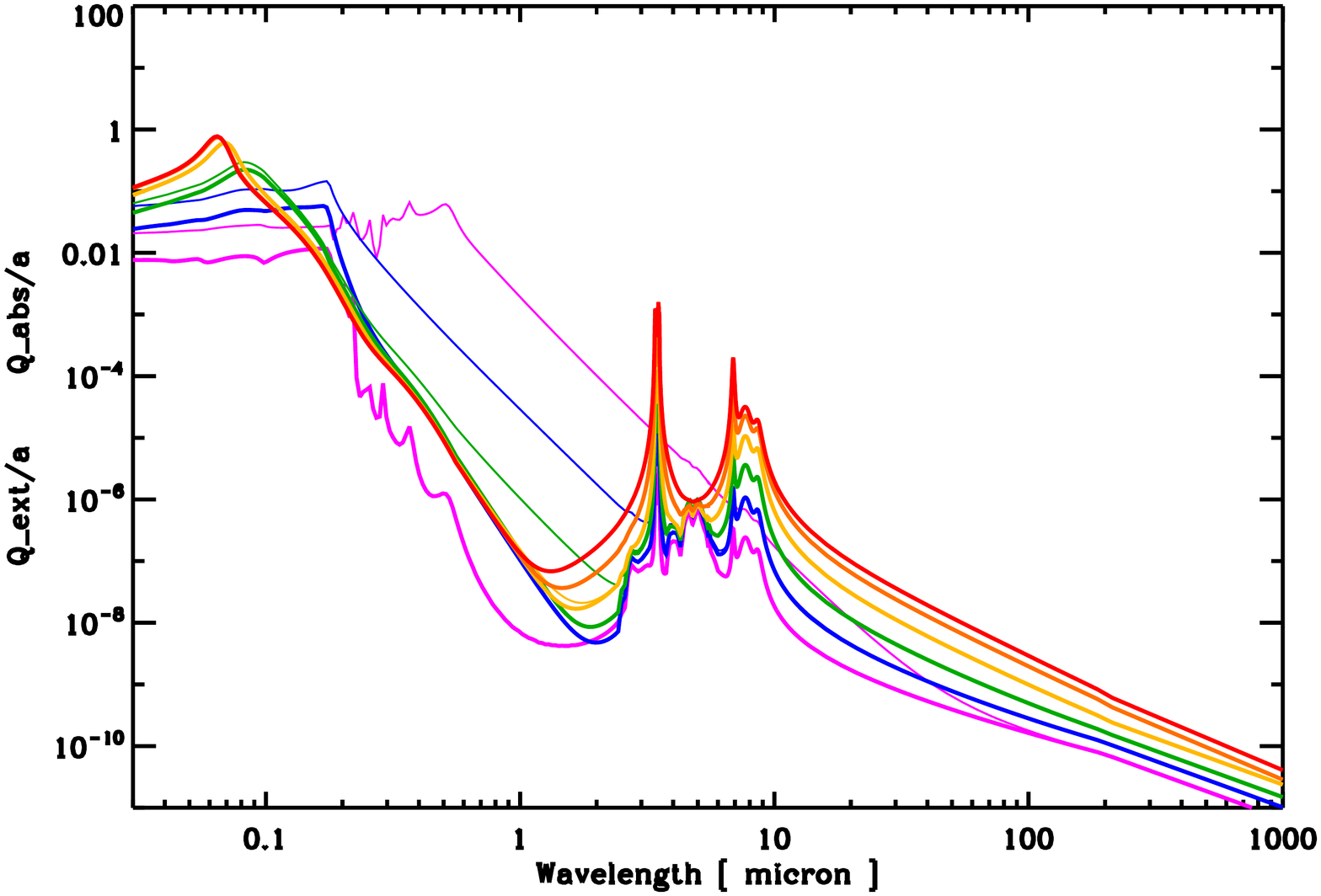}
   \includegraphics[width=9.0cm]{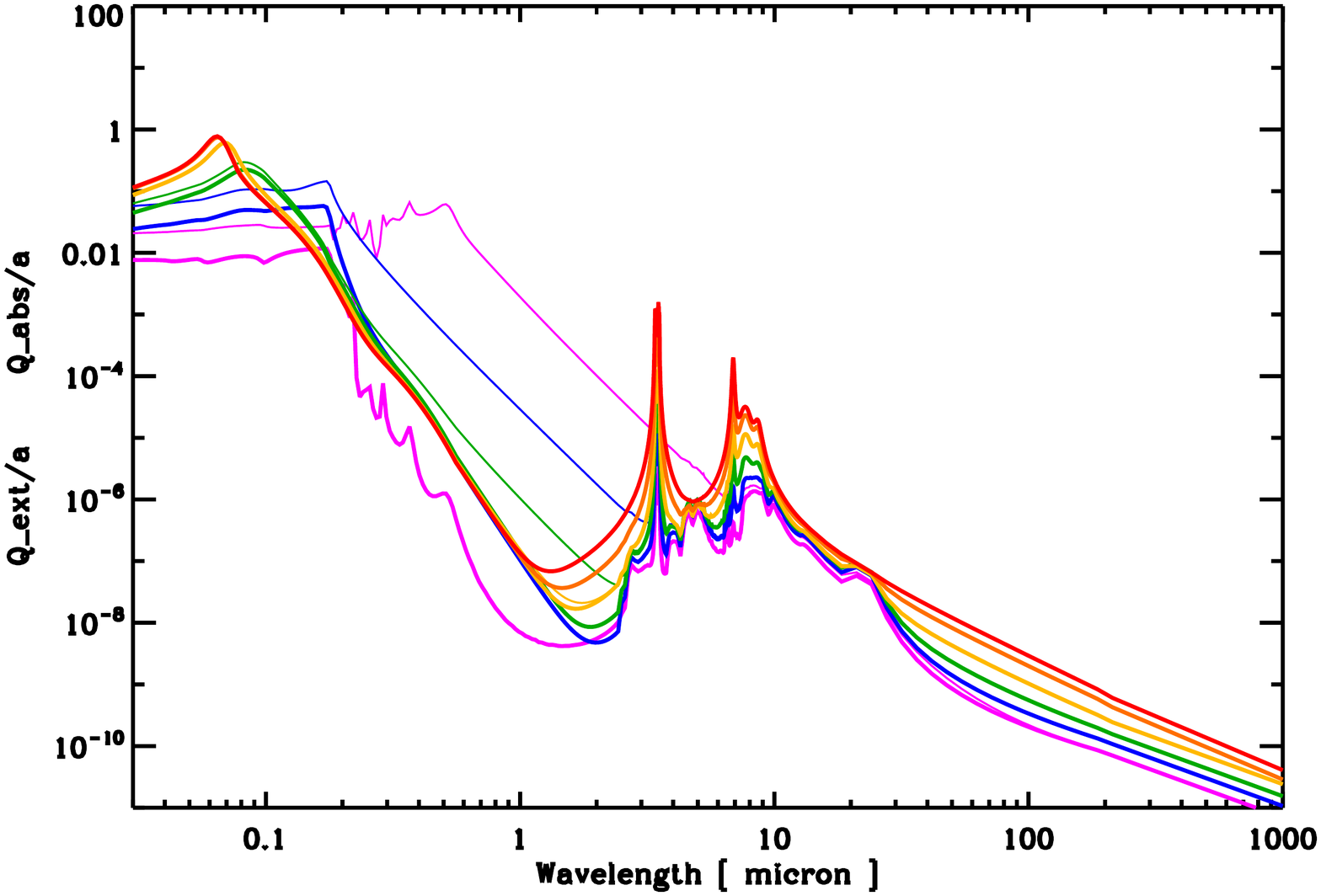}
\end{array} $
\end{center}
\vspace{-1.75cm}
\hspace*{-0.9cm}
\begin{center} $
\begin{array}{cc}
   \includegraphics[width=9.0cm]{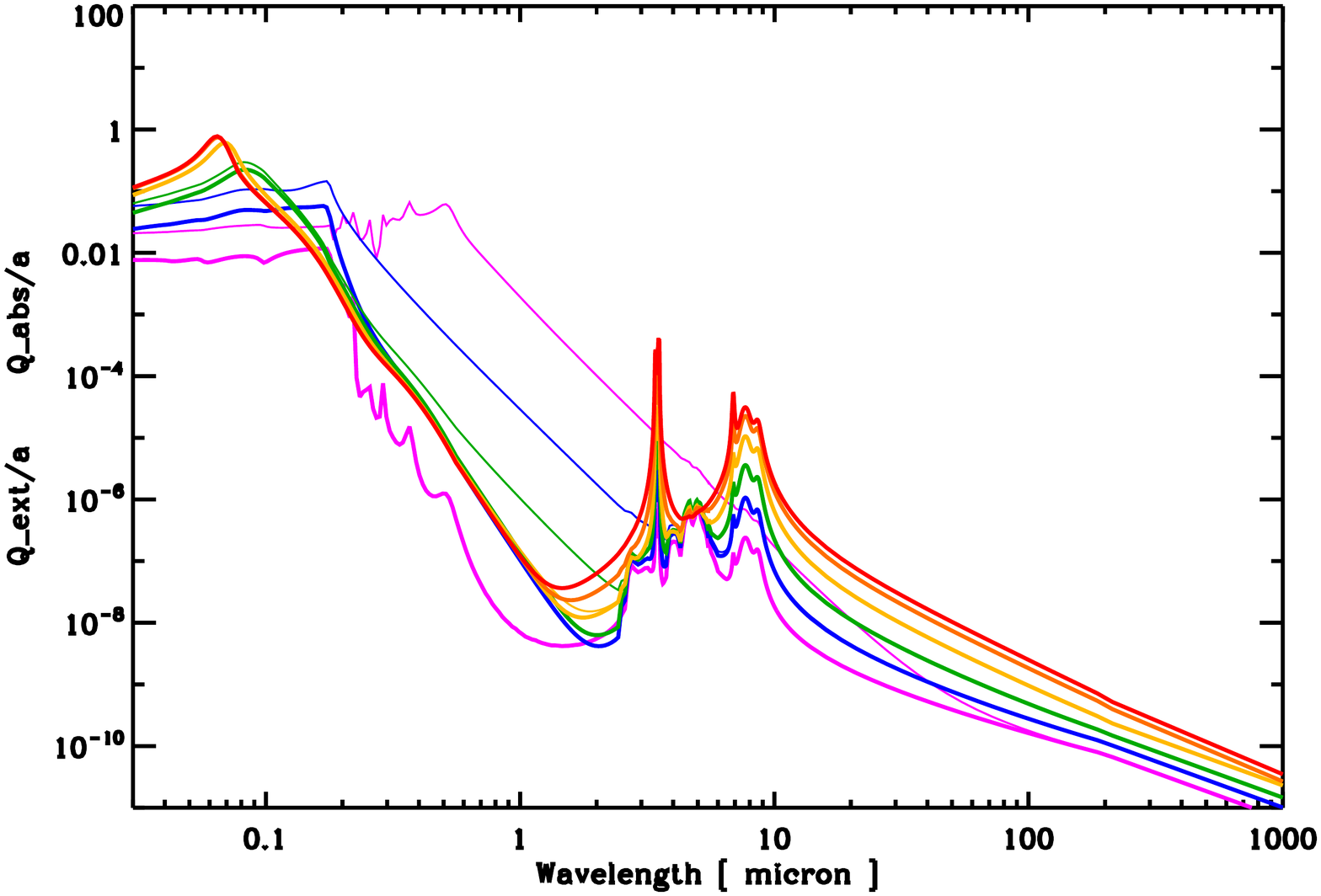}
   \includegraphics[width=9.0cm]{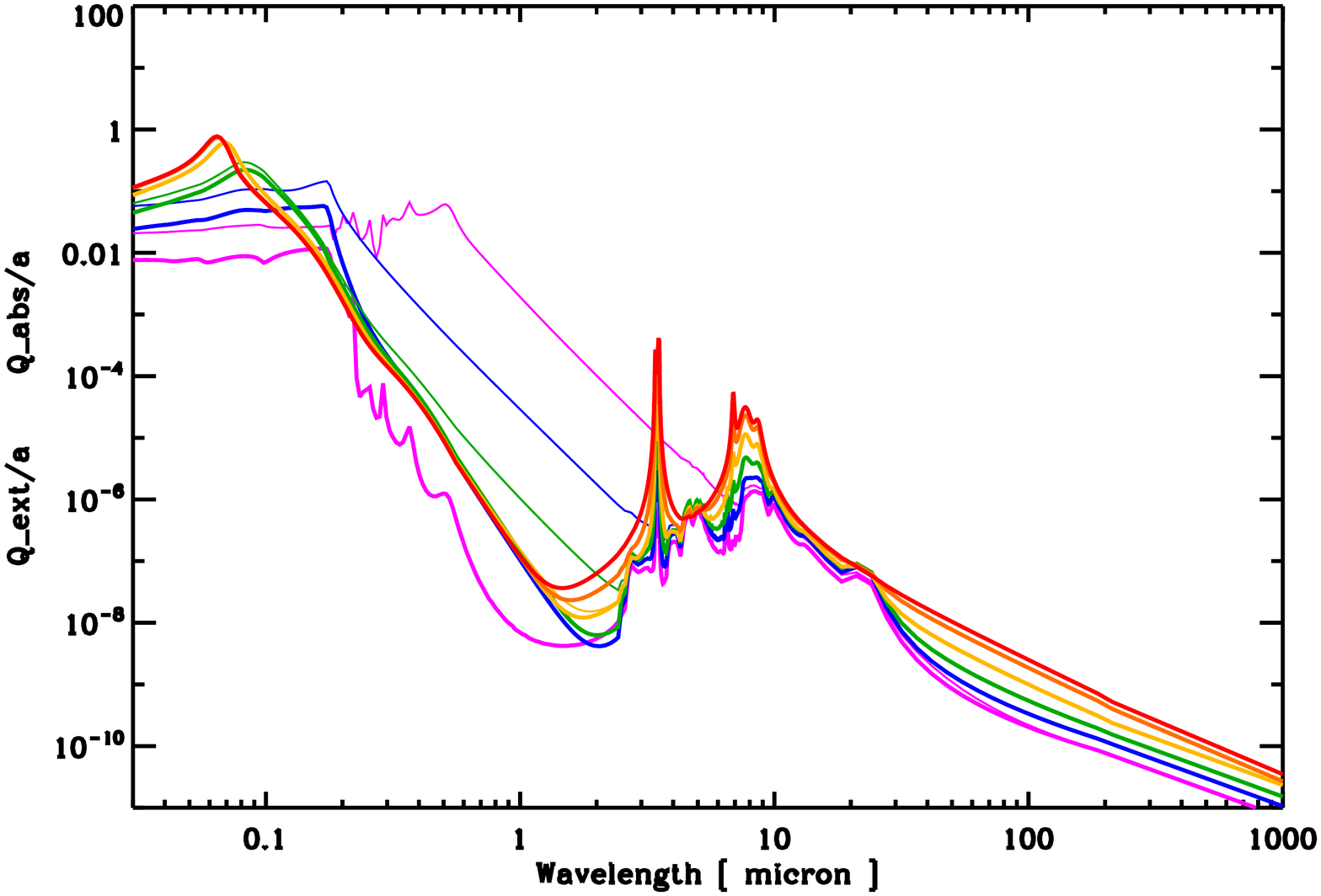}
\end{array} $
\end{center}
\vspace{-1.75cm}
\begin{center} $
\begin{array}{cc}
   \includegraphics[width=9.0cm]{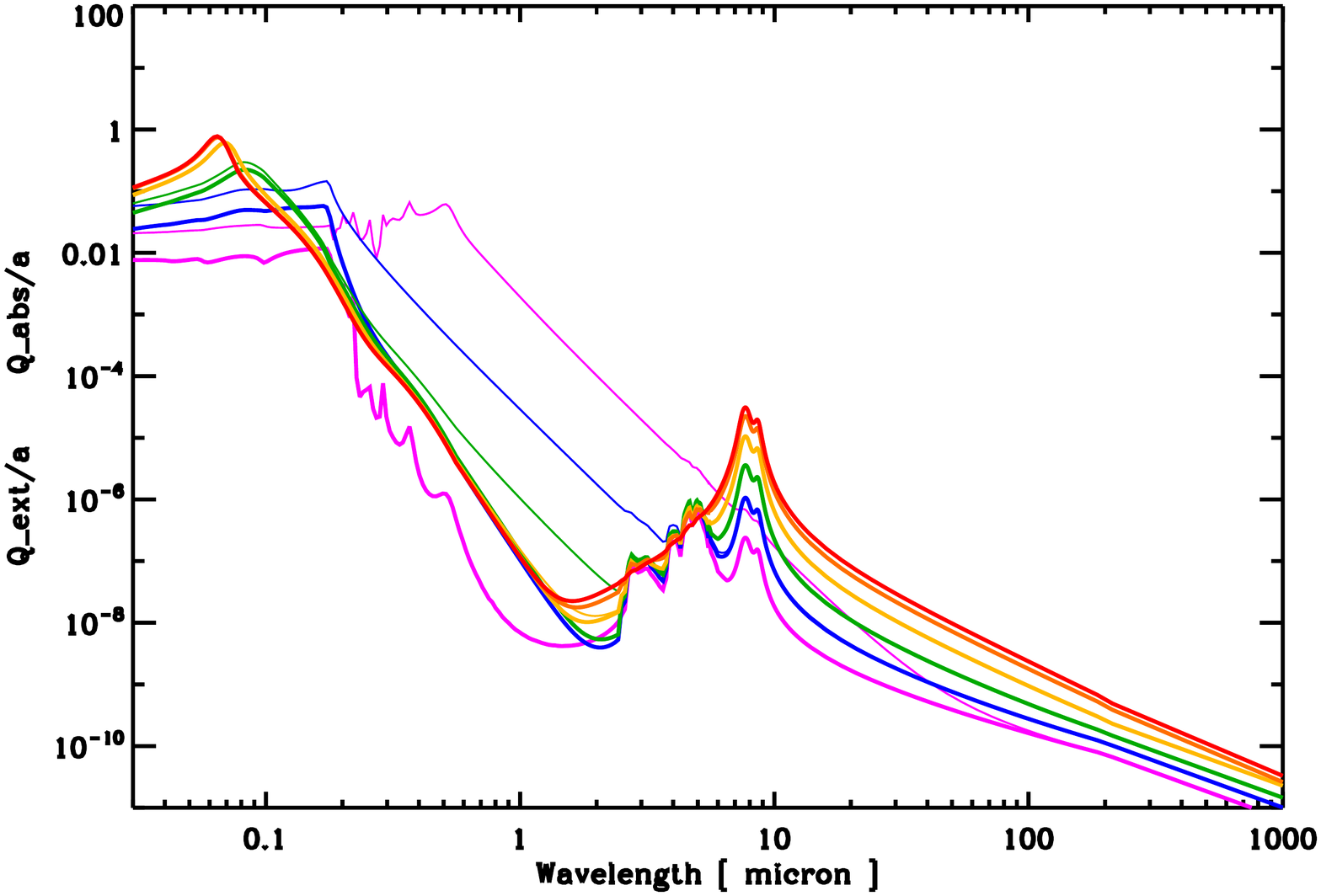} 
   \includegraphics[width=9.0cm]{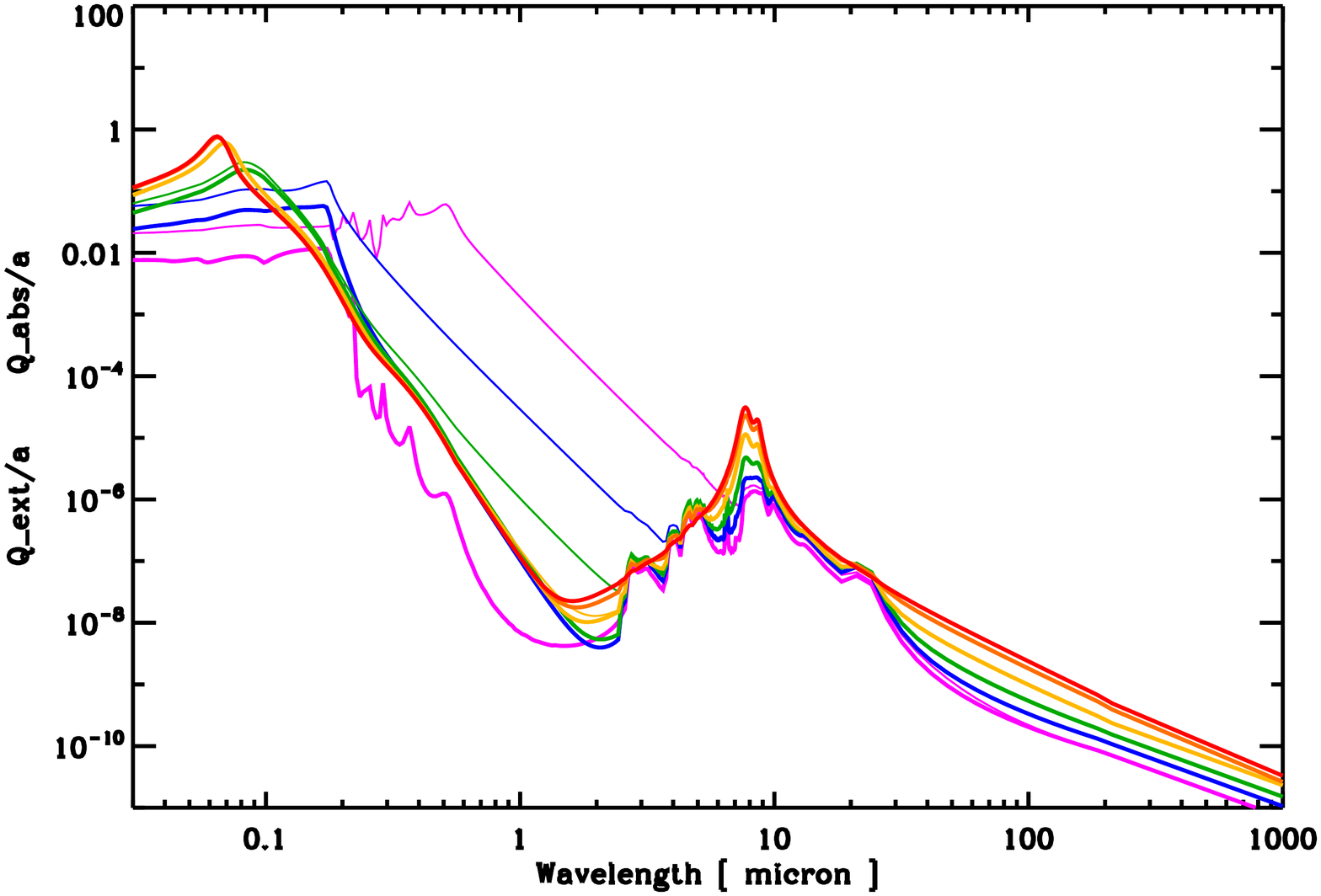}
\end{array} $
\end{center}
\vspace{-1.0cm}
      \caption{Nano-diamond optical properties (in units of nm$^{-1}$) plotted as $Q_{\rm ext}/a$ (thin lines) and $Q_{\rm abs}/a$ (thick lines). The lines and colour coding are the same as in Fig. \ref{fig_k_nanod}.}
      \label{fig_nanod_Qextabs}
\end{figure*}

Using the derived nano-diamond optical constants, $m(n,k)$, we have calculated the optical properties $Q_{\rm ext}$ and $Q_{\rm abs}$. These are plotted as $Q_{\rm ext}/a$ and $Q_{\rm abs}/a$ in units of nm$^{-1}$ in Fig. \ref{fig_nanod_Qextabs} for particle radii of 0.5, 1, 3, 10, 30, and 100\,nm (red, orange, yellow, green, blue, and violet lines, respectively). In this figure the absorption efficiency factor has been emphasised (thick lines) because it is this quantity, in both absorption and emission, {i.e.}, $Q_{\rm abs} = Q_{\rm em}$, that primarily determines the nano-diamond temperatures, their emission and therefore their observability in circumstellar and interstellar media. Indeed, nano-diamonds have only ever been observed in emission in astronomical objects.

What can be seen in the six panels of Fig. \ref{fig_nanod_Qextabs} is that in absorption (emission) the CH$_n$ band strengths in the $3-4\,\mu$m obviously decrease with increasing dehydrogenation (top to bottom) and that radiation damage induces bands longward of $10\,\mu$m (left to right). Interstellar dust equilibrium temperatures are typically $\simeq 20-50$\,K and the peak temperatures for the smallest stochastically-heated dust of the order of hundreds of degrees K. Equivalently, dust in circumstellar regions can be considerably warmer, exhibiting temperatures from hundreds to perhaps thousands of degrees K, be they thermal equilibrium or stochastic heating peak temperatures. These are also the typical temperatures that nano-diamonds will experience and so their emission will typically peak in the $50-150\,\mu$m ($20-50\,\mu$m) wavelength region for thermal equilibrium emission (stochastically-heated nano-dust) in the ISM and in the $1-10\,\mu$m wavelength region in circumstellar media. 

Looking at the $Q_{\rm abs}$ data it is clear that in the ISM nano-diamonds will tend to cool, {i.e.}, emit most, through their CC bands  and a relatively featureless mid-IR continuum emission. If the interior or bulk of the larger nano-diamonds ($a \sim 100$\,nm) is defective, and/or has a significant nitrogen hetero-atom content, then the nano-diamonds will also emit via these defect and/or nitrogen hetero-atom modes in the mid-IR, {i.e.}, in the $\sim 10-40\,\mu$m region. Meanwhile, the smallest nano-diamonds ($a \simeq 1$\,nm) will make narrow band contributions via their CH  bands in the $\sim 3-4\,\mu$m region and broader band contributions through their CC modes in the $\sim 6-8\,\mu$m region. These nano-diamond thermal emission properties are explored in detail in a follow-up paper.

We now use these $Q_{\rm abs}$ ($= Q_{\rm em}$) data to investigate the likely presence of nano-diamonds in the ambient diffuse ISM.

\section{Are nano-diamonds present in the ISM?}
\label{sect_nd_ISM}

To date nano-diamonds have only been clearly detected, via their CH$_n$ emission bands at $\simeq 3.43$ and $3.53\,\mu$m,  in the circumstellar discs around the hot stars HR\,4049, Elias\,1, and HD\,97048 \cite[{e.g.},][]{1999ApJ...521L.133G,2002A&A...384..568V,2004ApJ...614L.129H,2009ApJ...693..610G}. The above-derived nano-diamond $n$ and $k$ data can now be put to good use to investigate whether they could  possibly even exist in the diffuse ISM. Their processing and survival in the dics around the stars HR\,4049, Elias\,1, and HD\,97048 are explored in a following paper.

Based on the accepted premise that at least some of the analysed meteoritic nano-diamonds are pre-solar, \cite{2020_Jones_nd_CHn_ratios} suggested that nano-diamonds must then exist in the ISM because they were evidently formed far from the solar system,  associated with supernov\ae, and must have traversed the ISM to reach us. Using the abundance of the Xe atoms trapped within the pre-solar nano-diamonds, the so-called Xe-HL component, \cite{2020_Jones_nd_CHn_ratios} reasoned that up to a few percent of the cosmic carbon ought to be present in the ISM in the form of nano-diamonds. 
If this is indeed the case why do we not see any evidence of them there?

\begin{figure*}
\centering
\begin{center} $
\begin{array}{cc}
   \includegraphics[width=9.0cm]{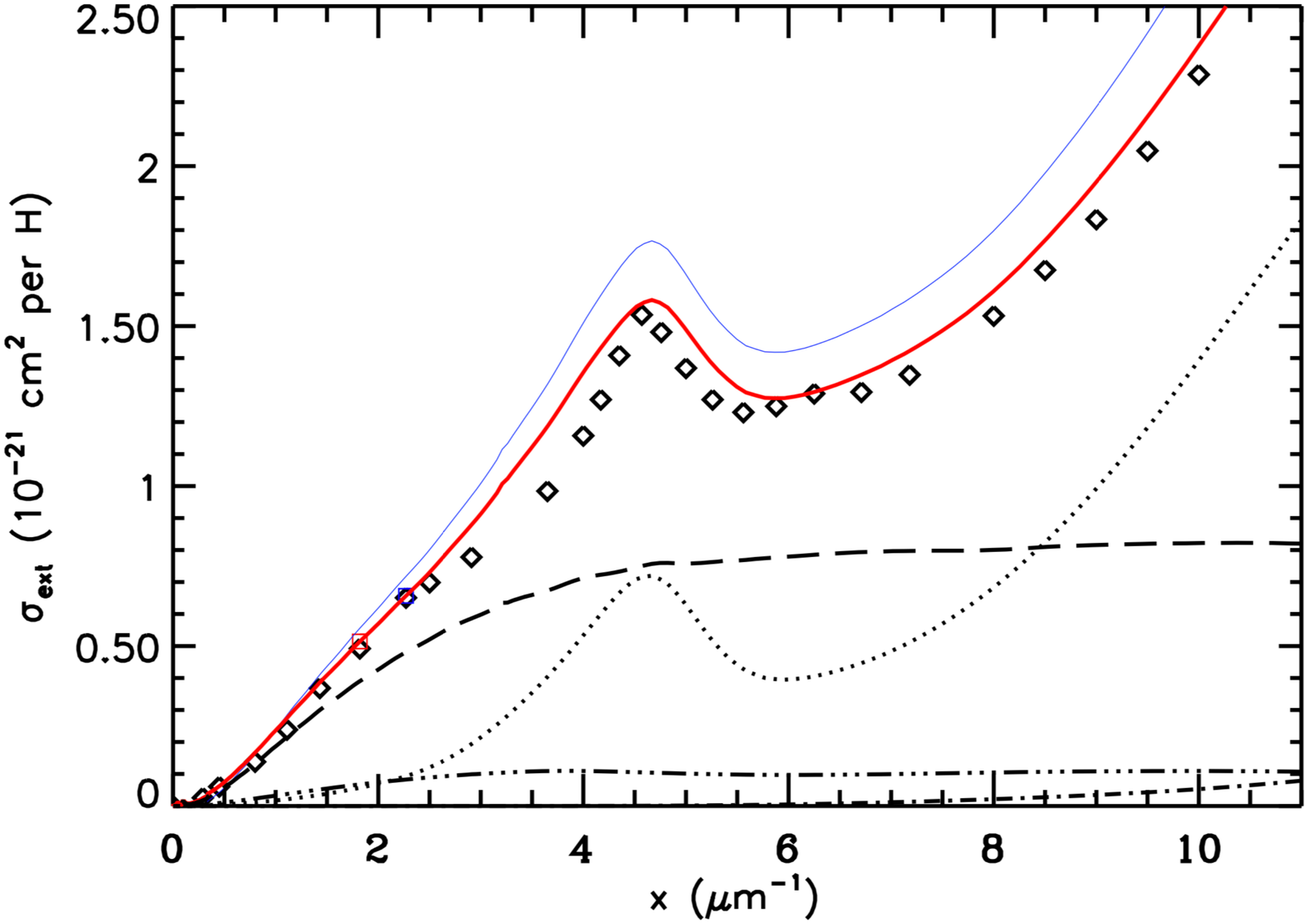}
   \includegraphics[width=9.0cm]{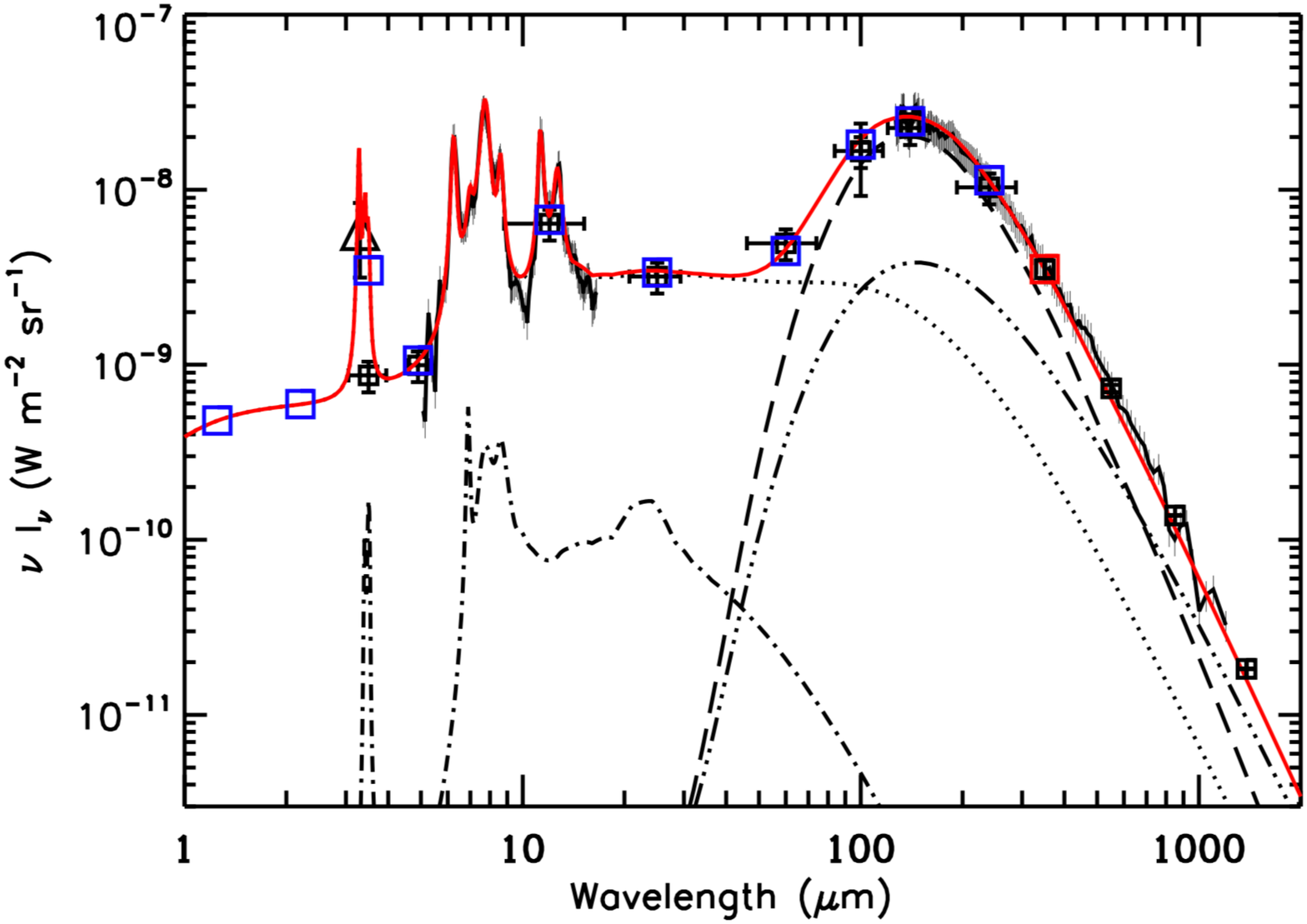}
\end{array} $
\end{center}
\hspace*{-0.9cm}
      \caption{Nano-diamond as a diffuse ISM dust component (dashed-dotted): extinction (left) and emission (right). In this model it is assumed that 17\,ppm of carbon is in the form of fully surface-hydrogenated, n-irradiated nano-diamonds \citep{2020_Jones_nd_CHn_ratios}. For comparison the standard diffuse ISM THEMIS CM dust constituents \citep{2017A&A...602A..46J} are also shown:  a-C (dotted), a-C:H/a-C (dashed-triple-dotted) and a-Sil$_{\rm Fe,FeS}$/a-C (long-dashed). The red lines shows the totals and the blue line (left) shows the E(B-V)-normalised extinction curve. These data were calculated using the DustEM tool \citep{2011A&A...525A.103C}.}
      \label{fig_nanod_THEMIS}
\end{figure*}

\begin{figure}
\centering
   \includegraphics[width=9.0cm]{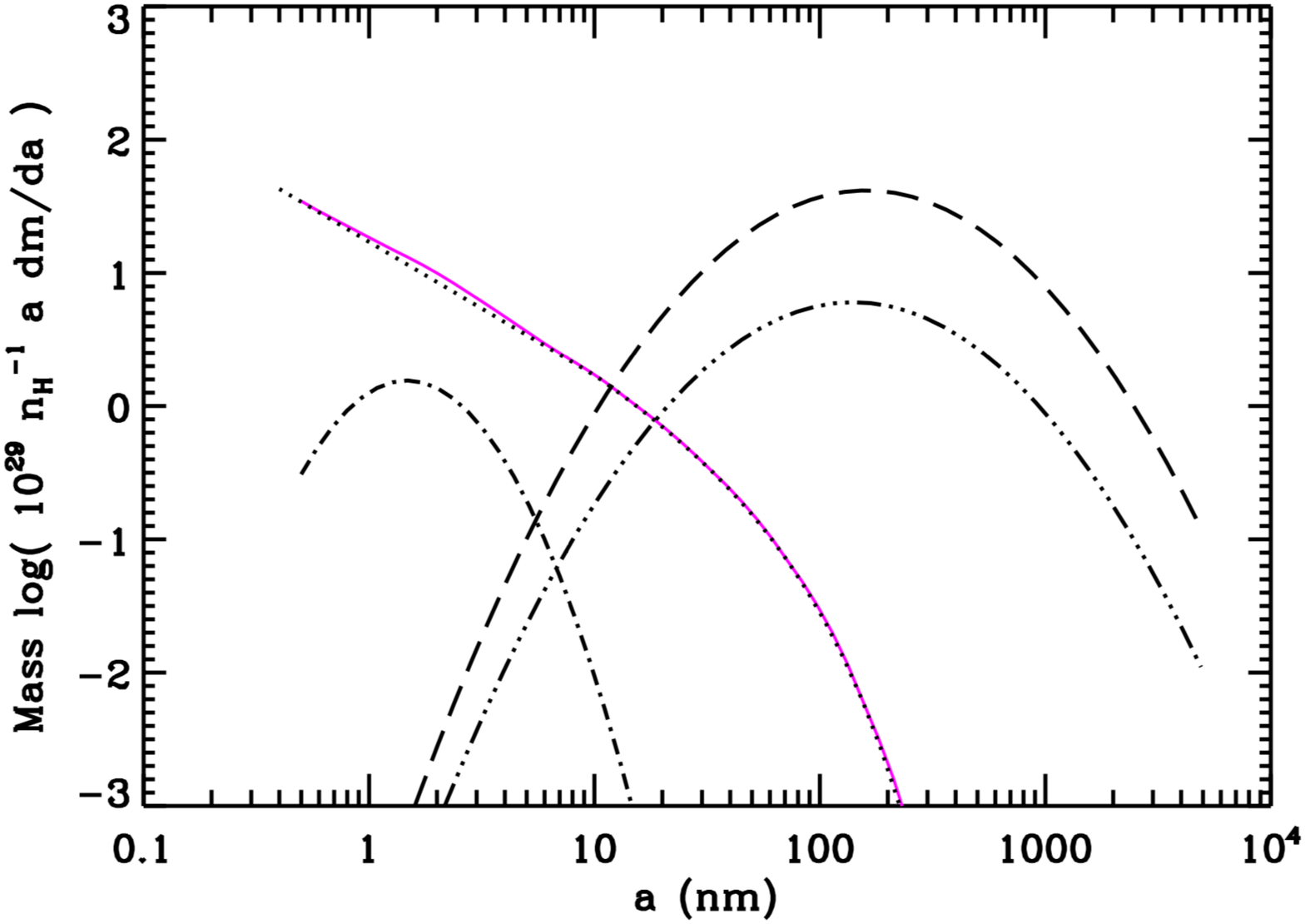}
      \caption{Nano-diamond size distribution assuming 17\,ppm of carbon in the form of fully surface-hydrogenated nano-diamonds \citep{2020_Jones_nd_CHn_ratios}, compared with the standard diffuse ISM THEMIS CM dust constituents \citep{2017A&A...602A..46J} are also shown:  a-C (dotted), a-C:H/a-C (dashed-triple-dotted) and a-Sil$_{\rm Fe,FeS}$/a-C (long-dashed).}
      \label{fig_nanod_sdist}
\end{figure}

The plausibility argument for nano-diamonds in the ISM is demonstrated by the results shown in Fig.~\ref{fig_nanod_THEMIS} obtained using the DustEM tool \citep{2011A&A...525A.103C},\footnote{https://www.ias.u-psud.fr/dustem/} where the THEMIS model \citep{2017A&A...602A..46J}\footnote{https://www.ias.u-psud.fr/themis/} has been augmented with 17\,ppm of carbon in the form of fully surface-hydrogenated nano-diamonds ($f_{\rm H} = 1$) and neutron-irradiated diamond \citep{1998A&A...336L..41H}, with a log-normal size distribution peaking at $\sim 1.5$\,nm (see Fig.~\ref{fig_nanod_sdist}),\footnote{The DustEM input parameters are:  a log-normal size distribution ({logn}) with 50 size bins, $\rho_{\rm nd} = 3.52$\,g\,cm$^{-3}$, $a_{min}$ : $a_{max}$ = 0.5\,nm : 1\,$\mu$m, $a_0 = 0.5$\,nm, $\sigma = 0.6$ and $Y_{\rm nd} = M_{\rm nd}/M_{\rm H} = 2 \times 10^{-4}$.} similar to that of pre-solar nano-diamonds \citep{1996GeCoA..60.4853D}.  This 17\,ppm of carbon represents $\approx 4$\% of the cosmically-available carbon in the form of nano-diamonds and is a factor of two larger than the lower limit of 2\% estimated by \cite{2020_Jones_nd_CHn_ratios} based on primitive meteorites abundances of up to $\simeq 1400$\,ppm \citep{1995GeCoA..59..115H}. The DustEM results in Fig.~\ref{fig_nanod_THEMIS} show that in the diffuse ISM the continuum thermal emission from nano-diamonds, as discussed above, occurs in the $5-50\,\mu$m wavelength region, equivalent to stochastically-averaged nano-diamond temperatures of the order of $100-200$\,K. In Fig.~\ref{fig_nanod_THEMIS} it can be seen that the narrow band CH and CC emission, and also that of the longer wavelength bands ($5-10\,\mu$m), is more than an order of magnitude below that of the THEMIS small a-C grains and that the $10-50\,\mu$m continuum emission from nano-diamonds is more than an order of magnitude below that of the standard diffuse ISM THEMIS model. Thus, at this level of abundance, which is likely to be an upper limit, it is clear that it would be extremely difficult to disentangle any nano-diamond emission features from the aromatic amorphous carbon, a-C, emission bands. On the other hand, if the nano-diamonds were to be de-hydrogenated and their cores of un-irradiated diamond, an unlikely scenario in the diffuse ISM, they would exhibit a practically featureless mid-IR continuum. In this case there is clearly little hope of identifying them through their continuum emission, unless they exhibit particularly strong defect or nitrogen hetero-atom bands. It might therefore prove worthwhile to survey diamond materials for such bands but we note that the nano-diamonds would still have to be significantly more abundant than pre-solar nano-diamonds, with respect to hydrocarbon dust, for there to be any hope of identifying them via mid-IR emission features. 

It is clear from Fig.~\ref{fig_nanod_THEMIS}, and the above analysis, that as much as 4\% of the available carbon could be present in the form of nano-diamonds in the diffuse ISM and not be observable. This figure shows that, even were nano-diamonds to be, hypothetically, this abundant they would make negligible contributions to the diffuse ISM dust extinction and emission (the dashed-dotted lines in Fig.~\ref{fig_nanod_THEMIS}). Thus, it appears to be entirely plausible that nano-diamonds, even were they to be a factor of two more abundant than the pre-solar nano-diamonds extracted from primitive meteorites, could exist in the ISM and remain practically undetectable there.

\begin{table*}[h]
\caption{Diamond optical features at room temperature due to defects and heteroatoms.}
\centering                          
\begin{tabular}{l l c c c c}        
\hline\hline \\[-0.25 cm]                 
Site & $\lambda_c$ [ nm ] & E [ eV ] & origin & fluorescence & occurence \\    
\hline \\[-0.25 cm]                        
ND1        & 393.6 & 3.150 & V$^-$ & blue & UV irrad. \\      
N3           & 415.2  & 2.985 & 3N+V & green & natural \\      
N2           & 477.2 (+ 423, 435, 452, 465) & 2.598 &      &      &       \\      
480          & 480 \ \ \ (broad band) & 2.580 & substituted O ? & green & natural \\      
H4           & {\bf 496.2} & 2.498 & 4N+2V & green & natural or UV irrad. \\      
3H           & {\bf 503.4} & 2.462 & interstitial C & --- & UV irrad. \\      
550          & 550 \ \ \ (broad band) & 2.250 & deformation? & --- & natural \\      
NV$^0$   & {\bf 575.1} (+ other bands) & 2.156 & NV$^0$ & red & natural or UV irrad. \\      
595          & {\bf 594.4} & 2.086 & N? & --- & natural or UV irrad. + annealing \\      
NV$^-$    & {\bf 637.5} (+ other bands) & 1.945 & NV$^-$ & red & natural or UV irrad. + annealing \\      
GR1        & 740.9, 744.4 (+ other bands) & 1.673, 1.665 & V$^0$ & --- & UV irrad. \\      
\hline                                   
\end{tabular}
\tablefoot{These data are  taken from \cite{shigley_table}, where V indicates a vacancy, N an impurity nitrogen atom, with their number preceding, and the 0 or $-$ superscript the charge state of the defect. The band centre positions, $\lambda_c$, boldfaced in column 2 indicate lines and features close to those observed in the Red Rectangle and/or amongst the diffuse interstellar bands (DIBs).}
\label{tab_diamond_optics}
\end{table*}

\subsection{The detection of nano-diamonds by other means}
\label{sect_nd_elsewhere}

Given that the IR bands of nano-diamonds that have been so extensively explored \cite[{e.g.},][]{2018ApJ...856L...9A,1998A&A...330.1080A,2002JChPh.116.1211C,1999ApJ...521L.133G,Jones:2004fu,1995MNRAS.277..986K,1995ApJ...454L.157M} are likely to be un-detectable or un-differentiable from a-C(:H) bands in the diffuse ISM, is it possible that their presence, other than in the few well-studied circumstellar sources, might show up elsewhere? 

Indeed, (nano-)diamond may well have been long-ago detected when \cite{1985MNRAS.215..259D} noted a remarkable similarity between some of the emission features in the Red Rectangle and sharp features observed in the luminescence spectra of terrestrial diamonds. These lines appear to be associated with the, normally forbidden, zero phonon lines of diamond and are superimposed on the broad extended red emission (ERE) that gives this object its characteristic colour. ERE-type emission may actually be a rather common occurrence in carbonaceous materials and is a viable source of the ERE in the Red Rectangle \citep{1997ApJ...482..866D}. More recently, and more specifically, \cite{2006ApJ...639L..63C} proposed nano-diamonds as a possible carrier of the ERE. 

Thus, we are perhaps overdue another long and careful look at the ERE and it would therefore be worthwhile performing high-resolution observations of the ERE to search for fine structure lines typical of diamond similar to those found in the Red Rectangle. 

Naturally-occurring diamond does exhibit optical lines and features, due to heteroatoms (predominantly N atoms) and/or vacancies (V), adding to their beauty and possibly making for observable optical lines, which might be worth searching for in the ISM and circumstellar sources. Table \ref{tab_diamond_optics} indicates some of these defect centre features and their origins \cite[data taken from][]{shigley_table}. Given the stability of nano-diamonds it is possible that they could be a source of some of the broader and more UV-resistant diffuse interstellar bands (DIBs). However, as noted by \cite{1985MNRAS.215..259D} the occurrence and strength of these lines depends upon the heteroatom content and the irradiation history of the diamond. It is also probable that as yet unknown nano-diamond optical line selection rules will apply, forbidding some lines while allowing other unexpected lines.  Hence, the positions of the bands in Table \ref{tab_diamond_optics} may only be indicative of where they might lie in interstellar nano-diamonds because the band positions are also likely to be particle size-dependent. 

All of the above brief considerations probably only serve to highlight how difficult the search for interstellar diamond has been and that it will likely remain so.

\section{Nano-diamond thermal processing}
\label{sect_nd_proc}

Given that intense and harsh radiation field environments, possibly with X-ray flares playing a key role \citep{2009ApJ...693..610G}, seem to be required to produce the characteristic (nano-)diamond IR emission spectra observed close to the hot stars HD\,97048 and Elias\,1, it is worth briefly exploring the effects of nano-diamond thermal processing. This is studied in more detail in a following paper within the context of nano-diamond heating in the highly excited inner regions of proto-planetary discs \citep{2020_Jones_nds_in_discs}. 

The lowest-temperature effect acting upon (nano-)diamonds is most likely dehydrogenation, which occurs at temperatures of the order of $\sim 1200-1500$\,K \citep[{e.g.},][]{FT9938903635}. Complete dehydrogenation  at temperatures $> 1500$\,K will obviously result in the loss of the identifying nano-diamond CH$_n$ spectral signatures in the $3-4\,\mu$m region. At higher temperatures yet, in the range $\sim 2000-2100$\,K, surface (and bulk) diamond reconstruction to sp$^2$ aromatic structures, {i.e.}, `graphitisation' or `aromatisation', will occur \citep[{e.g.},][]{Howes_1962,Fedoseev_etal_1986}.  Sustained at temperatures $> 2000$\,K nano-diamonds can no longer exist as such because they will be `graphitised' and if subsequently maintained at temperatures $\geqslant 2500$\,K the transformed `nano-diamonds' will undergo sublimation \citep[{e.g.},][]{Darken_Gurry_1953,Tsai_etal_2005}. 

The presence of olefinic and aromatic CC stretching and vibrational bands, in the $\lambda = 6.1-6.7\,\mu$m region, within nano-diamonds, as a result of aromatisation, will not change things significantly. This is because CC modes are, in general, about an order of magnitude weaker than the predominant CH$_n$ modes (Table \ref{spectral_bands}) and therefore cannot help to significantly cool hot nano-diamonds in the absence of surface hydrogenation. However, complete `graphitisation' will result in `protective' lower temperatures because aromatic materials are significantly more emissive at mid-IR wavelengths than (nano-)diamonds but then we are no longer dealing with nano-diamonds. 

To summarise, if nano-diamonds do exist in regions with harsh radiation fields, thermal processing will first lead to the dehydrogenation, followed by the re-construction of sp$^3$ diamond material to sp$^2$ aromatic carbon and eventually to their complete destruction by sublimation at temperatures $\geqslant 2500$\,K.

\section{Summary, brief discussion and cautions}
\label{sect_results}

It has recently been shown that regular euhedral nano-diamonds, except at the smallest sizes ($a < 1$\,nm), albeit with a large dispersion, seemingly have [CH]/[CH$_2$] ratios that are incompatible with observations and experimental results \citep{2020_Jones_nd_CHn_ratios}. In this same study spherical nano-diamonds with radii $< 2$\,nm were shown to exhibit large size-to-size variations in their [CH]/[CH$_2$] ratios for close-in-size particles. Nevertheless, the determined ratios were shown to be compatible with observations. Therefore it seems that we need to consider (quasi-)spherical nano-diamonds as the most commensurate with observations and, to this end, an analytical approximation to their [CH]/[CH$_2$] ratios provides a statistically-viable means to smooth out the large size-to-size variations for sub-2\,nm nano-diamonds. This is the approach that we have chosen to adopt here. Analysis from both a modelling and an observational point of view can therefore almost certainly only proceed by considering nano-diamonds in a statistical sense, an approach that we are unavoidably forced into in astrophysical nano-particle dust studies.

As expected the optical constants and optical properties of nano-diamonds derived here show little variation at UV wavelengths, as a function of diamond type or degree of surface hydrogenation, as compared to the IR and longer wavelengths. In the $2-4\,\mu$m spectral region the nano-diamond optical property variations are evidently driven by the surface hydrogenation state and to a lesser extent by  particles size. The spectra in the $3.35-3.55\,\mu$m wavelength range are particularly complex, consisting of at least four clear peaks and three shoulders on these peaks. 

There is inevitably some overlap in the nano-diamond IR band positions with those of a-C(:H) materials, although the bands of the latter generally peak at shorter wavelengths in the $3.35-3.55\,\mu$m window and are therefore generally distinct. The interpretation of the $\sim 3.4\,\mu$m band in any $3.35-3.55\,\mu$m spectra may not always be completely evident because it has a related origin in both aliphatic and nano-diamond sp$^3$ carbons. Thus, care should perhaps be exercised in interpreting observed $3.43\,\mu$m to $3.53\,\mu$m band ratios solely in terms of the $3.43\,\mu$m band originating in nano-diamonds. 

The $4-30\,\mu$m nano-diamond spectral variations are driven by the diamond bulk network structure, and any defects or lack thereof, with contributions at the shorter wavelengths in this region by near-surface carbon-carbon bonds and by any nitrogen atoms present in the structure. Note that the important and perhaps critical effects of nitrogen hetero-atoms are not considered in this work.

\section{Carbon in the dust budget}
\label{sect_C_budget}

In the following sub-sections we briefly consider and summarise our best estimates of the distribution of interstellar carbon within the currently-inferred and most probable dust components. The amount of carbon in all interstellar dust components is likely to be of the order of $206-390$\,ppm \citep{2019A&A...627A..38J}, with its incorporation degree depending upon environment \citep[{e.g.},][]{2016A&A...588A..43J,2015A&A...579A..15K,2016A&A...593C...4Y}. All percentages in the following refer to the fraction of carbon in dust, in the diffuse ISM, in that particular solid phase. Based on the THEMIS model we assume that about half of the cosmic carbon ({i.e.}, $\simeq 200$\,ppm) is in the form of dust in the diffuse ISM; the other half being in atoms, ions, radicals, and molecules. We note that the comparative carbon abundances in pre-solar grains are assumed to be lower limits because of sampling biases, incompleteness and dilution in the ISM. Table \ref{tab_C_in_dust} summarises our best estimates on the partition of carbon across the various dust phases, which, taken in the approximate order of their likely carbon take-up, are outlined in the following sub-sections.

\subsection{Carbon in the a-C(:H) family}
\label{sect_C_aCH}

It is noteworthy that carbonaceous constituents are a significant component of  primitive meteorites and that CI chondrites may contain up to $3-4$\% of carbon. A significant fraction of this carbon is in the form of organic matter, most of which is insoluble macromolecular organic matter with hydrogen and nitrogen isotopic anomalies indicating an interstellar origin or  formation in the cold, outer regions of the solar proto-planetary disc \citep[{e.g.},][]{2006Sci...312..727B}. This type of carbonaceous material is very close in nature to the hydrogenated amorphous carbons, a-C(:H), adopted in the THEMIS model and an intimate link has indeed been suggested between organic globules and the THEMIS large a-C:H/a-C core/mantle grains \citep{2016RSOS....360224J}. The THEMIS diffuse ISM dust model requires $\simeq 206-218$\,ppm of carbon in a-C(:H), with $\simeq 148-175$\,ppm and $\simeq 43-58$\,ppm of this in its H-poor a-C and H-rich a-C:H sub-components, respectively \citep{2017A&A...602A..46J}.

\subsection{Carbon in nano-diamonds}
\label{sect_C_nds}

In primitive meteorites the nano-diamonds are currently determined to be more abundant ($\simeq 1400$\,ppm) than the pre-solar silicates ($\gtrsim 200$\,ppm). However, it is not clear that all of these, perhaps anomalously-abundant, nano-diamonds are pre-solar grains and there has been much debate about their origins \cite[{e.g.},][]{1987ApJ...319L.109T,1989Natur.339..117L,1995GeCoA..59..115H,1996ApJ...463..344O,1996GeCoA..60.4853D,2011ApJ...738L..27S}. Given the findings of this work it is clear that free-flying nano-diamonds in the ISM cannot be as abundant, relative to silicate grains, as they are amongst the pre-solar grains because if they were they would easily outshine all other dust emission at IR wavelengths. However, if we now normalise the nano-diamond abundance (0.14\%) to that of the carbon in the organic matter ($3-4$\%) in primitive meteorites we estimate that of this carbonaceous matter some $3-5$\% is in the form nano-diamonds, which makes for a better comparison with our ISM nano-diamond abundance estimates ($\lesssim 7$\% of carbon in nano-diamonds). 

Thus, based on the present work, we can at best only provide an upper limit to the abundance of free-flying nano-diamonds, {i.e.}, they must contain $\lesssim 20$\,ppm of carbon. However, some nano-diamonds could still exist in the ISM but be un-observable because they are incorporated within much larger aggregate particles. This possibility will probably remain unresolvable because it is doubtful that nano-diamonds in aggregates will exhibit any observable features.

\subsection{Carbon in silicon carbide}
\label{sect_C_SiC}

Silicon carbide grains, SiC, are found at a level of $\simeq 50$\,ppm in pre-solar grains, which means that they contain $\lesssim 25$\,ppm of carbon in the ISM, {i.e.}, $\lesssim 8$\% of carbon in dust. It should be noted that carbon has not yet been observed in SiC grains in the diffuse ISM most probably because of dilution and confusion with other stronger signature IR bands, {e.g.}, the amorphous silicate $\sim 10$ and $\sim 18\,\mu$m bands will smother the equally broad and featureless SiC 11.3\,$\mu$m band.

\subsection{Carbon in graphite}
\label{sect_C_graphite}

Carbon in the form of graphite is found amongst the pre-solar grains incorporated into the Solar System but it is a rather rare material, originating from, principally, AGB stars and supernov\ae\ and present within primitive meteorites at a level $\simeq 10$\,ppm or $\lesssim 3$\% of carbonaceous dust and therefore of extremely low importance amongst the diffuse ISM dust components.

\subsection{Carbon in PAs and PAHs}
\label{sect_C_PAs}

Set as we are in our use of the THEMIS model, which has no need of PAHs in the strictest sense but does contain multi-aromatic domains as a key element of 3D contiguous carbon nano-particle network structures \cite[{e.g.}, see Figs. 3 and 4 of][]{2012ApJ...761...35M}, it is hard to estimate the fraction of carbon that could exist in strict PAHs or fullerenes. This is because in THEMIS the polycyclic aromatic (PA) fullerenes and the polycyclic aromatic hydrocarbons (PAHs) are assumed to form an integral component of the a-C nano-particles. Nevertheless, this model does not preclude the existence of free-flying PA(H) species, which are considered to be the break-down products of a-C nano-particle photo-dissociation \citep[{e.g.},][]{2012A&A...542A..98J,2012ApJ...761...35M,2013A&A...558A..62J,2015A&A...581A..92J,2016RSOS....360223J}. 

Although with THEMIS we cannot define a strict ``PAH" fraction, following \cite{2021A&A...649A..18G}, we can define the parameter $q_{AF}$, which is the mass fraction in the sub-1.5\,nm radius aromatic-rich, a-C nano-particles that are responsible for the infrared to mid-infrared emission bands ($3-15\,\mu$m). For the THEMIS diffuse ISM model 14\% of the carbon in dust ($\simeq 7$\% of the cosmic carbon) is in the emission band carriers. For comparison, dust models with an ``interstellar PAH" component generally require of the order of $10-15$\% of the carbon in dust to be in the form of PAHs \citep[{e.g.},][]{2004ApJS..152..211Z,2008ARA&A..46..289T,2014A&A...561A..82S,2021ApJ...917....3D}, which is in line with our estimate for the mass of carbon in nano-particles required to explain the emission bands with THEMIS.  

Nevertheless, we would remind the reader that in the strictest sense PAHs do not exhibit a $3.4\,\mu$m aliphatic band and note that the interstellar $3.3\,\mu$m emission band, attributed to an aromatic CH stretching mode, is almost always observed along with an adjacent suite of bands in the $3.4-3.6\,\mu$m region, of which the $3.4\,\mu$m aliphatic CH stretching band is usually the most prominent \citep[{e.g.},][]{1991ApJ...380..452T,1997ApJ...474..735S,2001A&A...372..981V,2004A&A...423..549D,2007A&A...463..635D,2012A&A...541A..10Y,2012ApJ...751L..18K}. Hence, most of the emission band emitters must be a-C(:H) materials rather than PAHs in the strictest sense of the definition. Clearly, attaching carbon chains to or super-hydrogenating a strict PAH makes it into a member of the large family of a-C(:H) nano-particles, which are not planar species but shell-like, self-folded structures, as pointed out above. 

Based upon the above discussion we hypothesise that it is probable that of the order of only one percent of the interstellar carbon in dust is likely to be found in strict, pure, planar and fully aromatic PAH molecules, or in fullerenes (PAs), in the diffuse ISM and that they are the end-of-the-road, top-down, photo-dissociation products of larger carbonaceous, a-C(:H), particles \citep[{e.g.},][]{2006FaDi..133..415D,2012A&A...542A..98J,2013A&A...558A..62J,2013ApJ...773...42O,2014ApJ...797L..30Z,2015A&A...581A..92J,2015ApJ...800L..33G,2016RSOS....360223J}.

\begin{table}
\caption{The likely partition of carbon in the diffuse ISM dust phases.}
\centering                          
\begin{tabular}{l c}        
\hline\hline \\[-0.25 cm]                 
Carbon phase & abundance  \\    
\hline \\[-0.35 cm]                        
 & \multicolumn{1}{l}{primitive meteorites [ \% ]} \\
\hline \\[-0.25 cm]                        
organic matter & 3-4  \\      
\hline \\[-0.35 cm]                        
 & \multicolumn{1}{l}{pre-solar grains [ ppm ]} \\
\hline \\[-0.25 cm]                        
nano-diamonds  & \ \ \ \ \ \ \ \ \ \ \ 1400 $\equiv 0.14$\% \\      
SiC  & 50  \\      
graphite  & 10  \\      

\hline \\[-0.35 cm]                        
 & \multicolumn{1}{l}{diffuse ISM abundances} \\
 & \multicolumn{1}{l}{ [ ppm and \% of C in dust] } \\
\hline \\[-0.25 cm]                        
a-C:H                      & \ \ $\gtrsim 60$\,ppm \ $\equiv \ \ \gtrsim 19$\% \\      
a-C                          & $\sim 150$\,ppm     \ $\equiv \ \ \sim 49$\%  \\      
nano-diamonds       & $\lesssim 20$\,ppm \ $\equiv \ \ \lesssim 7$\%  \\      
C in SiC                   & $\lesssim 25$\,ppm \ $\equiv \ \ \lesssim 8$\%  \\      
graphite                   & $\lesssim 10$\,ppm \ $\equiv \ \ \lesssim 3$\%  \\      
aromatic-rich a-C    & \ \ \ \, $\leq 5$\,ppm            \ $\equiv \ \ < 14$\%  \\      

\hline                                   
\end{tabular}
\tablefoot{These estimates of carbon in carbon-bearing dust phases in the diffuse ISM are based upon primitive meteorite abundances, the pre-solar grains extracted from primitive meteorites and THEMIS.}
\label{tab_C_in_dust}
\end{table}

\section{Conclusions}
\label{sect_conclusions}

We have derived the complex indices of refraction for nano-diamonds as a function of size, fully taking into account surface hydrogenation and the nature of diamond bulk material, which is important in the case of the larger particles with radii greater than a few tens of nm. For the latter case we consider the properties of low nitrogen diamond in its pristine and neutron-irradiated forms. 

Nano-diamonds have to date only been unequivocally detected in emission in less than a handful of sources ({e.g.}, HR\,4049, Elias\,1, and HD\,97048), which seems to argue against their ubiquity, unless it takes the especially extreme radiation environments, such as the X-ray flares in these objects, to make the nano-diamonds shine. 

If, however, nano-diamonds are indeed of extra-solar system origin, and observed in circumstellar disc regions,  then it would seem that they must have traversed the ISM to reach the solar system. Using the derived optical properties we show that nano-diamonds could in fact exist in the ISM in abundances compatible with their relative abundances in primitive meteorites and remain unobservable there. Thus, the question of the existence of nano-diamonds in the ISM remains an open one and suggests that there may be no need to form them in situ in the proto-planetary discs where they are observed. However, this possibility only shifts the sites of their formation to other and/or additional environments.

We conclude that, unlike in the few circumstellar sources, nano-diamonds are unlikely to be detectable in emission in the diffuse ISM through any of their infrared features because these will be drowned out by other carbonaceous dust emission features. Nevertheless, based upon the characteristics of bulk diamond, it appears that nano-diamonds could be observable through defect-related optical lines, which would then, unfortunately, be lost in and hard to disentangle from the forest of diffuse interstellar bands.

\begin{acknowledgements}
The authors wishes to thank Emilie Habart, Emmanuel Dartois and numerous other colleagues for interesting nano-diamond discussions.  \\ \\ 
APJ dedicates this work to Keith, his long-suffering brother of more than 63 years. Taken from us too soon he will remain forever in our hearts and minds. \\ Keith Edward Jones  ( 25$^{th}$ February 1957 $-$  29$^{th}$ October 2020 ) 
\end{acknowledgements}

\bibliographystyle{bibtex/aa} 
\bibliography{../../my_papers/Ant_bibliography} 

\appendix

\section{Nano-diamond heat capacities, $C_{\rm V}(T)$}
\label{appendix_C_V}

In Section \ref{sect_nd_ISM} we determined the nano-diamond contribution to the dust emission in the diffuse ISM (see Fig.~\ref{fig_nanod_THEMIS}), using the standard THEMIS model  \citep{2013A&A...558A..62J,2017A&A...602A..46J,2014A&A...565L...9K,2015A&A...577A.110Y}. In this environment nano-diamonds will be stochastically-heated and in calculating their temperature distributions we need to know their heat capacities, $C_{\rm V}(T)$. This appendix indicates how these were estimated.

It must by this point be clear that the physical properties of nano-diamonds can differ significantly from those of bulk diamond. It is therefore perhaps not surprising that their laboratory-measured heat capacities do not match those of bulk diamond.

\begin{figure*}
\centering
\includegraphics[width=9cm]{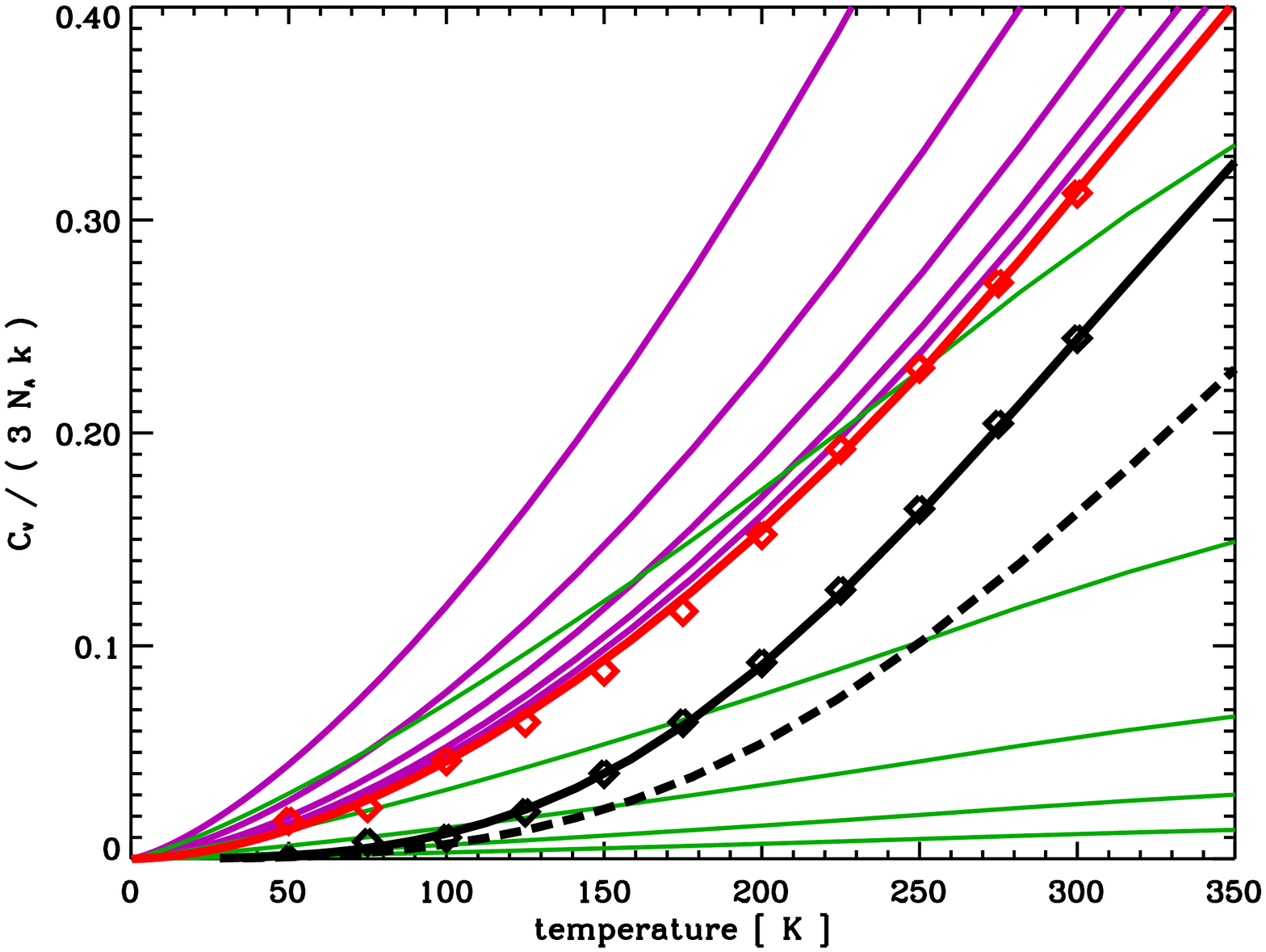}
\includegraphics[width=9cm]{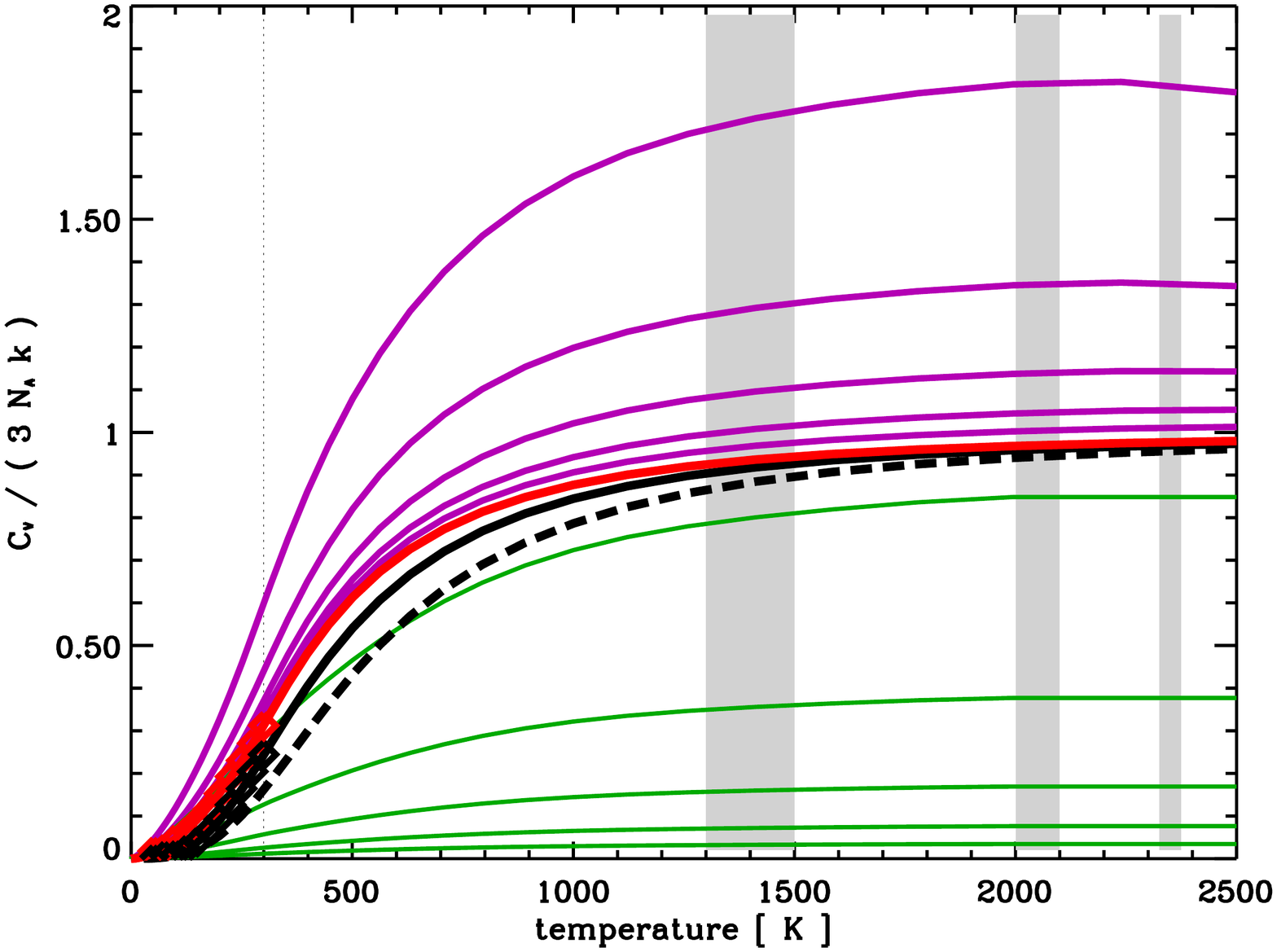}
\caption{Nano-diamond heat capacities, $C_{\rm V}(T)$, plotted as a function of temperature in the dimensionless units $C_{\rm V}(T)/(3 \, {\rm N_{\rm A}} \, {\rm k})$ where ${\rm N_{\rm A}}$ is the Avogadro number and k is the Botlzmann constant. The dashed black line shows the $C_{\rm V}(T)$ data for diamond assuming a standard Debye temperature, $T_{\rm D}$ of 2250\,K. The $50-300$\,K data points (left hand zoom plot) delineate the laboratory-measured (nano-)diamond heat capacities, as shown by  \cite{Vasiliev_etal_2010_R,Vasiliev_etal_2010,Vasiliev_etal_2011}, for detonation nano-diamonds (red diamonds) and synthetic diamond (black diamonds); the latter fitted with $T_{\rm D} = 1875$\,K (black line). The fit to the detonation nano-diamond data (red line) is described in the text. The other $C_{\rm V}(T)$ data shown are for: fully surface-hydrogenated nano-diamonds with radii of  0.54, 1.20, 2.66, 5.88, and  13.01\,nm (from top to bottom, solid purple lines). The green lines show the size-dependent contribution of the surface H atoms.}
\label{fig_C_V}
\centering
\end{figure*}

From solid state phonon physics and thermodynamical considerations the heat capacity of a material can be determined using the Debye model, which, expressed in dimensionless units, is 
\begin{equation}
\frac{C_{\rm V}}{(3 \, {\rm N_{\rm A}} \, {\rm k})} = 3 \left( \frac{T}{T_{\rm D}} \right)^3 \int_0^{T/T_{\rm D}} \frac{x^4 e^x}{(e^x-1)^2} \ dx
\label{eq_debeye}
\end{equation}
where ${\rm N_{\rm A}}$ is the Avogadro number, k is the Botlzmann constant and $T_{\rm D}$ is the Debye temperature. In the high-temperature limit ($T \gg T_{\rm D}$) $C_{\rm V}(T)/(3 \, {\rm N_{\rm A}} \, {\rm k}) = 1$ and in the low-temperature limit ($T \ll T_{\rm D}$) 
\begin{equation}
\frac{C_{\rm V}}{(3 \, {\rm N_{\rm A}} \, {\rm k})} = \frac{4 \pi^4}{5} \left( \frac{T}{T_{\rm D}} \right)^3. 
\label{eq_debeye_lowT}
\end{equation}
We remind the reader that this model is based on the phonon properties of solid materials and, as is now clear, the physical properties of these same materials, when incorporated into particles at nanometre scales, can diverge from those of the bulk. Currently there appear to be very few measurements of the heat capacity of diamond at nm sizes. One such set of work presents nano-diamond heat capacity measurements at low temperatures \citep[$50-300$\,K,][see Fig.\,\ref{fig_C_V}]{Vasiliev_etal_2010_R,Vasiliev_etal_2010,Vasiliev_etal_2011}. From the data in Fig.\,\ref{fig_C_V} we can indeed see that the heat capacities of detonation nano-diamonds and synthetic diamond \citep{Vasiliev_etal_2010_R,Vasiliev_etal_2010} differ significantly from that of natural diamond (from Eq.\,(\ref{eq_debeye}) with $T_{\rm D}$ of 2250\,K, dashed black line). The measured heat capacity of synthetic diamond (black diamonds) can still be fit with a Debye model, albeit with a Debye temperature of $T_{\rm D} = 1875$\,K (solid black line). However, that for detonation nano-diamonds (red diamonds) does not follow the Debye model $T^3$ dependence at low temperatures but can be fit with an empirically-modified Debye model as we now show. 

In fitting the detonation nano-diamond heat capacity of  \cite{Vasiliev_etal_2010_R} and \cite{Vasiliev_etal_2010}, shown in Fig.\,\ref{fig_C_V}, we adopt the standard bulk diamond density $\rho = 3.52$\,g\,cm$^{-3}$, a Debye temperature of $T_{\rm D} = 1650$\,K at high temperatures, a modified power-law dependence ($\frac{7}{4}$ instead of 3) in the low-temperature regime ($T \leq 300$\,K) and make the assumption that the nano-diamonds are not surface-hydrogenated. Thus, Eq.\,(\ref{eq_debeye_lowT}) becomes 
\begin{equation}
\frac{C_{\rm V,C}({\rm low \, T})}{(3 \, {\rm N_{\rm A}} \, {\rm k})} = 0.0795 \ \frac{4 \pi^4}{5} \left( \frac{T}{T_{\rm D}} \right)^{\frac{7}{4}},  
\label{eq_debeye_mod}
\end{equation}
where the normalising factor 0.0795 ensures a smooth transition to a high-temperature ($T > 300$\,K) Debye model heat capacity, $C_{\rm V,C}({\rm high \, T})/(3 \, {\rm N_{\rm A}} \, {\rm k})$, using Eq.\,(\ref{eq_debeye_lowT}) with $T_{\rm D} = 1650$\,K. This fit is the red line in the plots in Fig.\,\ref{fig_C_V}, which indicates an increased heat capacity for nano-diamonds with respect to both synthetic and natural diamond. 

Following \cite{2020_Jones_nd_CHn_ratios} we assume that nano-diamond surfaces are passivated with CH and CH$_2$ groups  and that we therefore need to take their contribution to the heat capacity into account, which can be significant for the smaller particles. For example, for the nano-diamond radii shown in Fig.\,\ref{fig_C_V}, {i.e.}, $a_{\rm nd} =$ 0.54, 1.20, 2.66, 5.88, and  13.01\,nm, and assuming full CH$_n$ surface-passivation, the [H]/[C] abundance ratios are 0.88, 0.39,  0.18, 0.08, and 0.04, respectively. In other words, for small nano-diamonds ($a \lesssim 0.5$\,nm) half, or more, of the atoms constituting the particle are H atoms. In calculating their contribution to the heat capacity we approximate the nano-diamond CH bond heat capacity following the approach of  \cite{1997ApJ...475..565D} in their appendix A.1 for CH bonds, in their group A designation, on the periphery of polycyclic aromatic hydrocarbons. 
 
We now define some key nano-diamond surface and bulk atom properties. The number of carbon atoms per nano-diamond is 
 \begin{equation}
N_{\rm C} =  \frac{ \frac{4}{3} \pi a_{\rm nd}^3  \, \rho }{ 12 \, m_{\rm H} }, 
\end{equation}
where $m_{\rm H}$ is the mass of a hydrogen atom. We take the surface [CH]/[CH$_2$] ratio from \citep{2020_Jones_nd_CHn_ratios}, {i.e.},  
\begin{equation}
\frac{[{\rm CH}]}{[{\rm CH_2}]} = 2.265  \, a_{\rm nd}^{0.03} - \left( \frac{1}{ 2.5 \, a_{\rm nd}}  \right) = \Theta,
\end{equation}
where the radius, $a_{\rm nd}$, is expressed in nm, and the CH${_n}$ fractions are 
\begin{equation}
f_{\rm CH} = \frac{\Theta}{1+\Theta} \ \ \ \ \ \ \ \ \ {\rm and} \ \ \ \ \ \ \ \ \ f_{\rm CH_2} = \frac{1}{1+\Theta}.
\end{equation}
The number of H atoms on a nano-diamond surface is given by 
\begin{equation}
N_{\rm CH} = \pi a_{\rm nd}^2 \, \frac{ ( f_{\rm CH} + 2 f_{\rm CH_2} ) }{ ( f_{\rm CH} \, A_{\rm CH} + f_{\rm CH_2} \, A_{\rm CH_2} ) },  
\end{equation}
where $A_{\rm CH}   = 0.0412$\,nm$^2$ and $A_{\rm CH_2}  = 0.0635$\,nm$^2$ are the surface areas of CH and CH$_2$ groups on nano-diamonds, respectively \citep{2020_Jones_nd_CHn_ratios}.

Using the methodology of \cite{1997ApJ...475..565D} the CH bond contribution to $C_{\rm V}(T)$ in the temperature range $T = 300-3000$\,K, in dimensionless units, can be approximated by 
\begin{equation}
\frac{C_{\rm V,CH}({\rm high}\,T)}{(3 \, {\rm N_{\rm A}} \, {\rm k})} = \frac{ 10^7 }{ ({\rm 3 \, N_A} \, {\rm k} )} \ \sum_{n=0}^{6} c_n \, T^n, 
\label{eq_CV_CH_1}
\end{equation}
 with the coefficients $c_n = \{ -1.23, 4.9 \times 10^{-2}, 2.07 \times 10^{-5}, -6.93 \times 10^{-8}, 4.85  \times 10^{-11}, -1.44  \times 10^{-14}, 1.57  \times 10^{-18} \}$. At low temperatures ($T < 300$\,K) we connect to the heat capacities from Eq.\,(\ref{eq_CV_CH_1}) using an empirically-modified version of Eq.\,(\ref{eq_debeye_lowT}), {i.e.}, 
\begin{equation}
\frac{C_{\rm V,CH}({\rm low}\,T)^\prime}{(3 \, {\rm N_{\rm A}} \, {\rm k})} = 0.06 \ \frac{4 \pi^4}{5} \left( \frac{T}{T_{\rm D}} \right)^{\frac{5}{4}},  
\label{eq_CV_CH_2}
\end{equation}
with $T_{\rm D} = 1650$\,K and 0.06 a normalisation factor to enable a smooth fit with the high-temperature part. Note that in this case the low-temperature power law is rather flat, {i.e.}, $\frac{5}{4}$, compared to $\frac{7}{4}$ for the carbon in detonation nano-diamonds and the classical value of 3 for the Debye model (Eq.\,\ref{eq_debeye_lowT}). This CH bond heat capacity is then multiplied by the H/C ratio and scaled by a factor of 0.59 to ensure that the maximum value of the dimensionless heat capacity ($C_{\rm V}(T)/(3 \, {\rm N_{\rm A}} \, {\rm k})$) for surface H atoms does not exceed unity. We therefore have  
\begin{equation}
\frac{C_{\rm V,CH}({\rm low}\,T)}{(3 \, {\rm N_{\rm A}} \, {\rm k})} = 0.59 \ \frac{N_{\rm CH}}{N_{\rm C}} \times \frac{C_{\rm V,CH}({\rm low}\,T)^\prime}{(3 \, {\rm N_{\rm A}} \, {\rm k})},    
\label{eq_CV_CH_3}
\end{equation}
as indicated by the green lines in Fig.\,\ref{fig_C_V} for nano-diamond radii of 0.54, 1.20, 2.66, 5.88, and  13.01\,nm, from top to bottom. The total nano-diamond heat capacity, $C_{\rm V,nd}(T)/(3 \, {\rm N_{\rm A}} \, {\rm k})$, is then the sum of the matched low-temperature and high-temperature (${\rm low|high}$) C atom and CH bond contributions, {i.e.}, 
\begin{equation}
\frac{C_{\rm V,nd}(T)}{(3 \, {\rm N_{\rm A}} \, {\rm k})} =  \frac{C_{\rm V,C}({\rm low|high}\,T)}{(3 \, {\rm N_{\rm A}} \, {\rm k})} + \frac{C_{\rm V,CH}({\rm low|high}\,T)}{(3 \, {\rm N_{\rm A}} \, {\rm k})}, 
\label{eq_CV_nd}
\end{equation}
which are shown by the purple lines in Fig.\,\ref{fig_C_V} for fully surface-hydrogenated nano-diamond radii of 0.54, 1.20, 2.66, 5.88, and  13.01\,nm, from top to bottom. It should be noted that in this figure the dimensionless heat capacities for hydrogenated nano-diamonds clearly exceed unity. This apparent contradiction with the Debye model is because the surface H atoms add extra capacity to the thermal properties and are not counted in the total number of atoms in the particle, which is normalised to the number C atoms in the nano-diamond particle.

\end{document}